\documentclass[aps, pre, a4paper, floatfix,  twocolumn, longbibliography, footinbib, amssymb]{revtex4-2}

\usepackage[colorlinks=true, hypertexnames=false]{hyperref}
\PassOptionsToPackage{linktocpage}{hyperref} 
\hypersetup{
  colorlinks  = true, 
  urlcolor = magenta!80!black,
  linkcolor  = blue!60!black, 
  citecolor  = blue!60 
}

\usepackage[T1]{fontenc}
\usepackage[utf8]{inputenc}
\usepackage{epsfig}
\usepackage[english]{babel}
\usepackage[table]{xcolor}
\usepackage{graphicx}
\usepackage{amsmath}
\usepackage{relsize}
\usepackage[percent]{overpic}
\usepackage{bm}
\usepackage{hyperref}
\usepackage{siunitx}
\usepackage{accents}
\usepackage{mathtools}
\usepackage{grffile}
\usepackage[normalem]{ulem}
\usepackage{pgf}
\usepackage{tikz}
\usepackage{times}
\usepackage{makecell}

\usepackage[capitalize]{cleveref}

\usepackage{dsfont}  

\newcommand{\avg}[1]{{\left<#1\right>}}

\newcommand{\dd}{\mathrm{d}}
\newcommand{\ee}{\mathrm{e}}

\usepackage{mathtools}
\def\multiset#1#2{\ensuremath{\left(\kern-.3em\left(\genfrac{}{}{0pt}{}{#1}{#2}\right)\kern-.3em\right)}}

\usepackage{amsmath}

\newcommand{\A}{\bm{A}}

\newcommand{\bb}{\bm{b}}

\newcolumntype{P}[1]{>{\centering\arraybackslash}p{#1}}

\usepackage{verbatim}
\usepackage{overpic}
\usepackage{booktabs}
\usepackage{placeins}

\usepackage{siunitx}
\usepackage{multirow}

\begin{document}

\title{Implicit models, latent compression, intrinsic biases, and cheap
lunches in community detection}

\author{Tiago \surname{P. Peixoto}}
\email{peixotot@ceu.edu}
\affiliation{Department of Network and Data Science, Central European University, Vienna, Austria}

\author{Alec \surname{Kirkley}}
\email{alec.w.kirkley@gmail.com}
\affiliation{Institute of Data Science, University of Hong Kong, Hong Kong}
\affiliation{Department of Urban Planning and Design, University of Hong Kong, Hong Kong}
\affiliation{Urban Systems Institute, University of Hong Kong, Hong Kong}

\begin{abstract}
The task of community detection, which aims to partition a network into
clusters of nodes to summarize its large-scale structure, has spawned
the development of many competing algorithms with varying
objectives. Some community detection methods are inferential, explicitly
deriving the clustering objective through a probabilistic generative
model, while other methods are descriptive, dividing a network according
to an objective motivated by a particular application, making it
challenging to compare these methods on the same scale. Here we present
a solution to this problem that associates any community detection
objective, inferential or descriptive, with its corresponding implicit
network generative model. This allows us to compute the description
length of a network and its partition under arbitrary objectives, providing
a principled measure to compare the performance of different
algorithms without the need for ``ground truth'' labels. Our approach
also gives access to instances of the community detection problem that
are optimal to any given algorithm, and in this way reveals intrinsic
biases in popular descriptive methods, explaining their tendency to
overfit. Using our framework, we compare a number of community detection
methods on artificial networks, and on a corpus of over 500 structurally
diverse empirical networks. We find that more expressive community
detection methods exhibit consistently superior compression performance
on structured data instances, without having degraded performance on
a minority of situations where more specialized algorithms perform
optimally. Our results undermine the implications of the ``no free
lunch'' theorem for community detection, both conceptually and in
practice, since it is confined to unstructured data instances, unlike
relevant community detection problems which are structured by
requirement.
\end{abstract}

\maketitle

\section{Introduction}

Community detection methods~\cite{fortunato_community_2010} are a
cornerstone of network data analysis. They fulfill the need to digest an
otherwise intractable large-scale structure of a complex system into a
simpler coarse-grained description, where groups of items are clustered
together according to shared patterns of interactions. This
methodological ansatz has proved useful in countless applications in
biology, physics, engineering, computer science, the social sciences,
and other fields.

The research on community detection has evolved substantially in the
last 20 years~\cite{fortunato_20_2022}, spawning a large variety of
different approaches. Substantial effort in this area has been devoted
to the development of methods that behave well in practice --- both in
the quality of results and algorithmic efficiency
--- as well as to our theoretical understanding of their
behavior~\cite{fortunato_community_2016}. Despite these advances, what
perhaps continues to be one of the biggest difficulties when employing
community detection methods in practice is that the task itself is not
uniquely defined: what constitutes a good coarse-graining of a network
is intrinsically tied to an ultimate objective, of which there can be
many~\cite{schaub_many_2017}, resulting in algorithms that yield
different answers for the same
network~\cite{hric_community_2014,ghasemian_evaluating_2019}.

Most methods agree qualitatively on what constitutes community structure
--- groups of nodes that are more connected with themselves than with
the rest of the network, or more generally, groups of nodes that have
the same tendency of connecting to other groups of nodes --- but the
context in which this concept is evoked and the resulting mathematical
definitions can vary substantially, to the point where two algorithms
can yield radically different partitions of the same network despite
sharing an overall conceptual agreement~\footnote{In principle, this should not
constitute an obstacle, as one would need only to match the most
appropriate algorithm to a given objective supplied by the practitioner
within the context of a particular application. However, due to their
qualitative similarity, users often expect universal algorithms that
work well independently of context
--- an attitude which is also reflected on a variety of works that
benchmark competing methods against the same criterion, such as
recovering planted community structure in artificial
networks~\cite{lancichinetti_benchmark_2008,lancichinetti_benchmarks_2009},
prediction of node covariates~\cite{hric_community_2014}, or of missing
links~\cite{ghasemian_evaluating_2019}, regardless of their divergences
in motivation.}.

In order to better understand the discrepancies and similarities between
community detection methods, it is useful to divide them into two classes,
according to their stated objectives: inferential and
descriptive~\cite{peixoto_descriptive_2022}. Inferential methods evoke
explicitly the notion of probabilistic generative models,
i.e.  network formation mechanisms that define how a division of the
network into groups affect the probability with which the nodes are
connected. In this setting, the community detection task consists of
assuming that an observed network is an instance of this generative
procedure, and attempting to fit it to data in order to infer the hidden
partition --- or more generally, a set of partitions ranked according to
their posterior plausibility~\cite{peixoto_bayesian_2019}. In this
scenario, it is possible to assess the statistical significance and
uncertainties of our inferences, and to quantify precisely how
parsimonious the obtained coarse-grained representation is, allowing us
to detect overfitting and underfitting, as well as to perform model
selection. Furthermore, from the fitted model it is possible to make
statements about edge placement probabilities and to make
generalizations about unobserved
data~\cite{guimera_missing_2009,peixoto_reconstructing_2018}.

Descriptive methods, conversely, do not involve an explicit definition
of a generative model, and divide the network into groups according to
other, application-specific criteria. What are perhaps the oldest
instances of this class of methods are the various algorithms for graph
partitioning in computer science~\cite{catalyurek_more_2022}, motivated
in large part by circuit design and task scheduling problems, instead of
data analysis. In this setting, the desired network division is the one
that optimizes a task conditioned on a given network
--- such as the spatial placement of transistors or division of tasks
among processors.  Prominent descriptive methods also use network
clustering to characterize the behavior of dynamical processes that run
on the network, typically random walks. For example, the Infomap
method~\cite{rosvall_maps_2008} clusters nodes in a manner that
minimizes the information required to encode a random walk taking place
on a network, according to how often it leaves and enters individual
groups. In this case, the network is a parameter of a dynamical process,
and therefore its generation is not modelled
directly~\footnote{Therefore, if we take its stated objective at face
value, when a method such as Infomap clusters a maximally random network
into many groups, as it is prone to
do~\cite{lancichinetti_community_2009,kawamoto_comparative_2018}, it is
not meaningful to describe this as overfitting, since no model fit is
nominally being attempted. Indeed, if a random graph is sufficiently
sparse, then a random walk may genuinely get trapped into quenched
random structures, such as groups of nodes that are more internally
connected by chance alone~\cite{guimera_modularity_2004}, or other
structures such as dangling trees~\cite{krzakala_spectral_2013}, which
could be well characterized by the network division found. This is
precisely what these methods set out to identify, and whatever
consternation this may cause in a particular application likely
indicates a mismatch between the stated objective of the method and what
is in fact desired or more appropriate in context, instead of a problem
with the method itself.}. Arguably the most popular community detection
method, modularity maximization~\cite{newman_modularity_2006}, can also
be classified as descriptive. Although it was originally motivated
according to an explicit inferential criterion
--- namely the deviation from a null model
--- it is inconsistent with this
stated goal, since it notoriously finds spurious deviations on networks
sampled from its own null model~\cite{guimera_modularity_2004}. Despite
an approximate equivalence with the parametric inference of a restricted
version of the stochastic block model
(SBM)~\cite{newman_equivalence_2016}, valid only when the true number of
groups is known and the data obeys certain
symmetries~\cite{zhang_statistical_2020}, this method lacks an explicit
inferential interpretation in the nonparametric manner it is actually
employed in practice. For these and other descriptive methods in general, the
notions of uncertainty and statistical significance are not inherent or
explicitly evoked.

Despite these clear differences in stated objectives, descriptive
methods are often used in practice with inferential aims. For example,
communities found with descriptive methods are frequently interpreted as
being the result of a homophilic edge formation mechanism in social
networks~\cite{tyler_email_2003,
yuta_gap_2007,zhang_community_2008,red_comparing_2011,traud_social_2012}
and functional modules in biological networks~\cite{spirin_protein_2003,
chen_detecting_2006,lewis_function_2009,wilkinson_method_2004}, to name
only a few analyses in which a concern for statistical significance is
expressed. Furthermore, attempts to benchmark
community detection methods against each other typically involve
comparing their performance in terms of their ability at recovering
known partitions in artificial random networks sampled from generative
models --- thus being clearly an inferential criterion
--- as is the case of the popular LFR
benchmark~\cite{lancichinetti_benchmark_2008,lancichinetti_community_2009}. More
recently, the tendency of algorithms to under- or overfit in a link
prediction task was considered in Ref.~\cite{ghasemian_evaluating_2019},
which relies on a manifestly inferential criterion. One also finds
prominent claims in the
literature~\cite{fortunato_community_2010,fortunato_community_2016,newman_modularity_2006}
that it would be undesirable for an arbitrary community detection method
to cluster a maximally random network sampled either from the
Erd\H{o}s-Rényi or configuration model into more than one group, since
this division would unveil purely random fluctuations in the placement
of the edges, and thus would amount to overfitting. Because of this,
very often results of descriptive community detection methods are
compared to what is obtained with randomized versions of the data, in an
attempt to quantify statistical significance~\cite{reichardt_when_2006}.
Such a comparison with a ``null model'' is evidently an inferential
concern, since it amounts to assessing the generative process underlying
the network formation.

\begin{figure*}[t]
  \includegraphics[width=\textwidth]{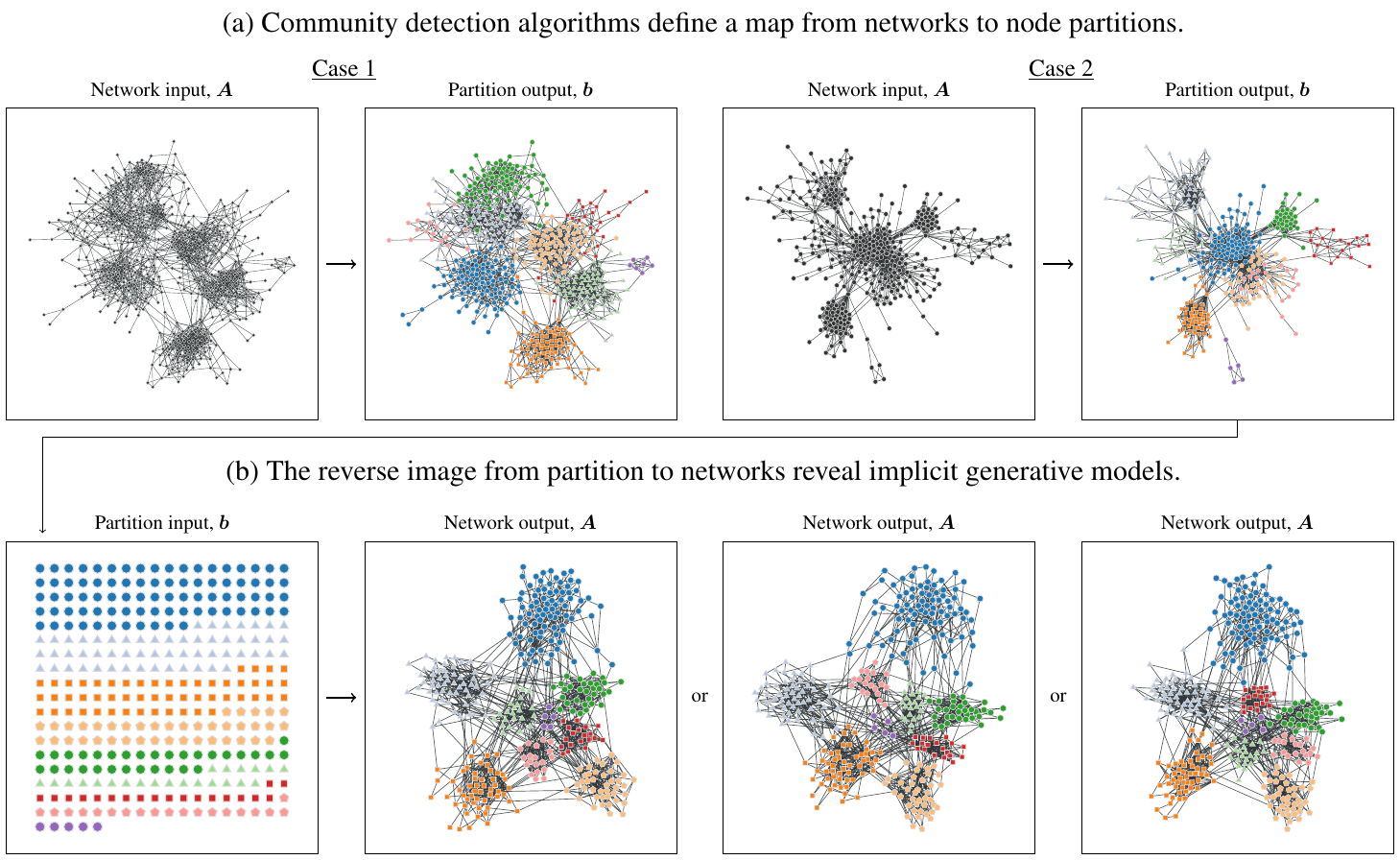}
  \caption{Diagrammatic illustration of the inverse problem we consider in this
    work. (a) A community detection algorithm provides a mapping
    $\hat\bb(\A)=\bb$ of a network $\A$ to a partition $\bb$ of its nodes. (b)
    This mapping can always be inverted, such that for any given partition $\bb$
    of the nodes we can consider the set of all networks $\A$ such that
    $\hat\bb(\A)=\bb$. This set of networks reveals the implicit generative
    model compatible with the community detection algorithm under consideration
    (we show three independent samples drawn uniformly from this set). The arrow
    from case 2 in panel (a) to (b) indicates that the same partition is
    considered in both examples. The networks generated by the implicit model in
    panel (b) are markedly different from the network in case 2 in (a), which
    would be generated only with a very low probability under this model. This
    happens because the mixing between groups tends to be homogeneous for
    networks sampled from the model, whereas in the network in case 2 in (a) the
    groups connect preferentially to a central group (in blue) and they have
    more heterogeneous densities. This mismatch indicates that the underlying
    model is in fact a poor representation of the network structure --- which
    would be impossible to determine from the results of panel (a) alone.
    Therefore, characterizing the implicit models hidden behind community
    detection methods allows us to evaluate their ability to faithfully capture
    network structure in a systematic manner and reveal their intrinsic biases towards 
    particular kinds of structure. This framing also allows us to compare different 
    community detection methods on equal grounds according to their latent compression of the data
    --- i.e., their description length for a given network $\bm{A}$ 
    and partition $\bm{b}$ --- which is a universal model selection criterion that
    removes the need for ground truth labels for method comparison. \label{fig:diagram}}
\end{figure*}

We explore in this work the fact there is no formal mathematical distinction
between inferential and descriptive methods, since, as we show, the definition
of any community detection model necessarily implies the existence of an
implicit generative model which yields an inference procedure identical to any
descriptive approach (see Fig.~\ref{fig:diagram} for an illustration of our
framework). The characterization of these implicit generative models allows us
to perform Bayesian model comparisons and assess statistical significance,
closing the gap between inferential and descriptive methods by evaluating
descriptive methods from a generative perspective. The same implicit generative
models also unveil the intrinsic biases present in arbitrary community detection
methods --- in other words, what kind of structure they expect to encounter a
priori, even when this is not explicitly articulated in the motivation of the
method --- which cause over- or underfitting of the data. Furthermore, we show
how we can use our method to appropriately tune parameters of algorithms to
mitigate these biases, simultaneously removing existing resolution limits and
the identification of spurious communities in maximally random networks from
arbitrary community detection methods.

Here we show that a broad class of methods, which includes the widely used 
modularity maximization~\cite{newman_modularity_2006} and
Infomap~\cite{rosvall_maps_2008}, are equivalent to special cases of an
assortative SBM with groups having uniform size and density (a.k.a. the
planted partition model~\cite{condon_algorithms_2001}), where the number
of groups and assortativity strength are determined directly by the
expected value of the quality function. We show that the prior
distribution of the number of groups is typically bimodal, concentrating
simultaneously on a low and a large value, where the latter is on the
order of the number of nodes in the network. This bimodality induces
discontinuous transitions in the statistical properties of typical
problem instances, preventing networks with moderate community structure
and a wide range of the number of groups from being generated. This a
priori bias towards particular kinds of uniform, but strong, community
structure gives new clarity to the observed behavior of these methods in
practice, and their tendency to find communities of equal size and
density and in maximally random networks.

Our framework allows us to perform a comparison between algorithms in their
capacity of uncovering community structure sampled from instances that are
optimal for a different method. In particular, we consider optimal instances
generated by modularity and the nested stochastic block model (NSBM) --- a more
expressive, hierarchical parametrization of the SBM which is a priori agnostic
about the actual mixing patterns between groups. We demonstrate that ---
according to compressibility and accuracy in community recovery --- there are
substantial asymmetries between methods, where the more general NSBM does just
as well (but no better) for instances that are optimal for the other algorithms,
but where other algorithms perform significantly worse on instances that are
optimal for the NSBM. We also perform a systematic comparison of methods on a
corpus of over 500 diverse empirical networks, finding that the NSBM provides a
better compression for the vast majority of them. This provides evidence for
``cheap lunches'' in community detection --- more versatile, but appropriately
regularized approaches tend to yield systematically better results over
structured problem instances. This result reveals a practical and conceptual
caveat to the ``no free lunch'' (NFL) theorem for community
detection~\cite{peel_ground_2017}, which states that when averaged over all
instances of ``community detection problems'' (i.e. arbitrary pairings of a
network and a node partition) all conceivable algorithms must yield the same
performance. This is because the majority of possible problems are unstructured
instances where community labels have no correlation with network structure.

This paper is divided as follows. We begin in Sec.~\ref{sec:generative}
with a discussion of implicit models for community detection algorithms,
describing how to compute their corresponding description lengths and
implicit priors analytically for a broad class of methods.
In Sec.~\ref{sec:optimal-instances}, we
follow up on this discussion by demonstrating the correspondence between
this class of community detection objectives and restricted instances
of stochastic block models, showing that these methods implicitly assume
networks with very limited structure. Next, in
Sec.~\ref{sec:cheap-lunches} we discuss how our framework can provide
insights into the shortcomings of the NFL theorem for community
detection when applied to structured problem instances by revealing
asymmetries in algorithm compression performance. Finally, in
Sec.~\ref{sec:empirical} we apply our method to compare the description
lengths associated with fitting a range of community detection
algorithms to a wide variety of empirical networks, finding that a small
number of more expressive algorithms have systematically better
performance, and that in the small minority of cases where a more
specialized algorithm yields better performance, its result does not
deviate substantially from what is obtained with the more general
approach. We finalize in Sec.~\ref{sec:conclusion} with a discussion.

\section{Generative models from community detection methods}\label{sec:generative}

Let us consider an arbitrary deterministic community detection algorithm that
for a given network finds a unique partition of its nodes into nonoverlapping
communities. More formally, such an algorithm defines a specific mapping
\begin{equation}
  \hat\bb(\A)=\bb
\end{equation}
of a network $\A$ to a partition $\bb$ of its nodes, where $\A=\{A_{ij}\}$ is
the adjacency matrix of an undirected simple graph of $N$ nodes, with entries
$A_{ij}\in \{0,1\}$, and $\bb=\{b_i\}$ is a partition of the nodes into $B$
groups, with $b_i\in [1,\dots,B]$ being the group membership of node $i$.

Our central observation (illustrated in Fig.~\ref{fig:diagram}) is that any such
mapping can be inverted, so that for some partition $\bb$, we can consider the
set of all possible networks $\A$ that obey $\hat\bb(\A)=\bb$ --- i.e., all
networks that yield a given partition $\bb$ as the result of the community
detection algorithm being considered. Selecting between these networks uniformly
at random defines a precise generative model with probability
\begin{equation}
  P(\A | \bb) = \frac{\delta_{\hat\bb(\A),\bb}}{Z(\bb)},
\end{equation}
where $Z(\bb)=\sum_{\A}\delta_{\hat\bb(\A),\bb}$ counts all networks in this
set. According to this model, the original community detection algorithm can be
equivalently formulated as a maximum a posteriori (MAP) estimate of the
following posterior distribution:
\begin{equation}
  P(\bb | \A) = \frac{P(\A | \bb)P(\bb)}{P(\A)} = \delta_{\bb,\hat\bb(\A)},
\end{equation}
with $P(\bb)$ being any nonzero prior probability. Trivially, we have that
\begin{equation}
  \hat\bb(\A)=\underset{\bb}{\operatorname{arg\ max}}\; P(\bb | \A).
\end{equation}
Therefore, there is no mathematical distinction between performing a Bayesian
inference of this implicit model and whatever procedure motivates the original
community detection algorithm in the first place.

Because of this general equivalence, if we interpret the results of any
community detection algorithm in an inferential way (e.g. by assuming that
communities capture homophily, or any statistically significant structure), then
we are unavoidably incorporating in our analysis the generative assumptions that
are inherent to this implicit model.

Unfortunately, although such implicit models must always exist for any
conceivable community detection algorithm, they may be challenging to
characterize explicitly, requiring a computationally expensive inversion
procedure, which in the worst case needs to be performed exhaustively. This
poses an outstanding problem, since otherwise it becomes impossible to evaluate
the hidden inferential assumptions associated with a particular method.

In this work, we make substantial progress with this inverse problem by
considering a representative subset of community detection algorithms that are
based on the maximization of an arbitrary quality function
$W(\A, \bb) \in \mathbb{R}$,
\begin{equation}\label{eq:Wopt}
  \hat\bb(\A) = \underset{\bb}{\operatorname{arg\ max}}\; W(\A, \bb).
\end{equation}
(Some community detection methods, including the popular label propagation and
its variants~\cite{raghavan_near_2007}, are neither deterministic nor involve
explicit quality functions, but nevertheless can also be cast into our reverse
inferential framework. See Appendix~\ref{app:others} for a discussion.)

In this case, a direct connection with an inference procedure is obtained by
noting that the above optimization is equivalent to a MAP estimate of the
following family of posterior distributions:
\begin{equation}
  P(\bb|\A,g) = \frac{\ee^{g(W(\A,\bb))}}{Z(\A,g)},
\end{equation}
with $Z(\A,g)=\sum_{\bb}\ee^{g(W(\A,\bb))}$ being a normalization
constant, and where $g(x)$ is any function that preserves the
optimization, i.e.
\begin{equation}
  \underset{\bb}{\operatorname{arg\ max}}\; g(W(\A, \bb)) = \underset{\bb}{\operatorname{arg\ max}}\; W(\A, \bb),
\end{equation}
for every $\bm A$, which in general means that $g(x)$ needs to be
invertible and strictly increasing. Going one step further, we observe
that the above posterior can be obtained from a general joint
distribution given by
\begin{equation}\label{eq:general}
  P(\A,\bb|g,f) = \frac{\ee^{g(W(\A,\bb)) + f(\A)}}{Z(g,f)},
\end{equation}
with $Z(g,f)=\sum_{\A,\bb}\ee^{g(W(\A,\bb))+f(\A)}$, and $f(\A)$ being
an arbitrary weight attributed to a given network, independent of how
its nodes are partitioned.

The above shows us that, although the quality function $W(\A,\bb)$ imposes very
particular constraints on the generative models that are compatible with it ---
specifically how the partitions can affect the network structure --- they are by
no means unique, since they are constrained only up to an invertible function
$g(x)$ and an arbitrary partition-independent weight $f(\A)$. Therefore, at
least at first, it seems that both $g(x)$ and $f(\A)$ are ``free'' modelling
choices that are not directly specified by the quality function $W(\A,\bb)$.
This is analogous to how descriptive statistics on numeric data such as the
population mean can serve as sufficient statistics for the estimation of
parameters of different generative models, e.g. the mean of geometric and
Poisson distributions in the case of non-negative integers.

However, there are two fundamental points that we can make to resolve this
degeneracy. First, as we demonstrate in Appendix~\ref{app:g_x}, the distribution
of Eq.~\ref{eq:general} is asymptotically invariant to any choice of the
function $g(x)$, since it becomes equivalent to the microcanonical ensemble given by
\begin{equation}
  P(\A,\bb|f) = \frac{\delta_{g(W^{*}),g(W(\A,\bb))}\ee^{f(\A)}}{Z(f)},
\end{equation}
for some value $W^{*}$, which clearly does not depend on how $g(x)$ is chosen.

Secondly, when considering a potential degeneracy of this kind, a reasonable
starting point is to consider all compatible generative models on equal footing.
We can formalize this lack of additional information about the data generating
process by employing the principle of maximum
entropy~\cite{jaynes_probability_2003}, subject to a minimal set of constraints.
Considering the expected value of the quality function itself as the only
parameter of the model, i.e.
\begin{align}
  \sum_{\A,\bb}W(\A,\bb)P(\A,\bb|g,f) &= \avg{W},
\end{align}
and maximizing the entropy $-\sum_{\A,\bb}P(\A,\bb)\ln P(\A,\bb)$
subject to the above constraint, we obtain
\begin{equation}\label{eq:Abjoint}
  P(\A,\bb|\beta) = \frac{\ee^{\beta W(\A,\bb)}}{Z(\beta)},
\end{equation}
with $Z(\beta)=\sum_{\A,\bb}\ee^{\beta W(\A,\bb)}$, and $\beta$ being an
``inverse temperature'' Lagrange multiplier. Thus, the maximum entropy ansatz
amounts to a choice $g(x) = \beta x$ and $f(\A)$ being an arbitrary constant. We
emphasize once more that the choice of $g(x)$ is not crucial for analysis --- in
fact it has no significant effect whatsoever in our calculations, as we
demonstrate in Appendix~\ref{app:g_x}. Because of this, it will be more
convenient henceforth to use $g(x)=\beta x$, but without any loss in generality.
We will return to the choice of $f(\A)$ in Sec.~\ref{sec:f} --- let us
momentarily abide by the maximum entropy choice.

The above joint distribution yields a posterior probability for
partitions,
\begin{equation}\label{eq:b_post}
  P(\bb|\A, \beta) = \frac{\ee^{\beta W(\A,\bb)}}{\sum_{\bb'}\ee^{\beta W(\A,\bb')}},
\end{equation}
which has been used before by Massen and
Doye~\cite{massen_thermodynamics_2006} and Zhang and
Moore~\cite{zhang_scalable_2014}, for the particular case of modularity,
to investigate the ensemble of all competing partitions, rather than the
single one that optimizes the quality function. Here we are more
directly interested in the joint distribution of Eq.~\ref{eq:Abjoint},
for two reasons. The first one is that it generates problem instances
for which the original community detection method is optimal. More
specifically, if we consider an estimator $\hat\bb(\A)$ for the
partition of a network $\A$, and the average of the error
$\epsilon(\bb',\bb)$ between the true and inferred partitions over all
problem instances,
\begin{equation}
  \Lambda = \sum_{\A,\bb}\epsilon(\bb,\hat\bb(\A))P(\A,\bb|\beta),
\end{equation}
then the estimator is optimal if it minimizes $\Lambda$, in which case
it must correspond to
\begin{equation}
  \hat\bb(\A) = \underset{\bb}{\operatorname{argmin}}\;\sum_{\bb'}\epsilon(\bb,\bb')P(\bb'|\A,\beta),
\end{equation}
which is the estimate that minimizes the error over the posterior
distribution conditioned on $\A$. In particular, for the ``zero-one''
error, $\epsilon(\bb,\bb') = 1-\prod_i\delta_{b_i,b_i'}$, which simply
identifies the correct answer and ignores all other ones, we recover the
original optimization
\begin{align}
  \hat\bb(\A) &= \underset{\bb}{\operatorname{arg\ max}}\;P(\bb|\A,\beta)\\
  &=\underset{\bb}{\operatorname{arg\ max}}\; W(\A, \bb).
\end{align}
Therefore, according to this error criterion~\footnote{The optimal
estimate of the partition will always depend on which criterion we use
to judge performance, namely the particular choice of the error function
$\epsilon(\bb,\bb')$. The choice of error function is an
application-dependent decision, and other choices will lead to
estimates that are different from
Eq.~\ref{eq:Wopt}~\cite{peixoto_revealing_2021}, although they will
always involve the posterior of Eq.~\ref{eq:b_post}.}, for the problem
instances sampled from Eq.~\ref{eq:Abjoint} there exists no algorithm
that can perform on average better than one that corresponds to the
optimization of Eq.~\ref{eq:Wopt} (although it is still possible for
alternative algorithms to perform just as well on the same
instances). This gives us access to problem instances for which, in a
formal sense, the results obtained with an arbitrary community detection
algorithm are maximally correct. As we will show, we can use this
information to investigate the implicit expected instances of arbitrary
community detection algorithms.

\subsection{Model selection and the description length}
In addition to the above, our second reason to focus on the joint
distribution of Eq.~\ref{eq:Abjoint} is that it can be used to assess the
overall statistical evidence for a particular partition of the network,
and to enable comparison with alternative models. More precisely, from
Eq.~\ref{eq:Abjoint} we can compute the so-called \emph{description
length}~\cite{rissanen_information_2010,grunwald_minimum_2007} of the
data, defined as
\begin{align}
  \Sigma(\A,\bb|\beta) &= -\log_2 P(\A,\bb|\beta)\\
  &=\underbrace{-\log_2 P(\A|\bb,\beta)}_{\mathcal S} \underbrace{- \log_2 P(\bb|\beta)}_{\mathcal L}.
\end{align}
The description length measures the size of the shortest binary message
required to transmit both the partition $\bb$ (with length $\mathcal L$)
and network $\A$ (with length $\mathcal S$) over a noiseless channel, in
such a manner that they can both be decoded from the message without
errors, and assuming that the value of $\beta$ is already known to the
decoder. This connection exposes a fundamental equivalence between
inference and compression, where the most likely model [largest
$P(\A,\bb|\beta)$] is also the most compressive [smallest
$\Sigma(\A,\bb|\beta)$]. The description length measures the degree of
parsimony of the obtained network partition, allowing us to compare with
alternative ones in what amounts to a formalization of Occam's razor. In
the context of the SBM, the description length has been used as a
criterion to perform
order~\cite{rosvall_information-theoretic_2007,peixoto_parsimonious_2013}
and
model~\cite{peixoto_model_2015,zhang_statistical_2020,peixoto_disentangling_2022}
selection, and here we extend this concept to arbitrary community
detection algorithms.

From Eq.~\ref{eq:Abjoint}, we can obtain the description length for an
arbitrary $W(\A,\bb)$ as follows (for convenience of notation, we will
henceforth compute the description length using the natural base instead
of base two, yielding values in nats instead of bits):
\begin{align}\label{eq:dl-general}
  \Sigma(\A,\bb|\beta)
    &= -\beta W(\A,\bb) + \ln Z(\beta)\\
    &= -\beta W(\A,\bb) + \ln\sum_{\A',\bb'}e^{\beta W(\A',\bb')}.
\end{align}
(Note that we will always have $\Sigma(\A,\bb|\beta) > 0$, regardless of
our choice of $W(\A,\bb)$ and $\beta$.)
The parameter $\beta$ is important since it determines the expected
value of the quality function, so we will consider its optimal value with
\begin{equation}
  \Sigma(\bm{A},\bm{b}) = \min_{\beta}\; \Sigma(\bm{A},\bm{b}|\beta).
\end{equation}
(Strictly speaking, for the description length to be complete we would
need to include the amount of information required to transmit the value
of $\beta$ up to a desired precision as well --- but since this is a
single global parameter, this will amount to an overall small constant
that we can neglect.)

The difficulty in obtaining
$\Sigma(\bm A, \bb)$ lies in computing $Z(\beta)$, which is in general
intractable analytically, since it involves a sum over all networks and
partitions. As we show in Appendix~\ref{app:g_x}, we can obtain an
asymptotic approximation of the description length given by
\begin{align}\label{eq:dl-approx-logXi}
  \Sigma(\A,\bb) \approx \ln \Xi(W(\A, \bb)),
\end{align}
where $\ln\Xi(W)$ is the \emph{entropic density} of the quality
function, obtained via \emph{the density of states},
\begin{equation}
  \Xi(W) = \sum_{\A, \bb}\delta_{W(\A,\bb), W}.
\end{equation}
This is a general result that allows us to compute the description
length for any quality function $W(\A,\bb)$, provided the density of
states can be estimated. In general, this may be done numerically with
Monte Carlo, using algorithms such as
Wang-Landau~\cite{landau_guide_2005}, or other thermodynamic integration
methods. Importantly, however we choose to perform this computation, it
does not affect the time required to obtain a result from the original
community detection algorithm of Eq.~\ref{eq:Wopt} --- we need either to
perform the computation only once for the value of $W(\A,\bb)$ obtained
as the output, or from a pre-computed table with enough resolution.

In this work, we will be able to obtain accurate analytical approximations of
the description length (which do not make direct use of the approximation of
Eq.~\ref{eq:dl-approx-logXi}) for a fairly wide class of quality functions
$W(\A, \bb)$ that can be expressed as a function of the edge counts between
groups and the group sizes, i.e.
\begin{equation}\label{eq:Wsbm}
  W(\A, \bb) = W(\bm e, \bm n),
\end{equation}
with $e_{rs}=\sum_{ij}A_{ij}\delta_{b_i,r}\delta_{b_j,s}$ and
$n_r=\sum_i\delta_{b_i,r}$. In this case we can perform the following
change of variables,
\begin{equation}
  \sum_{\A,\bb}\ee^{\beta W(\A,\bb)} = \sum_{B, \bm e, \bm n}\ee^{\beta W(\bm e, \bm n)}\Omega(\bm e, \bm n, B),
\end{equation}
with $\Omega(\bm e, \bm n, B)$ being the microcanonical partition
function of the SBM~\cite{peixoto_entropy_2012}
\begin{align}
  \Omega(\bm e, \bm n,B) &=
  \sum_{\A,\bb}\prod_{r\le s}\delta_{\sum_{ij}A_{ij}\delta_{b_i,r}\delta_{b_j,s},e_{rs}}\times \prod_r\delta_{\sum_i\delta_{b_i,r}, n_r}\\
  &=\prod_{r<s}{n_rn_s\choose e_{rs}}\prod_r{{n_r\choose 2}\choose e_{rr}/2}\times \frac{N!}{\prod_r{n_r!}}.
\end{align}
The above computation makes it clear that whenever Eq.~\ref{eq:Wsbm}
holds, which happens to be true for many popular quality functions, then
the overall approach can be seen as equivalent to the inference of a
particular version of the SBM, with a specific weighting factor given by
$W(\bm e, \bm n)$. We will focus on the class of methods where
Eq.~\ref{eq:Wsbm} holds for our further analyses, as they permit simple
analytical treatment. (In Appendix~\ref{app:others} we consider in more
detail situations not covered by our main ansatz, including when the
community detection algorithm is not the result of an optimization.)

Based on this parametrization, we can now decompose $Z(\beta)$ as
\begin{align}
  Z(\beta)&=\sum_{B,\bm e, \bm n}\ee^{\beta W(\bm e, \bm n)}\Omega(\bm e, \bm n,B)\\
  &= \int \ee^{\beta W}\Xi(W)\,\dd W,
\end{align}
with
\begin{equation}
  \Xi(W) = \sum_B\Xi(W,B),
\end{equation}
being the $\beta$-independent density of states, where
\begin{equation}
  \Xi(W,B) = \sum_{\bm e, \bm n}\Omega(\bm e, \bm n, B)\delta(W(\bm e, \bm n)-W),
\end{equation}
is the contribution for a particular number of groups $B$.

With $\Xi(W)$ at hand, the description length is then computed as
\begin{equation}
  \Sigma(\bm{A},\bm{b}|\beta) = -\beta W(\A, \bb) + \ln \int \ee^{\beta W}\Xi(W)\,\dd W.\label{eq:W_dl}
\end{equation}

The computation above allows us to ascribe a description length to an arbitrary
quality function $W(\A,\bb)$, and hence compare it with any other generative
model in its relative ability to provide a plausible account for the
data~\footnote{We make the code used to perform these computations for the
  modularity and Infomap objectives freely available as part of the
  \texttt{graph-tool} Python library~\cite{peixoto_graph-tool_2014}.}.

We note that if the quality function being used is already the joint
log-likelihood of a generative model, i.e.
\begin{equation}
  W(\A,\bb) = \ln P(\A,\bb),
\end{equation}
then the above procedure will recover the original description length
$\Sigma(\A,\bb) = -\ln P(\A,\bb)$ for $\beta=1$. The optimization of the
parameter $\beta$ may yield a marginal compression, which will vanish
asymptotically if the data happens to be sampled from the same model.

\subsection{Partition-independent compression}\label{sec:f}

Any given posterior distribution $P(\bb|\A)$ is not uniquely associated
with a description length, since the latter depends also on modelling
choices that are independent of the relationship between network and
partition. In fact, for any generative model $P(\A,\bb)$, we can devise
an entire family of model alternatives determined up to an arbitrary
exponential weight $f(\A)$, i.e.
\begin{equation}\label{eq:Pprime}
  P'(\A,\bb) = \frac{P(\A,\bb)\,\ee^{f(\A)}}{\sum_{\A',\bb'}P(\A',\bb')\,\ee^{f(\A')}},
\end{equation}
all of which will result in the same posterior distribution for the
partitions, $P(\bb|\A)=P'(\A,\bb)/P'(\A)=P(\A,\bb)/P(\A)$, independent
of $f(\A)$. Therefore, the choice of $f(\A)$ will affect the
description length (as well as predictive tasks such as link
prediction~\cite{guimera_missing_2009,peixoto_reconstructing_2018,ghasemian_evaluating_2019}),
but not the posterior for the node partitions, despite the corresponding
model generating different networks. It is important to emphasize that
the choice of $f(\A)$ cannot significantly alter the community
structure of the networks generated. We can see this by formulating the
sampling of an instance of the model of Eq.~\ref{eq:Pprime} with the
following rejection algorithm:
\begin{enumerate}
\item A pair $(\A,\bb)$ is sampled from the original $P(\A,\bb)$.
\item With probability $\ee^{f(\A) - f^*}$, where $f^* = \max_{\A}
      f(\A)$, the sample is accepted, otherwise it is rejected and we go
      to step 1.
\end{enumerate}
Therefore, the re-weighting of Eq.~\ref{eq:Pprime} will only suppress
networks from the original ensemble in a manner that cannot take into
account the node partition $\bb$.

Although all models in the above family generate networks with the same
kind of community structure, they can deviate with respect to other
attributes that are uncoupled from this property. If these attributes
happen to match more closely an observed network, this can be used to
compress it further.

In the calculation of the previous section we used the principle of
maximum entropy to fill this modelling gap, which yielded a constant
value for $f(\A)$. However, it is possible to deviate from this
principle, and improve the description length by including properties we
know to be ubiquitous. For example, we can introduce the exact number of
edges $E$ as an additional hard constraint,
\begin{equation}
  \sum_{\A,\bb}P(\A,\bb|g,f)\delta_{\sum_{i<j}A_{ij},M} = \delta_{M,E}\label{eq:E},
\end{equation}
which if added to the entropy maximization yields
\begin{equation}
  P(\A,\bb|\beta,E) = \frac{\ee^{\beta W(\A,\bb)}\delta_{\sum_{i<j}A_{ij},E}}{Z(\beta,E)},
\end{equation}
with $Z(\beta,E)=\sum_{\A,\bb}\ee^{\beta
W(\A,\bb)}\delta_{\sum_{i<j}A_{ij},E}$. To remove the parameter $E$ we
must introduce a uniform prior,
\begin{equation}
  P(E) = \frac{1}{{N \choose 2} + 1},
\end{equation}
obtaining thus an alternative joint likelihood via marginalization,
\begin{align}\label{eq:Abjoint_sparse}
  P(\A,\bb|\beta) &= \sum_E P(\A,\bb|\beta,E)P(E)\\
  &= \frac{\ee^{\beta W(\A,\bb)}}{Z(\beta,\sum_{i<j}A_{ij})\left[{N \choose 2} + 1\right]}.
\end{align}
The description length obtained with the joint distribution above will
almost always be significantly shorter than what is obtained with
Eq.~\ref{eq:Abjoint}, since the latter will sample networks which will
tend to be dense --- as long as the values of $W(\A,\bb)$ are not
affected directly by the network density. We will use
Eq.~\ref{eq:Abjoint_sparse} in our ensuing analysis, instead of
Eq.~\ref{eq:Abjoint}, since we will be considering only sparse
networks. The density of states in this case is computed in the same
manner as before, but keeping the total number of edges fixed,
\begin{equation}
  \Xi(W,B,E) = \sum_{\bm e, \bm n}\Omega(\bm e, \bm n, B)\delta(W(\bm e, \bm n)-W)\delta_{\sum_{rs}e_{rs},2E}.
\end{equation}

We can follow this route further and seek additional constraints that
condition $f(\A)$ to favor network patterns that are more likely to be
encountered. For example, instead of constraining only the total number
of edges, we can fix the entire degree sequence $\bm k = \{k_i\}$, where
$k_i=\sum_jA_{ij}$ is degree of node $i$, i.e.
\begin{align}
  \sum_{\A,\bb}P(\A,\bb|g,f)\delta_{\sum_{j}A_{ij},m_i} &= \delta_{m_i,k_i},
\end{align}
which will lead to
\begin{equation}
  P(\A,\bb|\beta,\bm k) = \frac{\ee^{\beta W(\A,\bb)}\prod_i\delta_{\sum_jA_{ij},k_i}}{Z(\beta,\bm k)},
\end{equation}
with $Z(\beta, \bm k) = \sum_{\A, \bb}\ee^{\beta
W(\A,\bb)}\prod_i\delta_{\sum_jA_{ij},k_i}$. Note that, now, instead of
a single parameter, we have $N+1$. To retain the same number of
parameters as before, we need a prior for the degree sequence $\bm
k$. One choice is a uniform model with
\begin{equation}\label{eq:kprior_flat}
  P(\bm k | E) = \multiset{N}{E}^{-1},
\end{equation}
where $\multiset{n}{m} = {n+m-1\choose m}$ is the number of
$n$-tuples of non-negative integers whose sum is $m$. Another choice is
a deeper Bayesian hierarchy with
\begin{align}
  P(\bm k | E) = P(\bm k | \bm\eta) P(\bm\eta | E),
\end{align}
where $\bm\eta=\{\eta_k\}$ are the degree counts, i.e. $\eta_k = \sum_i\delta_{k_i,k}$, such that 
\begin{equation}
  P(\bm k | \bm\eta) = \frac{\prod_k\eta_k!}{N!},\quad
  P(\bm\eta | E) = q(2E, N)^{-1},
\end{equation}
where $q(m,n)$ is the number of possible partitions of integer $m$ into
at most $n$ parts, which can be calculated exactly via a recursion, or
approximated accurately for large arguments, as described in
Ref.~\cite{peixoto_nonparametric_2017}. The latter choice tends to
provide a more parsimonious model for most empirical degree sequences,
as long as they deviate sufficiently from a geometric degree
distribution, which is (marginally) better described by
Eq.~\ref{eq:kprior_flat} (see Ref.~\cite{peixoto_nonparametric_2017} for
a discussion). With this prior in place, the final joint distribution
becomes,
\begin{align}
  P(\A,\bb|\beta) &= \sum_{\bm k, E}P(\A,\bb|\beta,\bm k)P(\bm k|E)P(E),\\
  &= \frac{\ee^{\beta W(\A,\bb)}\prod_k\hat\eta_k!}{Z(\beta,\hat{\bm k})q(\sum_{ij}A_{ij},N)\left[{N \choose 2} + 1\right]N!},
\end{align}
where $\hat k_i = \sum_jA_{ij}$ and $\hat\eta_k = \sum_i\delta_{\hat
  k_i, k}$. In this case the SBM partition function is given by
\begin{multline}
  \Omega(\bm e, \bm n, \bm k, B) =
  \sum_{\A,\bb}\prod_{r\le s}\delta_{\sum_{ij}A_{ij}\delta_{b_i,r}\delta_{b_j,s},e_{rs}}\times\\
  \prod_r\delta_{\sum_i\delta_{b_i,r}, n_r}\times\prod_i\delta_{\sum_jA_{ij},k_i},
\end{multline}
which is unfortunately intractable~\cite{bender_asymptotic_1978-1}. However, it
can be approximated by counting
configurations~\cite{peixoto_entropy_2012,peixoto_nonparametric_2017},
\begin{equation}\label{eq:DCSBM-dos}
  \Omega(\bm e, \bm n, \bm k, B) \approx
  \frac{\prod_re_r!}{\prod_{r<s}e_{rs}!\prod_re_{rr}!!\prod_ik_i!} \times \frac{N!}{\prod_r{n_r!}}.
\end{equation}
which will yield an asymptotically exact enumeration as long as $k_i\ll
\sqrt{N/B}$, and a still useful approximation otherwise.

The above alternative yields ``degree-corrected'' variants for the
description length, which we will use in our analysis as well. Note that
the above modification is different from the degree correction of the
SBM~\cite{karrer_stochastic_2011}, which correlates the degrees with the
group memberships, and hence alters the posterior
distribution~\cite{peixoto_nonparametric_2017}. The correction above
changes the description length, but not the posterior distribution of
partitions --- all of the variations above remain fully equivalent to
the original community detection ansatz of Eq.~\ref{eq:Wopt}.

One could in principle proceed indefinitely with adding
partition-independent constraints that influence $f(\A)$, together with
prior distributions that keep the final distribution nonparametric ---
however, these quickly become very difficult to compute as soon as
higher-order structures are considered. But more importantly, as
Eq.~\ref{eq:Pprime} shows, these kinds of modelling refinements can be
imposed on any generative model. The above choices that impose sparsity
and degree-correction already attempt to extract the largest amount of
compression, on par with what is done with state-of-the-art inferential
methods based on the SBM~\cite{peixoto_bayesian_2019}. Therefore, if
further partition-independent improvements are possible, these can be
employed systematically and on equal grounds for every model considered
in this work.

\subsection{Implicit priors and the role of the inverse temperature}

From the joint distribution of Eq.~\ref{eq:Abjoint} we can recover
implicit priors via marginalization. For example, the marginal distribution
for the value of the quality function is
\begin{align}
  P(W|\beta) &= \sum_{\A,\bb}\delta(W(\A,\bb)-W)P(\A,\bb|\beta)\\
  &=\frac{\mathrm{e}^{\beta W}\Xi(W)}{Z(\beta)}.
\end{align}
Likewise, the prior for the number of groups can be obtained via
\begin{align}
  P(B|\beta) &= \sum_{\A,\bb}\delta_{B(\bb),B}P(\A,\bb|\beta)\\
  &=\frac{\int \mathrm{e}^{\beta W}\Xi(W,B)\,\dd W}{Z(\beta)}.
\end{align}
From the above equations we see that the inverse temperature $\beta$
will influence both the expected number of groups, as well as the values
of the quality function. Notably, the conditional prior
\begin{equation}
  P(B|W) = \frac{P(W,B|\beta)}{P(W|\beta)} = \frac{\Xi(W,B)}{\Xi(W)}
\end{equation}
is $\beta$-independent.

For inferential methods based on the SBM~\cite{peixoto_bayesian_2019}
the priors above are set explicitly, usually in a non-informative manner
to avoid biases during inference. Instead, for a given $W(\A,\bb)$ these
need to be reverse-engineered via the above computations.

We proceed now to the application of the above method for the
generalized modularity quality
function~\cite{reichardt_statistical_2006}, which we will use as our
example to illustrate the insights we can obtain by casting community
detection objectives into our inferential framework. We perform an
analogous analysis of Infomap in Appendix~\ref{app:infomap}, and discuss
further objective functions in Appendix~\ref{app:objectives}.

\begin{figure*}
  \begin{tabular}{ccc}
    \multicolumn{3}{c}{(a) Density of states}\\
    \includegraphics[width=.3\textwidth]{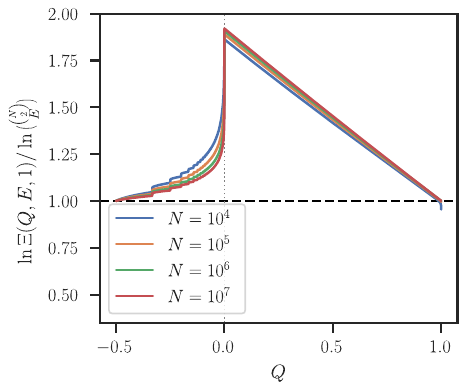} &
    \includegraphics[width=.3\textwidth]{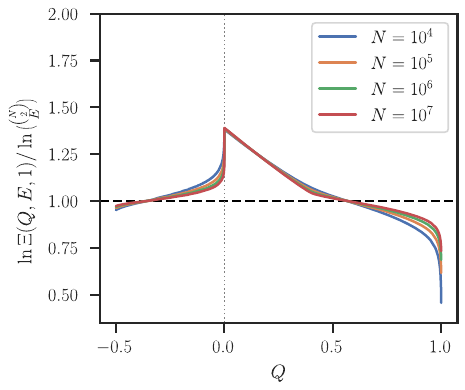} &
    \includegraphics[width=.3\textwidth]{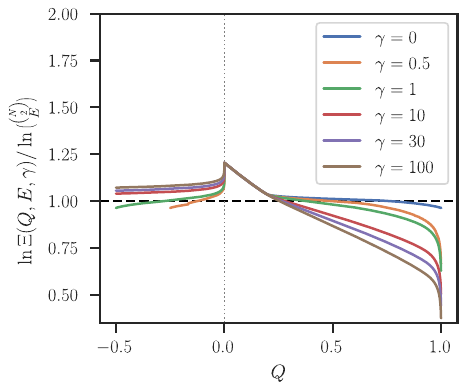}\\
    \multicolumn{3}{c}{(b) Description length}\\
    \includegraphics[width=.3\textwidth]{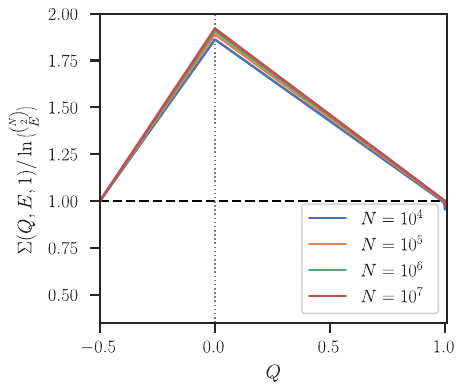} &
    \includegraphics[width=.3\textwidth]{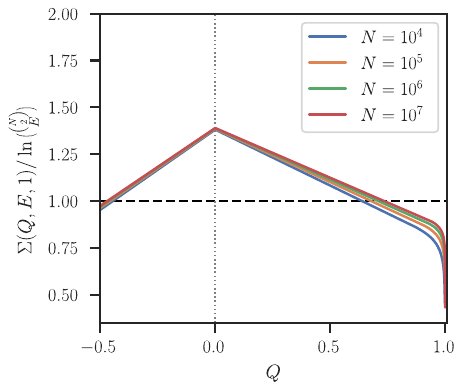}&
    \includegraphics[width=.3\textwidth]{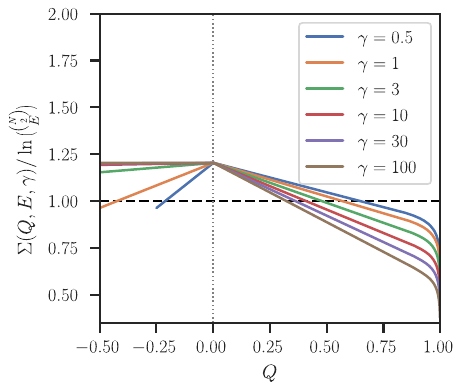}\\
    $\avg{k}=2$ & $\avg{k}=5$ & $\avg{k}=10$
  \end{tabular} \caption{(a) Density of states $\Xi(Q, E, \gamma)$ and
  (b) description length $\Sigma(Q, E, \gamma)$, as a function of the value
  of modularity $Q$, for different number of nodes $N$ and average
  degree values, $\avg{k}=2, 5, 10$, from left to right. The values are
  shown relative to the ER baseline. The description length in
  particular tells us what should be considered a statistically
  significant modularity value.\label{fig:Qdl}}
\end{figure*}

\subsection{Modularity maximization}\label{sec:modularity}

The generalized modularity quality function is given by
\begin{equation}\label{eq:modularity}
Q(\A, \bb,\gamma) = \frac{1}{2E}\sum_r{e_{rr} - \gamma\frac{e_r^2}{2E}},
\end{equation}
where $\gamma$ is the so-called resolution parameter. The method of
modularity maximization~\cite{newman_modularity_2006} consists of
finding the partition that maximizes this quantity, typically with
$\gamma=1$.

As is required for our computation, modularity can be written solely as
a function of the microcanonical SBM parameters, i.e. $Q(\A, \bb,\gamma)
= Q(\bm e,\bm n, \gamma)$, and we are interested in obtaining the
density of states,
\begin{equation}\label{eq:Q-dos-exact}
  \Xi(Q, E)= \sum_{\bm e, \bm n}\Omega(\bm e, \bm n)\delta(Q(\bm e, \bm n,\gamma) - Q)\delta_{2E,\textstyle\sum_{rs}e_{rs}}.
\end{equation}
As we show in Appendix~\ref{app:max-ent}, the dominating terms of the above sum
will correspond to a uniform planted partition model with $e_r = 2E/B$ and $n_r=N/B$
for which we can write
\begin{align}
  Q(\A, \bb,\gamma) = \frac{E_{\text{in}}}{E} - \frac{\gamma}{B},
\end{align}
with $E_{\text{in}}=\sum_re_{rr}/2$ being the edges internal to
communities. Based on this, we can write
\begin{equation}
  \Xi(Q, E) \ge \sum_{B}\Omega(E, E_{\text{in}}(Q, E, B,\gamma), B),\label{eq:xi_approx}
\end{equation}
with
\begin{equation}
  E_{\text{in}}(Q, E, B, \gamma) = E(Q + \gamma/B),
\end{equation}
where Eq.~\ref{eq:xi_approx} accounts for the number of partitioned networks
with exactly $E_{\text{in}}$ edges between nodes of the same group, which can be
computed as
\begin{align}
  &\Omega(E, E_{\text{in}}, B)\nonumber\\
  &= \sum_{\bm e, \bm n}\Omega(\bm e, \bm n, B)\delta_{\sum_re_{rr}/2,E_{\text{in}}}\delta_{\sum_{rs}e_{rs},2E}\prod_r\delta_{n_r,N/B}  \label{eq:lb}\\ \nonumber
  &= \sum_{\bm e}\left[\prod_{r<s}{N^2/B^2\choose e_{rs}}\prod_r{{N/B\choose 2}\choose e_{rr}/2}\times\frac{N!}{[(N/B)!]^B}\right. \\
  &\qquad\qquad\left.\times\delta_{\sum_re_{rr}/2,E_{\text{in}}}\delta_{\sum_{r<s}e_{rs},E-E_{\text{in}}}\vphantom{\prod_{r<s}{N^2/B^2\choose e_{rs}}}\right]\label{eq:lb2}\\
  & = {B{N/B\choose 2}\choose E_{\text{in}}}{\frac{N^2}{B^2}{B\choose 2} \choose E-E_{\text{in}}}\frac{N!}{[(N/B)!]^B}.\label{eq:omega}
\end{align}
To
obtain Eq.~\ref{eq:omega} from Eq.~\ref{eq:lb2} we simply used the
generalized Vandermonde's identity,
\begin{equation}
  \sum_{k_1+\cdots+k_p = m} {n_1\choose k_1}\cdots{n_p\choose k_p} = {n_1+\cdots+n_p\choose m}.
\end{equation}

It is important to reiterate that Eq.~\ref{eq:lb} allows us to obtain a
strict lower bound on the density of states $\Xi(Q,E)$, since it
accounts only for partitions with equal size. However it will
asymptotically dominate the exact sum for large networks as we show in
Appendix~\ref{app:max-ent}. Nevertheless, in the pre-asymptotic regime,
the above calculations will therefore yield a \emph{strict lower-bound}
on the resulting description length, since the exact final values of
$Z(\beta,E)$ can only be larger than what is obtained via the above
computation.

Analogously, for the degree-corrected version of modularity we have
instead
\begin{align}
  &\Omega(E, E_{\text{in}}, \bm k, B)\nonumber\\
  &= \sum_{\bm e, \bm n}\Omega(\bm e, \bm n, \bm k, B)\delta_{\sum_re_{rr}/2,E_{\text{in}}}\delta_{\sum_{rs}e_{rs},2E}\prod_r\delta_{n_r,N/B}  \label{eq:lb_k}\\
  &= \sum_{\bm e}\left[\frac{[(2E/B)!]^B}{\prod_{r<s}e_{rs}!\prod_re_{rr}!!\prod_ik_i!}  \times \frac{N!}{[(N/B)!]^B}\right.\nonumber\\
  &\qquad\qquad \times \left.\delta_{\sum_re_{rr}/2,E_{\text{in}}}\delta_{\sum_{r<s}e_{rs},E-E_{\text{in}}}\vphantom{\frac{[(2E/B)!]^B}{\prod_{r<s}e_{rs}!}}\right]\\
  & = \frac{[(2E/B)!]^BB^{E_{\text{in}}}{B\choose 2}^{E-E_{\text{in}}}N!}{(2E_{\text{in}})!! (E-E_{\text{in}})![(N/B)!]^B\prod_ik_i!},
  \label{eq:omega_dc}
\end{align}
where in the last step we have used the multinomial theorem,
\begin{equation}
  \sum_{k_1+\cdots+k_p = m} \frac{m!}{\prod_{i=1}^{p}k_i!}\prod_{i=1}^{p}x_i^{k_i}=\left(\sum_{i=1}^px_i\right)^m.
\end{equation}

With the density of states at hand, we can obtain the description length
according to Eq.~\ref{eq:W_dl}, which involves a sum over $B$ in
Eq.~\ref{eq:xi_approx} and an integral over $W=Q$, both of which can be
done efficiently numerically, up to an arbitrary precision.

In Fig.~\ref{fig:Qdl} we see the result of the above computation for
some network sizes and densities. [We focus for the moment on the
non-degree-corrected version, although the degree-corrected variants are
qualitatively very similar (not shown).] It shows the density of states
and description length values relative to the ER baseline
\begin{equation}
  \Sigma_{ER} = \ln {{N\choose 2}\choose E}.
\end{equation}
Therefore, a value smaller than this would amount to a compression
relative to a fully random model, pointing thus to statistically
significant structure. The values shown on the bottom row of
Fig.~\ref{fig:Qdl} offer us an important mapping from $Q$ values ---
which by themselves cannot be interpreted statistically --- to
description length values. The latter quantities allow for an
information-theoretical evaluation of the statistical significance and
degree parsimony for $Q$ values obtained with modularity maximization
algorithms. As can be seen in Fig.~\ref{fig:Qdl}, we often obtain
\emph{inflation} for intermediary values of $Q$ --- which therefore
would indicate overfitting --- and compression only for relatively high
values. The compression region becomes larger for denser networks (for
$\avg{k}=2$ compression is impossible for most $Q$ values), which is
also anticipated by higher values of the resolution parameter $\gamma$.

\begin{figure}
  \begin{tabular}{cc}
    \begin{overpic}[width=.5\columnwidth]{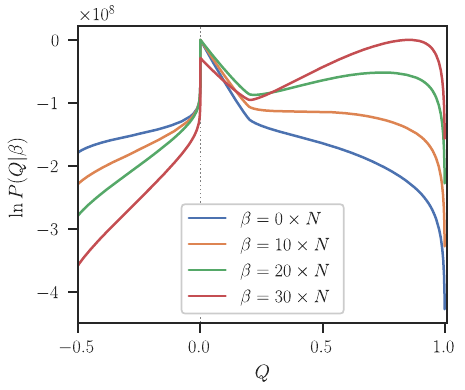}
      \put(0,85){\smaller (a)}
    \end{overpic}&
    \begin{overpic}[width=.5\columnwidth]{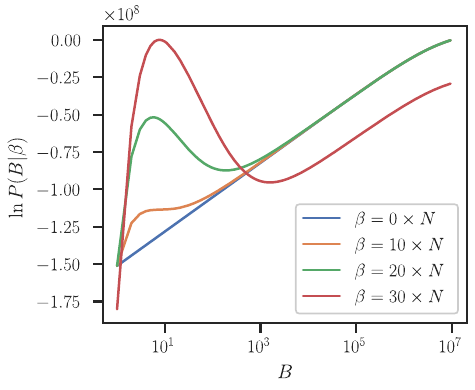}
      \put(0,85){\smaller (b)}
    \end{overpic}\\
    \begin{overpic}[width=.5\columnwidth]{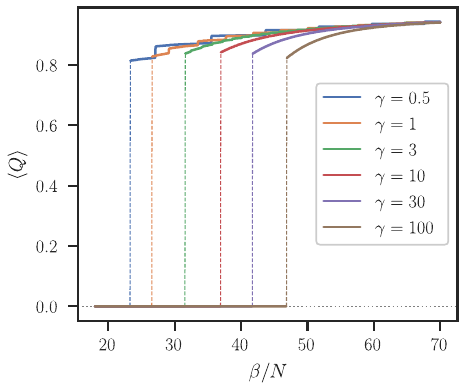}
      \put(0,85){\smaller (c)}
    \end{overpic}&
    \begin{overpic}[width=.5\columnwidth]{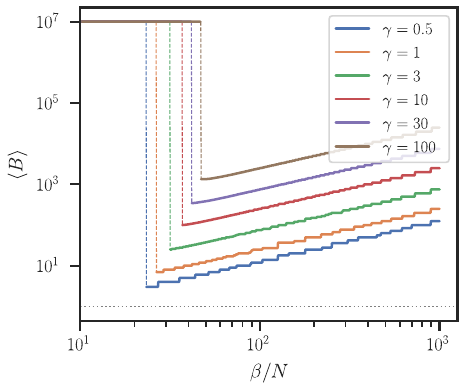}
      \put(0,85){\smaller (d)}
    \end{overpic}\\
    \begin{overpic}[width=.5\columnwidth]{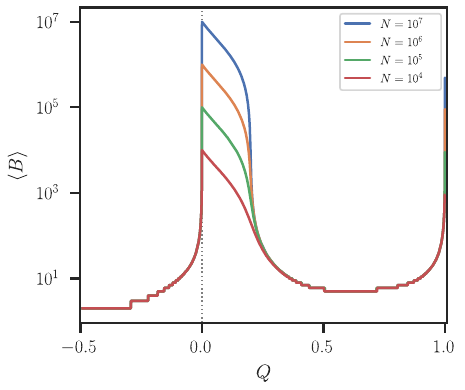}
      \put(0,85){\smaller (e)}
    \end{overpic}&
    \begin{overpic}[width=.5\columnwidth]{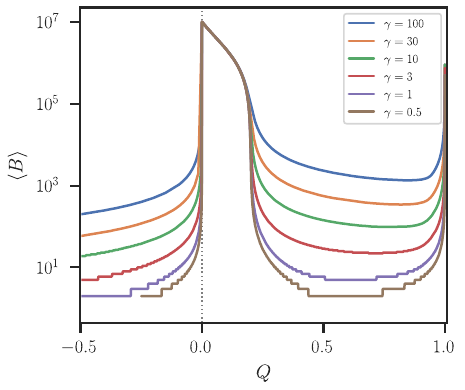}
      \put(0,85){\smaller (f)}
    \end{overpic}
  \end{tabular}

  \caption{Implicit priors for the method of modularity
  maximization. Top row: (a) Implicit prior distribution for the value
  of modularity $Q$, and (b) number of groups $B$, for different values
  of $\beta$, $\gamma=1$, $N=10^7$ and $\avg{k}=10$. Middle row: Average
  values of (c) $Q$ and (d) $B$, as a function of $\beta$, and different
  values of $\gamma$. Bottom row: (e) Average value of $B$ as a function
  of $Q$ for different values of $N$ and $\gamma=1$, and (f) the same as
  (e) but with $N=10^7$ only and different values of
  $\gamma$.\label{fig:Qprior}}
\end{figure}

An important aspect of our analysis is that it allow us to understand
the implicit prior assumptions that are intrinsic to modularity
maximization, as we show in Fig.~\ref{fig:Qprior}. As seen in panels (a)
and (b), both the prior for the modularity value, $P(Q|\beta)$, and the
number of groups, $P(B|\beta)$, are extremely informative and bimodal,
concentrating very strongly on particular high and low values. The value
of $\beta$ determines which mode dominates, inducing a discontinuous
transition at a particular value $\beta^*$ for the mean values $\avg{Q}$
and $\avg{B}$, as we can see in panels (c) and (d). This kind of
transition is reminiscent of the degeneracy encountered in exponential
random graphs models~\cite{park_solution_2004,park_solution_2005}, where
the ensemble mean of an enforced constraint results in bimodal
distributions, where no typical sample from the ensemble obeys the
enforced constraint. Importantly, this kind of prior assumption is
hardly justified in most applications in the absence of substantial
additional evidence supporting it. The case of strict modularity
maximization, where we are interested only in the partition that
maximizes the posterior of Eq.~\ref{eq:b_post}, amounts to the situation
$\beta\to\infty$, where prior modularity values concentrate on $Q=1$ and
$B \propto N$, explaining the tendency of the method to overfit, which
is only avoided only if the evidence in the data is sufficiently strong
to contradict the prior assumptions.

We can further understand the behavior of modularity maximization via
the conditional prior $P(B|Q)$, which is $\beta$-independent, seen in
Fig.~\ref{fig:Qprior}e. The range of large $Q$ values shows an intuitive
behavior: as $Q$ increases, so does the expected number of
groups. However, the same happens for low $Q$ values approaching
zero. This contradicts the intuition that low $Q$ values, specially
$Q=0$, would amount to small or negligible community structure. What is
occurring here is that for low $Q$ the density of states is dominated by
the contribution of the node partitions, which is largest for $B=O(N)$,
since there are many networks that admit a low $Q$ with an arbitrary
partition. As soon as $Q$ increases, the contribution of the actual
network structure dominates instead, since relatively fewer networks
allow for a high $Q$ partition, and forces the number of groups to
decrease, before increasing again. This tension between the partition
and network entropic contributions also explains the transitions between
the low $Q$ and divergent $B$, and high $Q$ and finite $B$ regimes
observed as a function of $\beta$.

\begin{figure}
  \begin{tabular}{cc}
    \multicolumn{2}{c}{
      \begin{overpic}[width=\columnwidth]{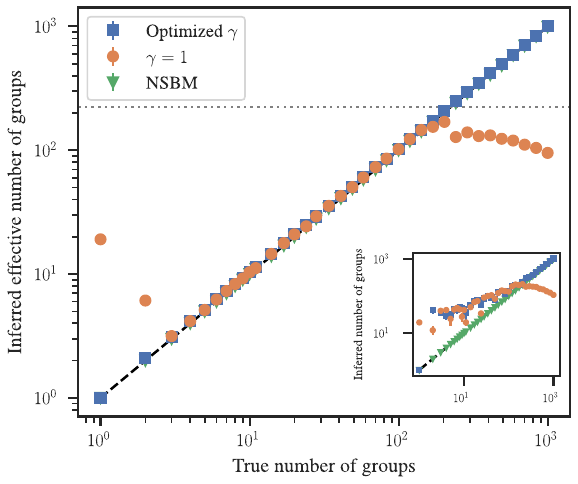}
        \put(0, 80){\smaller (a)}
      \end{overpic}}
    \\[1em]
    \begin{overpic}[width=.5\columnwidth]{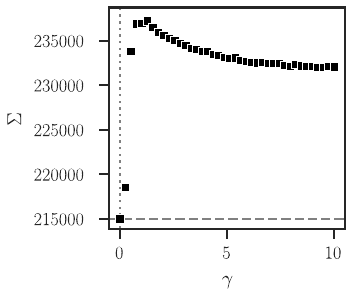}
      \put(0, 86){\smaller (b)}
      \end{overpic} &
    \begin{overpic}[width=.5\columnwidth]{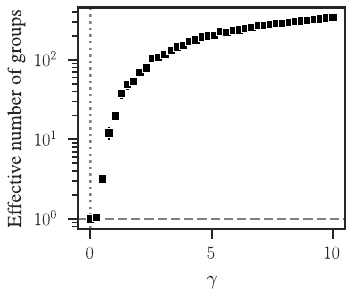}
      \put(0, 86){\smaller (c)}
      \end{overpic} \\
    \begin{overpic}[width=.5\columnwidth]{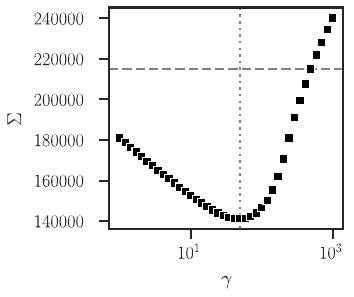}
      \put(0, 86){\smaller (d)}
      \end{overpic} &
    \begin{overpic}[width=.5\columnwidth]{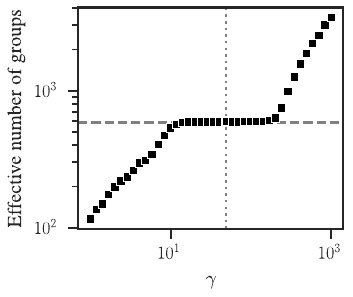}
      \put(0, 86){\smaller (e)}
      \end{overpic}
  \end{tabular} \caption{Computing the description length can alleviate
  the overfitting and underfitting (resolution limit) of the modularity
  maximization method. Panel (a) shows the inferred effective number of
  groups $B_e=\exp(-\sum_r\frac{n_r}{N}\ln \frac{n_r}{N})$, as a
  function of the true number of groups $B$, for networks sampled from a
  PP model with uniform group sizes, $E_{\text{in}} = E - (B-1)\avg{k}$,
  for $N=10^4$ and $\avg{k}=5$, obtained using modularity maximization
  with $\gamma=1$ and the value of $\gamma$ that minimizes the
  description length, as shown in the legend. It also shows the results
  obtained with the NSBM. The horizontal dashed line marks the value
  $\sqrt{2E}$. The inset shows the inferred number of non-empty groups,
  instead of the effective number. (b) Description length versus
  $\gamma$ for $B=1$, for the same networks as in (a). The dashed
  vertical line marks the value of $\gamma$ corresponding to the minimum
  description length, and the horizontal line the description
  length of the ER model. (c) Effective number of groups for the same
  networks as in (b). The horizontal line marks the planted value. The
  panels (d) and (e) are analogous to
  (b) and (c), but with $B=587$. \label{fig:dl_gamma}}
\end{figure}

The behavior above also explains the tendency of the modularity method
to simultaneously overfit (i.e. when it finds spurious communities) and
underfit, i.e. when the number of groups exceeds the $\sqrt{\gamma 2E}$
resolution limit~\cite{fortunato_resolution_2007} it merges groups
together. In fact, we can use the value of description length to correct
for both these effects via the parameter $\gamma$ by choosing the value
that most compresses the network, as shown in Fig.~\ref{fig:dl_gamma}.
While using a value of $\gamma=1$ finds spurious groups whenever the
true number of planted groups is small, and too few groups whenever the
true number lies above $\sqrt{2E}$, the most compressive $\gamma$ values
reveals the correct number throughout the entire range, thus removing a
long-standing limitation of this method.

Although the above approach serves as principled, unified and
non-parametric solution to the overfitting and resolution limit problems
of modularity maximization, we emphasize that are other problems
intrinsic to the method that remains. In particular, optimizing $\gamma$
yields an effective number of groups, computed as
\begin{equation}
  B_e=\exp\left(-\sum_r\frac{n_r}{N}\ln \frac{n_r}{N}\right),
\end{equation}
which lies very close to the true value, but the actual number of
inferred groups is often larger, as shown in the inset of
Fig.~\ref{fig:dl_gamma}a. This is because the value of $Q$, and as a
consequence its description length encoding, are insensitive to the
existence of very small groups, therefore some marginal amount of
overfitting cannot be fully removed. More importantly, the method will
still enforce a characteristic scale for the community sizes, and will
not behave well when communities of unequal sizes
exist~\cite{lancichinetti_limits_2011}. The computation and minimization
of the description length can be seen as a ``post-processing'' of the
results obtained with modularity maximization, and it can only influence
the intrinsic biases of the method via a free parameter like $\gamma$. A
more direct strategy to tackle the vices of the method involves a more
appropriate formulation the prior assumptions, precisely as is done with
the SBM-based approaches~\cite{peixoto_bayesian_2019, zhang_statistical_2020}. In
Fig.~\ref{fig:dl_gamma} we show the result obtained with the nested
stochastic block model
(NSBM)~\cite{peixoto_hierarchical_2014,peixoto_nonparametric_2017},
discussed in more detail in Sec.~\ref{sec:nsbm}, which has no difficulty
in finding not only the effective number of groups, but also its nominal
value.

\section{Optimal problem instances}\label{sec:optimal-instances}

As discussed previously, problem instances $(\A,\bb)$ sampled from the
distribution
\begin{equation}\label{eq:Wmodel}
  P(\A,\bb|\beta) = \frac{\ee^{\beta W(\A,\bb)}}{Z(\beta, \sum_{i<j}A_{ij})\left[{N \choose 2} + 1\right]},
\end{equation}
are optimal for a community detection algorithm that maximizes the
quality function $W(\A,\bb)$, since no other algorithm can achieve
better average performance on those instances. If a quality function can
be written in terms of the microcanonical SBM parameters
$W(\A,\bb)=W(\bm e, \bm n)$, then it can be interpreted as being
proportional to the log-likelihood of a particular constrained version
of the SBM. We can see this by approximating
\begin{equation}
  Z(\beta, E) = \int e^{\beta W}\Xi(W,E)\;\dd W \approx e^{\beta W^*}\Xi(W^*, E),
\end{equation}
with $W^* = \operatorname{arg\ max}_W e^{\beta W}\Xi(W,E)$, such that
\begin{equation}
  P(\A,\bb|\beta) \approx \frac{\ee^{\beta [W(\bm e, \bm n)- W^*]}}{\Xi(W^*, \sum_{i<j}A_{ij})}.
\end{equation}
Approximating further
\begin{equation}
  \Xi(W^*, E) = \sum_B\Xi(W^*, B, E) \approx \Xi(W^*, B^*, E)
\end{equation}
with $B^* = \operatorname{arg\ max}_B \Xi(W^*, B, E)$, and neglecting
finite-size fluctuations around the most typical samples with $W(\bm e,
\bm n) = W^*$, we can write the likelihood as
\begin{equation}
  P(\A,\bb|\beta) \approx \frac{\delta_{W(\sum_re_{rr}/2, \sum_{i<j}A_{ij}, B^*), W^*}}{\Xi(W^*, B^*, \sum_{i<j}A_{ij})\left[{N \choose 2} + 1\right]}.
\end{equation}
where $W(E_{\text{in}}, E, B)$ is the value of the quality function for exactly
$E_{\text{in}}$ edges internal to equal-sized communities. Re-arranging, we
have
\begin{equation}
  P(\A,\bb|\beta) \approx P(\A|E_{\text{in}}^*, E, \bb)P(\bb|B^*)P(E),
\end{equation}
where $E_{\text{in}}^*$ is the solution of
\begin{equation}
  W(E_{\text{in}}, E, B^*) = W^*,
\end{equation}
and
\begin{equation}
  P(\A|E_{\text{in}}, E, \bb) = \frac{\delta_{\sum_{i<j}A_{ij}\delta_{b_i,b_j},E_{\text{in}}}\delta_{\sum_{i<j}A_{ij},E}}
  {{\sum_r{n_r\choose 2}\choose E_{\text{in}}}{\sum_{r<s}n_rn_s \choose E-E_{\text{in}}}}
\end{equation}
is the likelihood of a microcanonical planted partition SBM with exactly
$E_{\text{in}}$ edges internal to communities,
and
\begin{equation}
  P(\bb | B) = \frac{\prod_r\delta_{n_r, N/B}}{N! / [(N/B)!]^B}
\end{equation}
is the likelihood of a random partition into $B$ groups of the same
size, and finally $P(E)=\left[{N \choose 2} + 1\right]^{-1}$. The values
of $W^*$ and $B^*$ are uniquely determined by $\beta$ with
\begin{align}
  W^* &= \operatorname{arg\ max}_W\; e^{\beta W}\Xi(W,E)\\
  B^* &= \operatorname{arg\ max}_B\; \Xi(W^*, B, E).
\end{align}
Therefore, the model of Eq.~\ref{eq:Wmodel} is asymptotically equivalent
to sampling a network from a planted partition SBM with the number of
groups and assortativity strength determined by the same $\beta$
parameter.

The above equivalence is a more general, but compatible nonparametric
version of the approximate one shown for modularity in
Ref.~\cite{newman_equivalence_2016}. That work showed that if both the
number of groups and the planted partition mixing parameter are known
and fixed, and if the partitions have equal size and
density~\cite{zhang_statistical_2020}, then the maximum likelihood of
the degree-corrected planted partition model is approximately the same
as the maximum modularity one with a particular value of $\gamma$. 
In contrast, the model we derive above is nonparametric, i.e. generates
in addition to the network also the number of groups, partition, and mixing
strength, and does not rely on any assumptions on the data. Crucially, unlike
the model of Ref.~\cite{newman_equivalence_2016}, from ours we can compute
the description length of the data.

\begin{figure}
  \begin{tabular}{ccc}
    \includegraphics[width=.33\columnwidth]{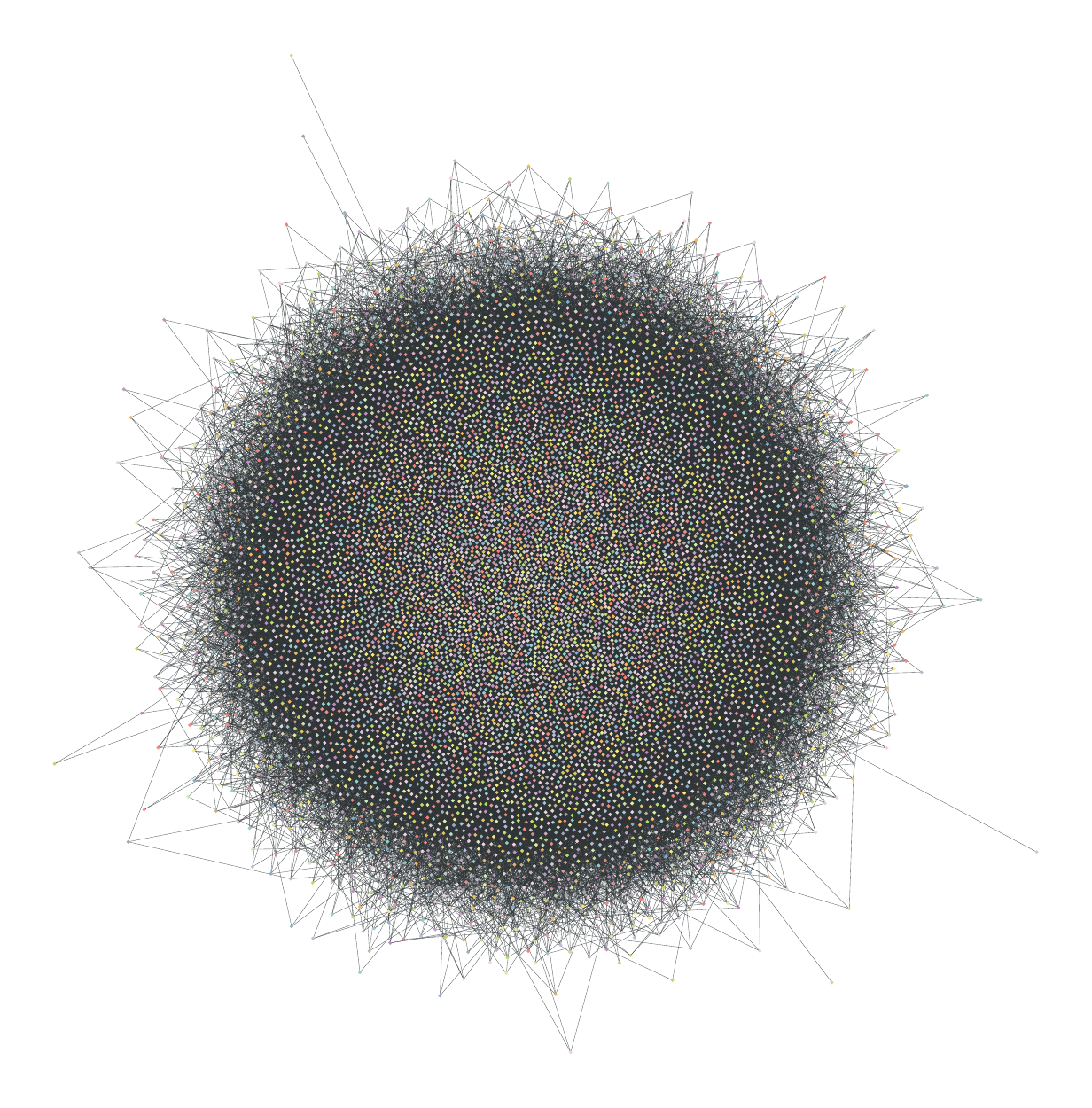} &
    \includegraphics[width=.33\columnwidth]{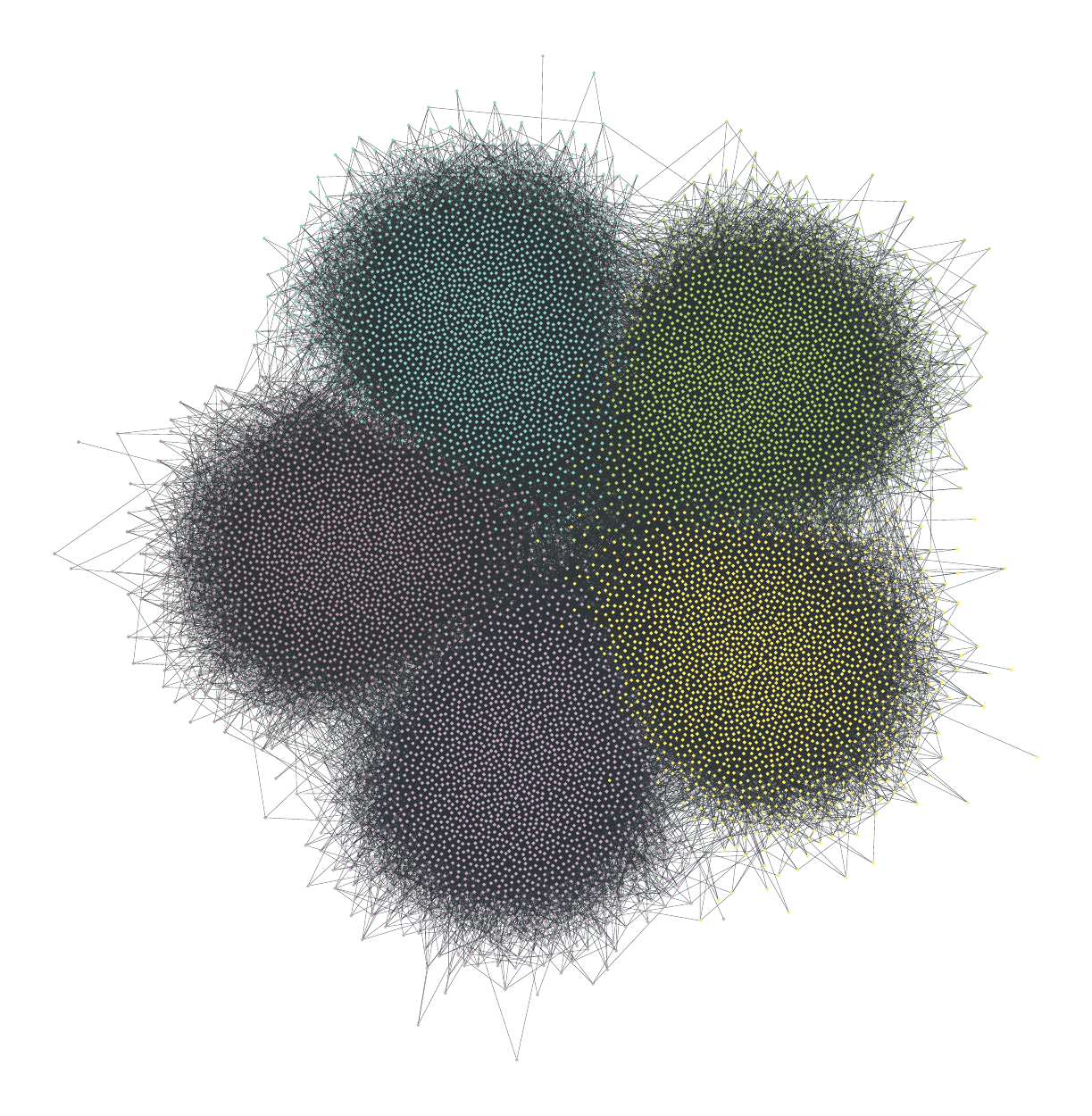} &
    \includegraphics[width=.33\columnwidth]{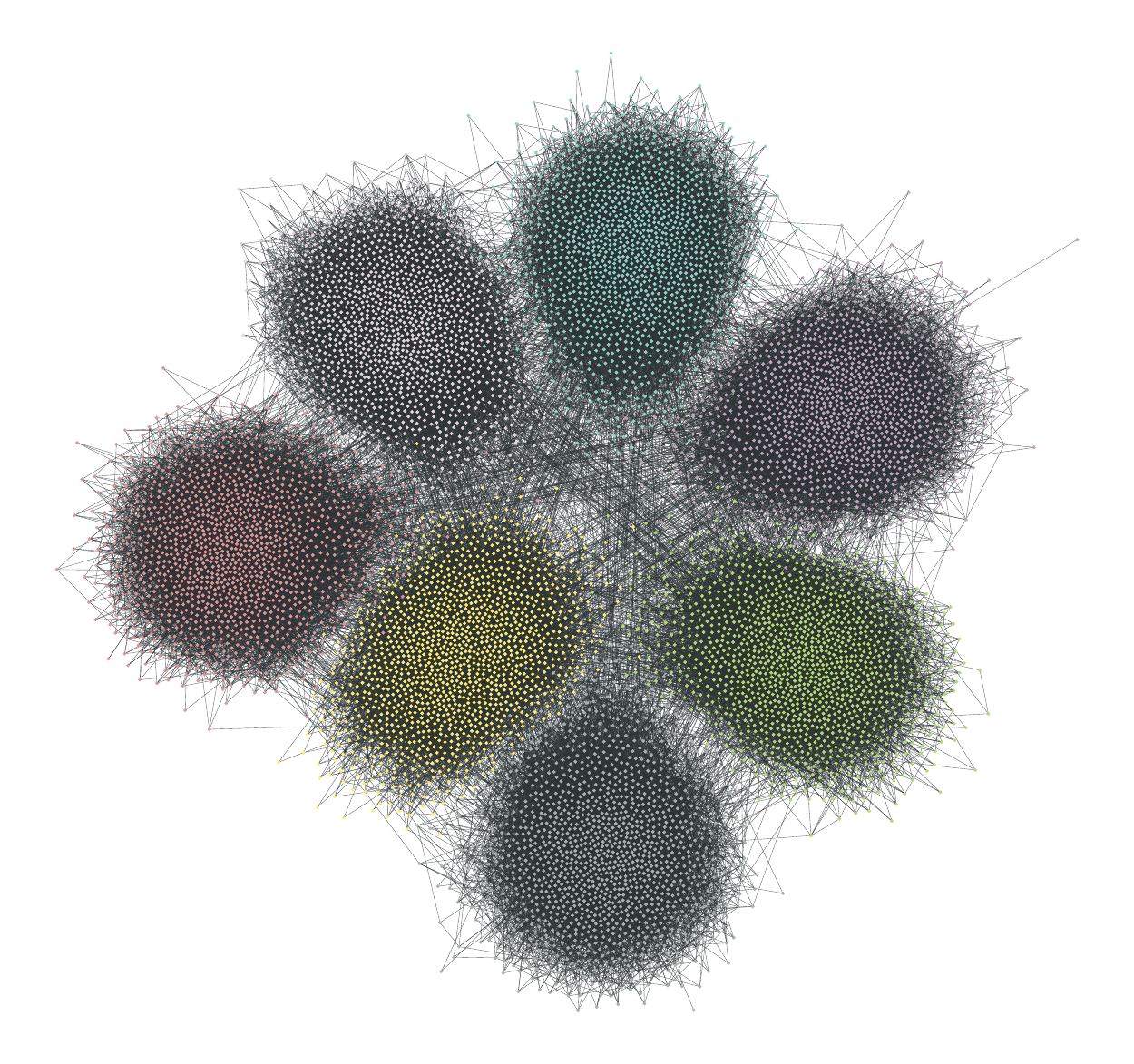} \\
    $\beta/N=17$ & $\beta/N=18$ & $\beta/N=25$ \\
    \includegraphics[width=.33\columnwidth]{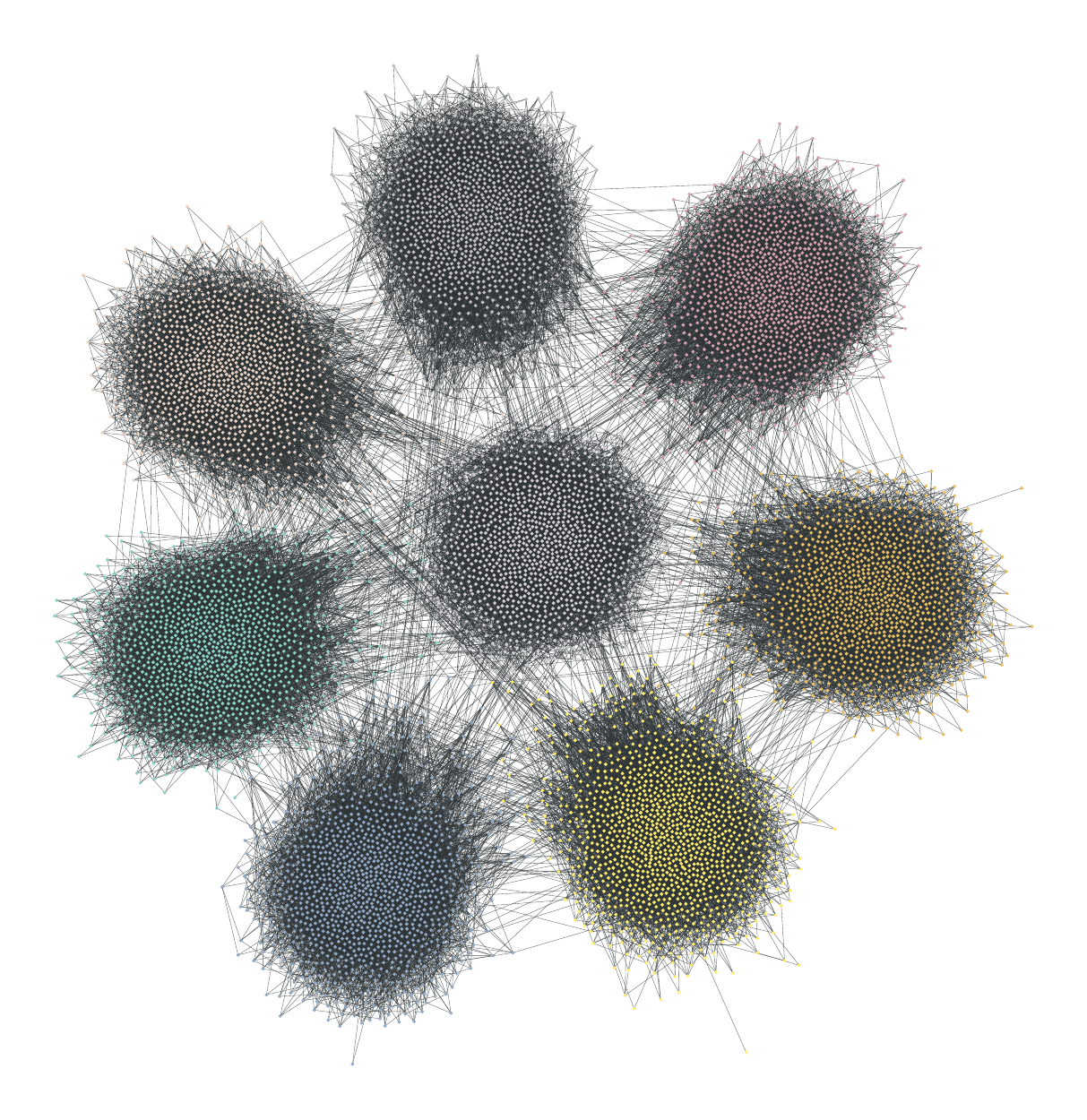} &
    \includegraphics[width=.33\columnwidth]{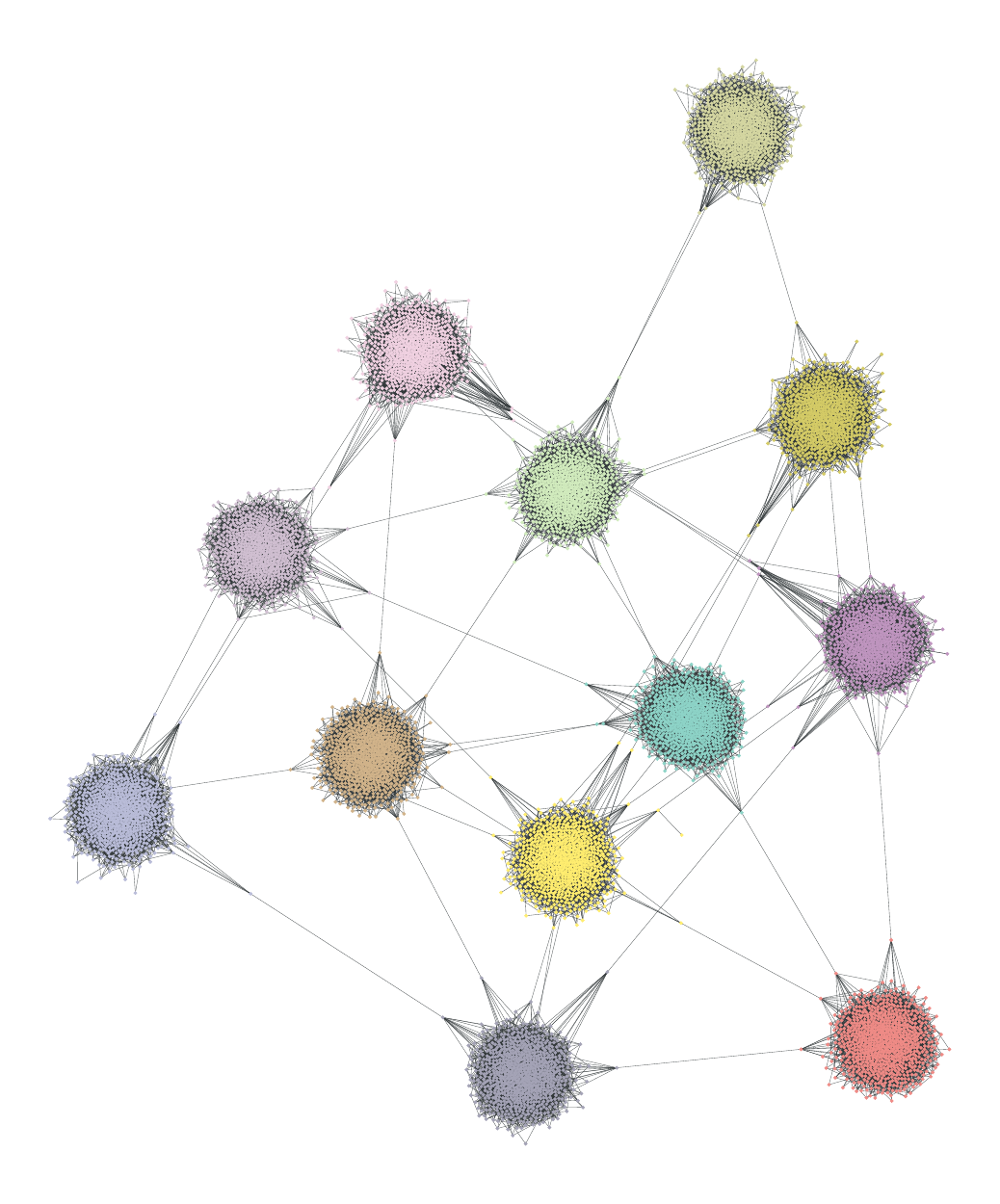} &
    \includegraphics[width=.33\columnwidth]{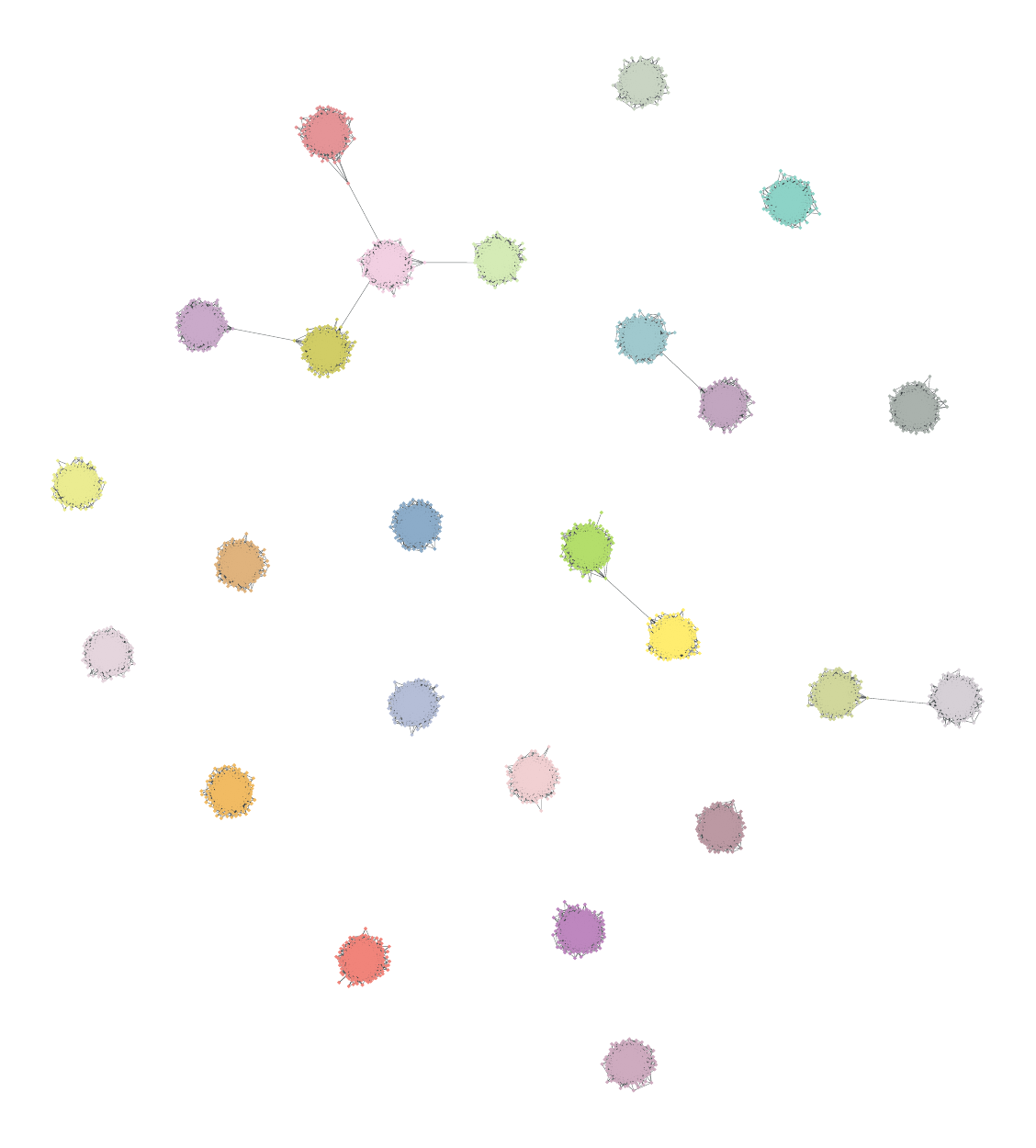} \\
    $\beta/N=30$ & $\beta/N=50$ & $\beta/N=100$
  \end{tabular}

  \caption{Samples from the implicit generative model behind modularity
  maximization with $\gamma=1$, for different inverse temperature values
  $\beta$, $N=10^4$ and $\avg{k}=10$. The colors indicate the sampled
  network partitions. For these problem instances, the method of
  modularity maximization is Bayes-optimal.\label{fig:Qsamples}}
\end{figure}

\begin{figure}
  \includegraphics[width=\columnwidth]{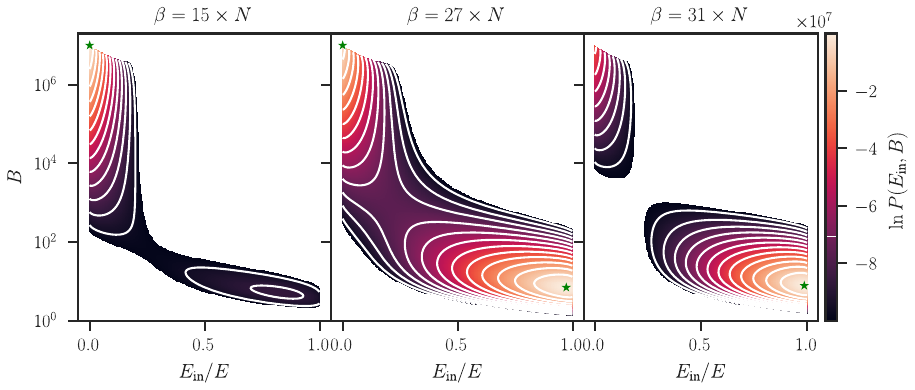}

  \caption{Joint probability $P(E_{\text{in}}, B|\beta)$ for the
    modularity model, with $N=10^7$,
    $\avg{k}=10$, $\gamma=1$, and different values of $\beta$. The global
    maxima of the distribution are marked with star symbols. As the
    value of $\beta$ increases, the global maximum changes abruptly from
    a value close to $(E_{\text{in}},B)=(0,N)$ to a value with large
    $E_{\text{in}}$ and finite $B$. \label{fig:Q_transition}}
\end{figure}

In Fig.~\ref{fig:Qsamples} we show some example networks sampled from
the optimal model for modularity maximization, for various values of
$\beta$. As discussed previously, for a small value of $\beta$ the model
concentrates on low $Q$ values with diverging $B \propto N$, and
undergoes a discontinuous transition at value $\beta=\beta^*$, after
which it concentrates on high $Q$ values with a finite $B$. An example
of this transition is shown in Fig.~\ref{fig:Q_transition} via the joint
probability $P(E_{\text{in}}, B|\beta) =\ee^{\beta Q(E_{\text{in}}, E,
\gamma, B)}\Xi(W,B,E)/Z(\beta)$.

\begin{figure}
  \includegraphics[width=\columnwidth]{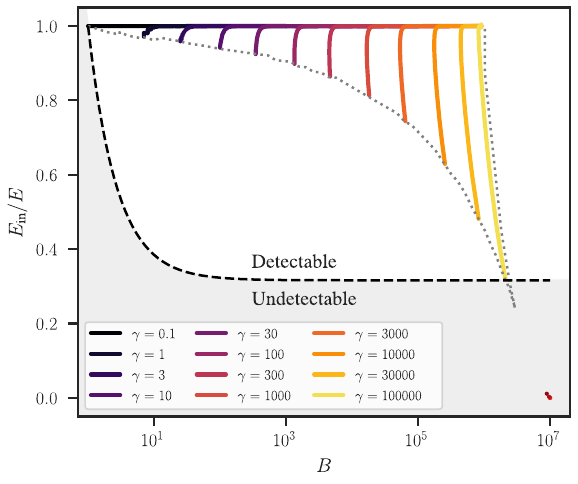}

  \caption{Feasible realizations of the modularity model. Each curve
  corresponds to the $(E_{\text{in}}, B)$ values achieved with $\beta$
  in the range $[0,\infty]$ for a specific value of $\gamma$, as
  indicated in the legend, $N=10^7$, and $\avg{k}=10$. The dotted line
  delineates the feasible region for any parameter value. The dashed
  line marks the detectability transition of
  Eq.~\ref{eq:detect}.\label{fig:Q_feasible}}
\end{figure}

Note that for a single value of $\gamma$ there is no way to
independently control the number of groups and strength of community
structure. However, we might imagine that setting the value of the
resolution parameter $\gamma$ would allow for a precise tuning of the
strength of assortativity $E_{\text{in}}$ together with any arbitrary
number of groups $B$ --- in other words, we could expect a bijection
between $(\beta,\gamma)$ and $(E_{\text{in}},B)$, up to
discretization. In reality, however, a wide range of $(E_{\text{in}},B)$
values is not achievable for any combination of $(\beta,\gamma)$, as we
show in Fig.~\ref{fig:Q_feasible}. Indeed, the model is only capable of
generating networks with quite strong community structure, far away from
the detectability threshold of the plated partition model, which lies at
\begin{equation}\label{eq:detect}
  \frac{E_{\text{in}}^{\star}}{E} = \frac{1}{B} + \frac{B-1}{B\sqrt{\avg{k}}}.
\end{equation}
For any network sampled from the PP model with $E_{\text{in}} <
E_{\text{in}}^{\star}$, it is not possible with any algorithm to recover
any information about the true
partition~\cite{decelle_asymptotic_2011}. As we see in
Fig.~\ref{fig:Q_feasible}, the optimal model for modularity only
generates networks with $E_{\text{in}}$ much larger than
$E_{\text{in}}^{\star}$ --- except for a small fraction of
$(\beta,\gamma)$ combinations that lead to very large $B$
values. However, the undetectable regime (and hence also the
detectability transition) only exists in the limit $B/N \to 0$, and the
values of $B$ for which we obtain $E_{\text{in}} <
E_{\text{in}}^{\star}$ scale proportionally with $N$ as it increases
(not shown). Therefore, it is not possible to generate an undetectable
community structure with this model, other than by setting $\beta <
\beta^*$, in which case the networks generated are maximally random and
uncorrelated with the node partitions.

The result above is not entirely surprising, since it is known that
modularity maximization is not an optimal algorithm for networks sampled
close to the detectability transition of the PP model, since it already
fails for easier problem instances~\cite{krzakala_spectral_2013}. If it
were possible to generate such hard realizations with the above optimal
model for modularity, it would lead to a contradiction.

Overall, we see that the optimal instances for modularity maximization are quite
contrived, and composed of unrealistically strong and uniform community
structure, resulting in relatively easy labelling tasks, as we will see in the
following section. (We demonstrate similar results for the Infomap objective in
Appendix~\ref{app:infomap}.) These problem instances are also unrealistic in
their regularity, with a maximally homogeneous community structure composed of
equal-sized groups that also have the same density. Although modularity
maximization is optimal for these instances, it is very likely that other
algorithms will work just as well for them too. In the following, we demonstrate
that more general algorithms indeed perform just as well in these instances, but
the opposite is not true: modularity maximization does not perform well with
instances that are optimal to a more general algorithm.

\section{``Cheap lunches''}\label{sec:cheap-lunches}

Recently, the notion of universal algorithms for community detection has
been challenged by a ``no free lunch'' (NFL)
theorem~\cite{peel_ground_2017}, which states that when averaged over
all instances of community detection problems, all conceivable
algorithms must yield the same performance. This would mean, therefore,
that no algorithm can be truly universal, and that for one algorithm to
behave better than another on a subset of the problem instances, then it
must do worse on the remaining instances in a complementary
fashion. However, digging only slightly below the surface of the
statement of the NFL theorem of Ref.~\cite{peel_ground_2017} reveals
that it in fact tells us very little about the kind of problems that
virtually any community detection method attempts to solve. As stated
previously, despite their different mathematical definitions, most
methods attempt to divide networks into groups of nodes with more
internal than external connections, or more generally, according to
arbitrary preferences of connection between groups. In spite of this,
the class of problems considered in Ref.~\cite{peel_ground_2017}
completely violates this qualitative constraint, and considers instead
as equally valid instances of a community detection problem any
arbitrary pairing of a network and a true node partition that an
algorithm needs to find to be maximally accurate
--- regardless of how the nodes are actually divided in this partition
and how this division relates to the structure of the network. In fact,
most such problem instances are \emph{unstructured}, in a formal sense,
since they correspond to maximally random networks with nodes divided in
equally maximally random partitions, in violent disagreement with almost
every notion of community structure in the entire literature on the
topic~\cite{peixoto_descriptive_2022}.

In more detail, the NFL theorem states that, given an arbitrary
deterministic community detection algorithm indexed by $f$ which
ascribes a partition $\hat{\bm b}_f(\A)$ to a network $\A$, and an
appropriately chosen error function $\epsilon(\bm b, \bm b')$, then we
must have
\begin{equation}\label{eq:nfl}
  \sum_{\A,\bb}\epsilon(\hat{\bm b}_f(\A), \bm b) = \Lambda_{\epsilon},
\end{equation}
where $\Lambda_{\epsilon}$ is a constant that does not depend on the
chosen algorithm $f$, only on the error function $\epsilon(\bm b, \bm b')$.
In other words, when summed over all possible
pairs $(\A,\bb)$, all algorithms must have the same
performance. Crucially, the sum above does not necessarily involve pairs
$(\A,\bb)$ which correspond to a partitioned network with any
actual community structure --- regardless of how one defines it --- they
are entirely arbitrary. In fact, we can re-write the statement of the
theorem using a probabilistic language, thus
\begin{equation}
  \sum_{\A,\bb}P(\A,\bb)\epsilon(\hat{\bm b}_f(\A), \bm b) \propto \Lambda_{\epsilon},
\end{equation}
where the joint probability is trivially uniform and hence uncorrelated,
i.e.
\begin{align}
  P(\A,\bb) &= P(\A)P(\bb),\\
  P(\A) &\propto 1,\\
  P(\bb) &\propto 1.
\end{align}
Indeed, in this situation a uniformity between algorithms is entirely
unsurprising, since the posterior distribution is maximally uniform
$P(\bb|\A) = P(\bb) \propto 1$, and the Bayes-optimal algorithm amounts
to simply selecting a random partition uniformly at random, ignoring the
network altogether. The best possible algorithm will achieve a minimal
accuracy corresponding to a blind random guess, and hence
$\Lambda_{\epsilon}$ will correspond to the maximal possible value for
every algorithm. Since all algorithms perform maximally poorly, there is
no actual trade-off between them in this scenario~\cite{peixoto_descriptive_2022}
--- in contrast to how the NFL theorem is sometimes
interpreted~\cite{ghasemian_evaluating_2019,ghasemian_stacking_2020}.

The vast majority of problem instances sampled from the uniform
distribution are incompressible, i.e. cannot be described using fewer
bits than what is prescribed by the uniform distribution, and hence
correspond to \emph{unstructured} problem instances. Crucially, the
subset of structured problem instances, i.e. a network with actual
community structure --- again, regardless of how one precisely defines
it --- has an asymptotic measure of zero with respect to the set of all
instances,
i.e. the probability of encountering them when sampling from the uniform
distribution will vanish rapidly as the size of the data
increases~\cite{cover_elements_1991}.  Therefore, the statement of
Eq.~\ref{eq:nfl} tells us very little about actual community detection
problems, which in order to be structured, need to be compressible. (The
same can be said about other kinds of NFL theorems, outside of community
detection~\cite{streeter_two_2003,mcgregor_no_2006,everitt_universal_2013,lattimore_no_2013,schurz_humes_2019,hutter_universal_2007}.)

Importantly, the NFL theorem does not imply that there is a performance
equivalence between algorithms when they are faced with structured
problem instances. Using our understanding of the connection between
descriptive community detection objectives and implicit network
generative models, here we address this issue and demonstrate that for
structured problem instances, there are asymmetries where more general
approaches can outperform more specialized ones, without degrading the
performance in more specific instances.

Let us consider two alternative distributions of problem instances,
$P(\A,\bb)$ and $Q(\A,\bb)$. We can quantify the ability of model
$Q(\A,\bb)$ to capture the structure of instances sampled from a model
$P(\A,\bb)$ via the Kullback-Leibler (KL) divergence from $Q$ to $P$,
\begin{align}
  D_{\text{KL}}(P||Q) &= \sum_{\A,\bb}P(\A,\bb)\ln\frac{P(\A,\bb)}{Q(\A,\bb)}\\
                   &= \sum_{\A,\bb}P(\A,\bb)\left[\Sigma_Q(\A,\bb) - \Sigma_P(\A,\bb)\right],
\end{align}
which in this context measures the average description length difference
according to models $Q$ and $P$, for problem instances sampled from
$P$. Note that the KL divergence is strictly positive,
$D_{\text{KL}}(P||Q)\ge 0$, with the equality attainable only for
$P=Q$. Therefore, it is not possible on average to obtain improved
compression with a code optimized for $Q$ if the instances come from
$P\ne Q$. Crucially, the KL divergence is in general asymmetric,
i.e. $D_{\text{KL}}(P||Q) \ne D_{\text{KL}}(Q||P)$. Therefore, the
amount of information ``wasted'' by encoding data from $P$ with model
$Q$ is not the same as encoding from $Q$ with $P$. Indeed, this
indicates the possibility of more general models which not only compress
their own instances optimally (as every model does), but also do very
well for instances of other models, while the converse is not true.  A
concrete example of this is a general mixture given by
\begin{equation}
  Q(\A,\bb) = \sum_{m=1}^MP_m(\A,\bb)P(m),
\end{equation}
where the individual components $P_m(\A,\bb)$ are entirely arbitrary. In
this case, we have $\Sigma_Q(\A,\bb) \le \Sigma_m(\A,\bb) -\ln P(m)$ for every
$m$, and hence
\begin{equation}
  D_{\text{KL}}(P_m||Q) \le -\ln P(m),
\end{equation}
where $-\ln P(m) = O(\ln M)$ if the mixtures have similar probability, while
the reverse $D_{\text{KL}}(Q||P_m)$ can be arbitrarily large. In our
context, we can speak of a good alternative code $Q$ for $P$ if
$D_{\text{KL}}(P||Q) = O(\ln N)$, since in this case the encoding
``penalty'' of using $Q$ instead of $P$ will be much smaller than the
optimal $\Sigma_P$, which tends to scale as $O(N\ln N)$.  Therefore, in
the uniform case $P(m)=1/M$, the general mixture will provide a good
description for any of its components even if their number $M$ grows as
any polynomial in $N$.

Since the intrinsic model behind modularity maximization
amounts to a particular parametrization of the SBM, we can therefore posit
that a more general mixture will have a superior performance in most cases,
while still performing very well for instances that are optimal for
modularity maximization. Here we review one such mixture, the
nested stochastic block model
(NSBM)~\cite{peixoto_hierarchical_2014,peixoto_nonparametric_2017}, and
demonstrate that it indeed possesses this property.

\begin{figure}[h!]
  \resizebox{\columnwidth}{!}{
  \begin{tabular}{cccc}
    \multicolumn{4}{c}{(a) Samples from modularity's implicit model}\\
    \includegraphics[width=.25\columnwidth]{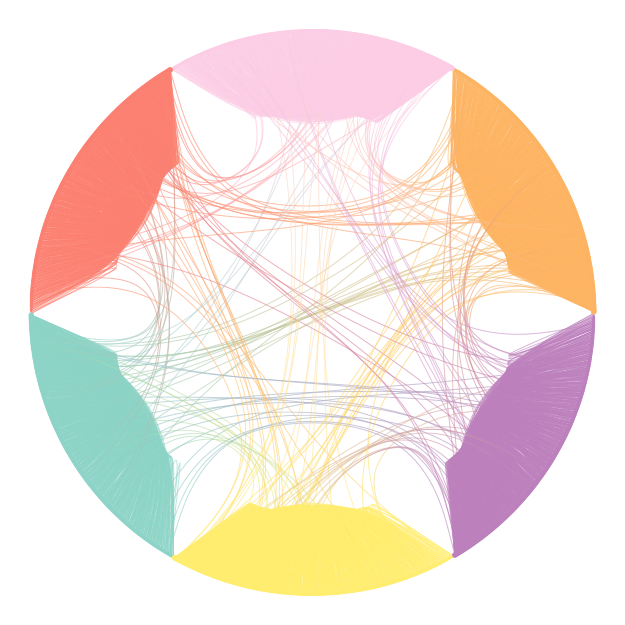} &
    \includegraphics[width=.25\columnwidth]{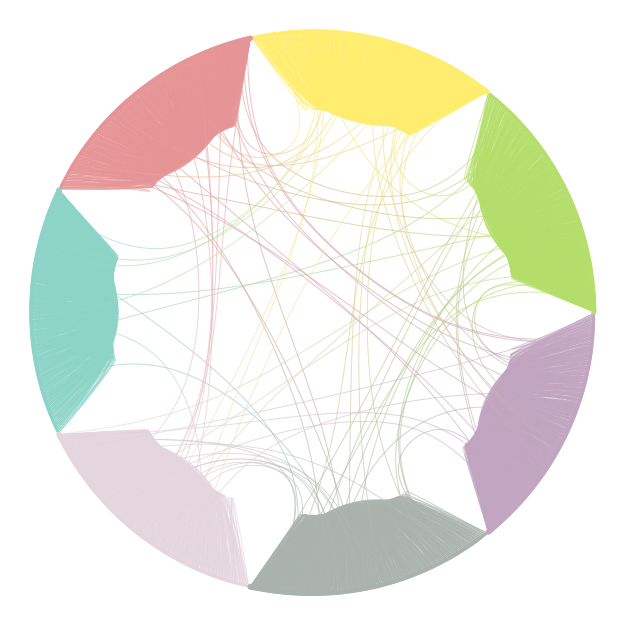} &
    \includegraphics[width=.25\columnwidth]{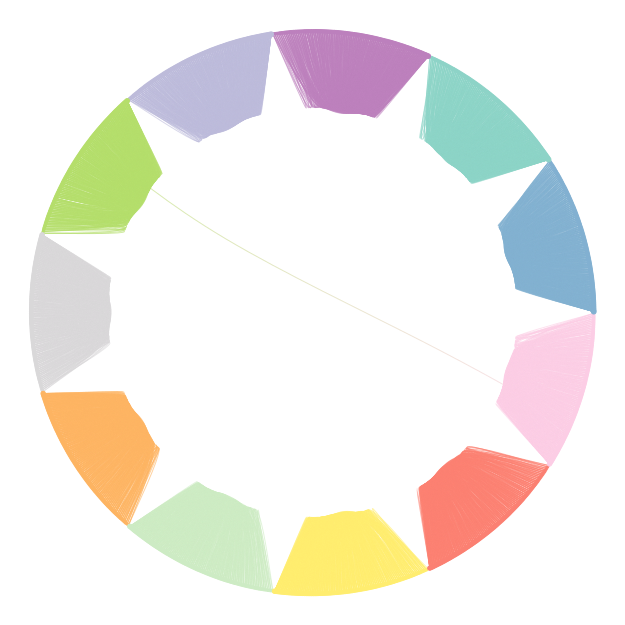} &
    \includegraphics[width=.25\columnwidth]{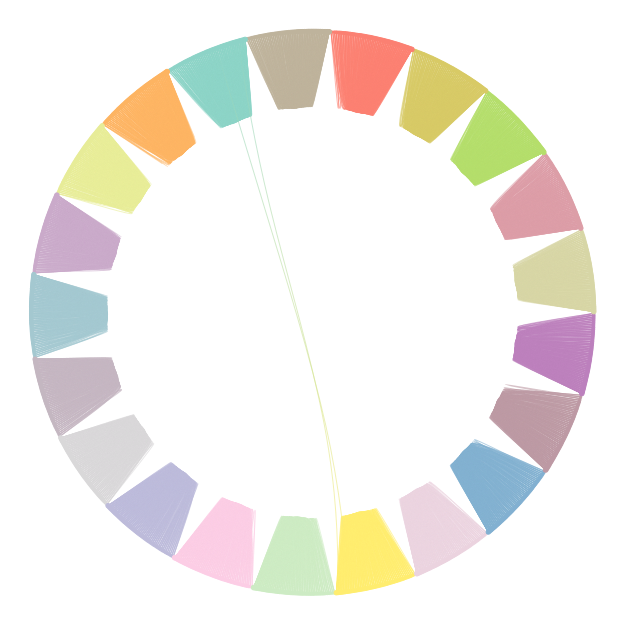} \\
    $\beta/N=25$ & $\beta/N=30$ & $\beta/N=50$ & $\beta/N=100$ \\[.5em]
    \multicolumn{4}{c}{(b) Samples from the NSBM}\\
    \includegraphics[width=.25\columnwidth]{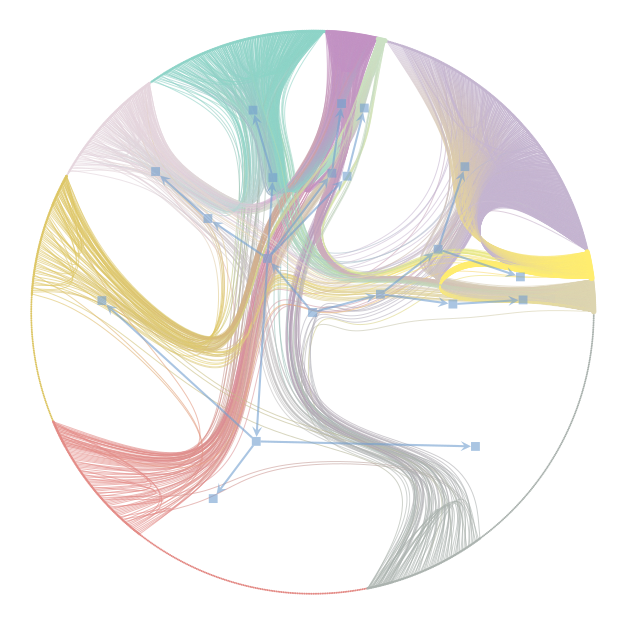} &
    \includegraphics[width=.25\columnwidth]{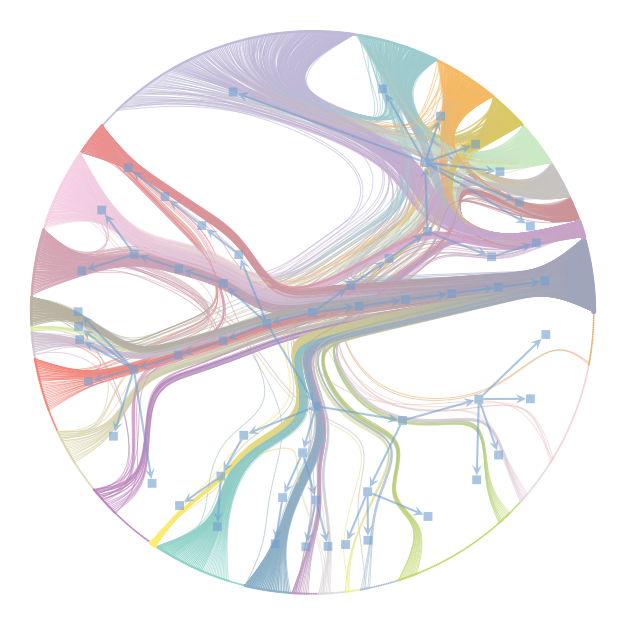} &
    \includegraphics[width=.25\columnwidth, angle=-90, origin=c]{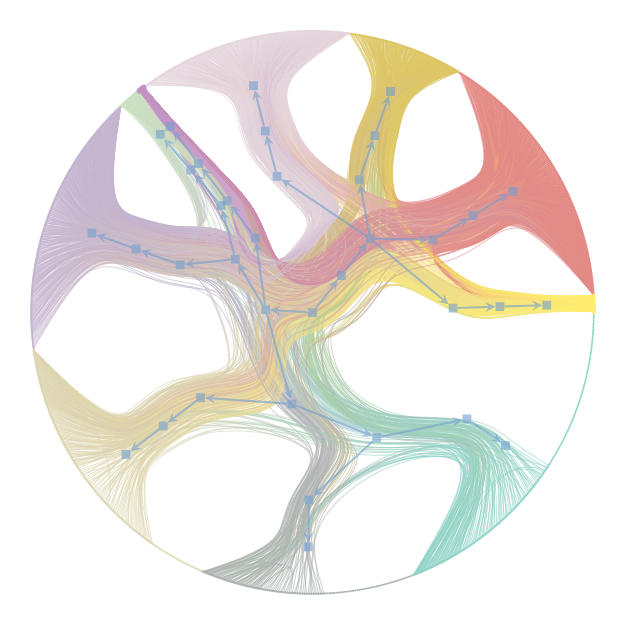} &
    \includegraphics[width=.25\columnwidth,angle=90,origin=c]{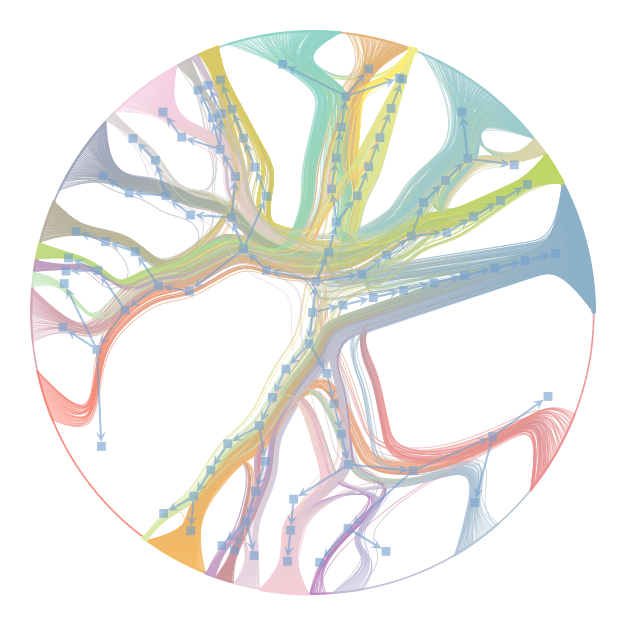}
  \end{tabular}}
  \resizebox{\columnwidth}{!}{
  \begin{tabular}{cc}
    \multicolumn{2}{c}{KL divergences between modularity and NSBM}\\
    \smaller[2] (c) Sampled from modularity's model & \smaller[2] (d) Sampled from NSBM \\
    \includegraphics[width=.5\columnwidth]{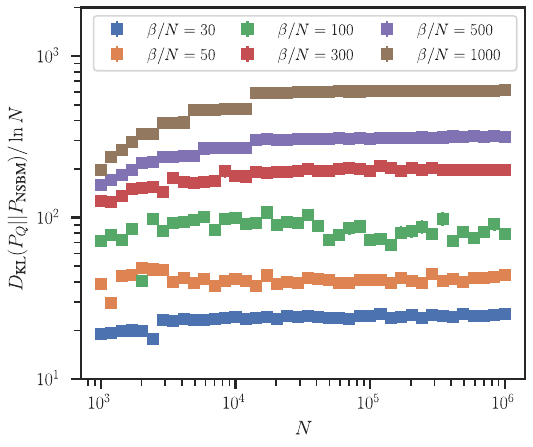} &
    \includegraphics[width=.5\columnwidth]{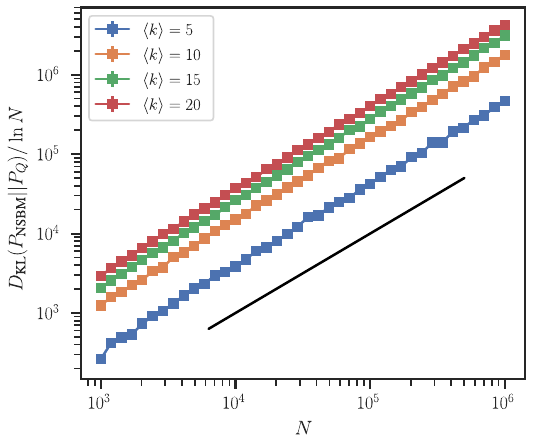}\\
    \multicolumn{2}{c}{Inference accuracy between modularity and NSBM}\\
    \smaller[2] (e) Sampled from modularity's model & \smaller[2] (f) Sampled from NSBM \\
    \includegraphics[width=.5\columnwidth]{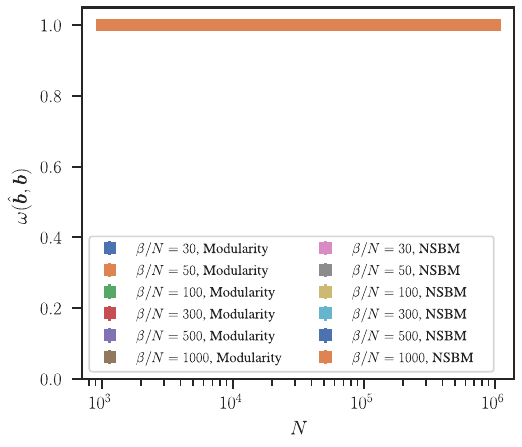} &
    \includegraphics[width=.5\columnwidth]{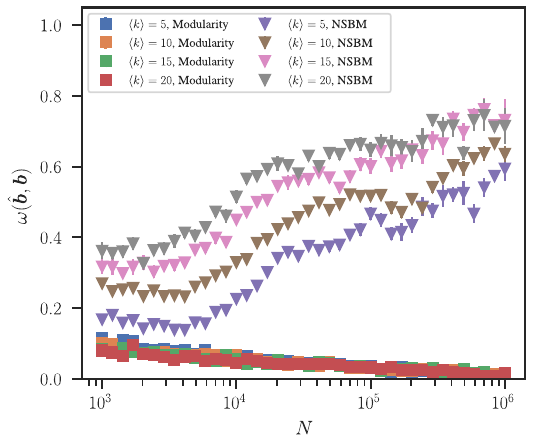}\\
  \end{tabular}}

  \caption{Asymmetric trade-off between the NSBM and the
  implicit model behind modularity maximization. In panel (a) we show
  samples from modularity's model for different values of $\beta$, and
  in (b) we show samples from the NSBM (which is nonparametric) --- in
  both cases visualized as chord diagrams. In (b) the corresponding
  hierarchical partitions are overlaid. In (c) and (d) we show the KL
  divergences, $D_{\text{KL}}(P_Q||P_{\text{NSBM}})$ and
  $D_{\text{KL}}(P_{\text{NSBM}}||P_Q)$ respectively, in both cases
  divided by $\ln N$, as a function of the number of nodes $N$. The
  solid line shows the linear slope. In (c) the networks are sampled
  from modularity's model with $\avg{k}=10$, and various values of
  $\beta$ as indicated in the legend. In (d) the networks are sampled
  from the NSBM, for various $\avg{k}$ as indicated in the legend. In
  (e) and (f) are shown the overlaps between the inferred and true
  partitions, for the same problem instances in (c) and (d),
  respectively, when inferred with modularity maximization and with the
  NSBM, as indicated in the legend.
  \label{fig:nsbm}}
\end{figure}

\subsection{The nested stochastic block model (NSBM)}\label{sec:nsbm}

The NSBM is based on a parametric formulation of the microcanonical SBM,
which is defined by a likelihood
\begin{equation}
  P(\A|\bm e, \bb),
\end{equation}
where $\bm e = \{e_{rs}\}$ is again the matrix of edge counts between
groups. The matrix $\bm e$ determines the mixing patterns between
groups, which is a free parameter. Clearly, we can realize optimal
instances of modularity by choosing $\bm e$ and $\bm b$ accordingly. The
NSBM consists of introducing a parametric prior for $\bm e$ which
depends on a partition $\bm b_2$ of the \emph{groups} of $\bm b$, and
another matrix of edge counts $\bm e_2 = \{e_{tu}^{(2)}\}$, with
elements $e_{tu}^{(2)}$ determining the number of edges between groups
of groups. As a result, we have a marginal likelihood
\begin{equation}
  P(\A|\bm e_2, \bb, \bb_2) = \sum_{\bm e}P(\A|\bm e, \bb)P(\bm e | \bm e_2, \bb_2),
\end{equation}
with the sum having trivially only one non-zero summand, due to the hard
constraints imposed. Naturally, we can proceed indefinitely up to $L$
hierarchical levels, where we enforce that on the last level $L+1$ there
is a trivial partition into one group, leading to a marginal likelihood
\begin{equation}
  P(\A|\bb, \bb_2, \dots, \bb_{L}).
\end{equation}
Choosing priors $P(\bb_{l})$ for the partitions leads to a nonparametric
joint distribution $P(\A,\bb, \bb_2, \dots, \bb_{L})$ and a description
length for the hierarchical partition given by
\begin{equation}
  \Sigma(\A,\bb, \bb_2, \dots, \bb_{L}) = -\ln P(\A,\bb, \bb_2, \dots, \bb_{L}).
\end{equation}
For further details on the derivation of the likelihoods, including the
degree-corrected variation (DC-NSBM), we refer to
Refs~\cite{peixoto_hierarchical_2014,peixoto_nonparametric_2017}. The
description length for the first-level partition is obtained by
marginalization,
\begin{align}
  \Sigma_{\text{NSBM}}(\A,\bb) &= -\ln\sum_{\bb_2,\bb_3,\dots,\bb_L}P(\A,\bb,\bb_2,\bb_3,\dots,\bb_L)\\
                            &\leq -\ln P(\A,\bb,\bb^*_2,\bb^*_3,\dots,\bb^*_L).
\end{align}
Although the sum over the higher-level partitions is intractable, the
marginal description length is upper bounded by any particular choice
$\{\bb^*_l\}$, as shown in the last line of the above equation. This
gives us an upper bound for $D_{\text{KL}}(P_Q||P_{\text{NSBM}})$ and a
lower bound for $D_{\text{KL}}(P_{\text{NSBM}}||P_Q)$, which are
sufficient for our analysis.

\begin{figure}[b!]
  \includegraphics[width=\columnwidth]{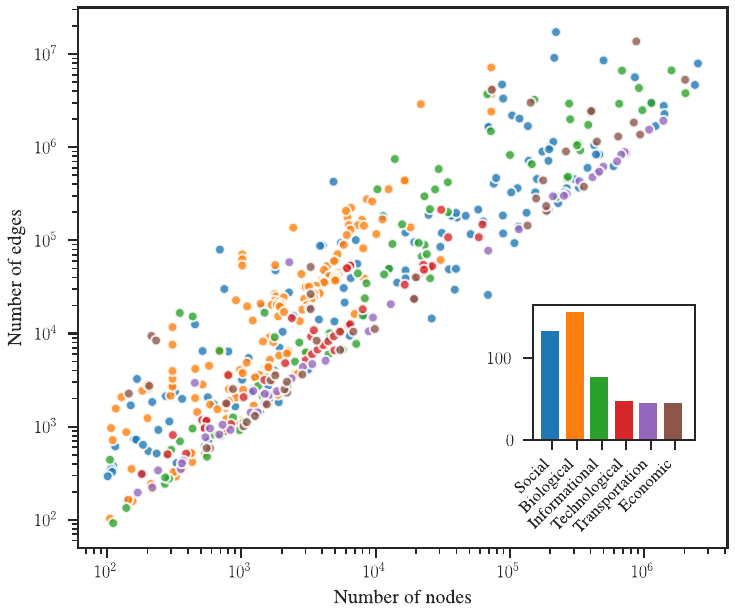} \caption{Number
  of nodes and edges, as well as distribution of domains (inset), for the
  509 empirical networks considered in this work, avaliable from the
  Netzschleuder repository~\cite{peixoto_netzschleuder_2020}. The symbol
  colors correspond to the network domain, as shown in the
  inset.\label{fig:empirical_desc}}
\end{figure}

\begin{figure*}
  \includegraphics[width=\textwidth]{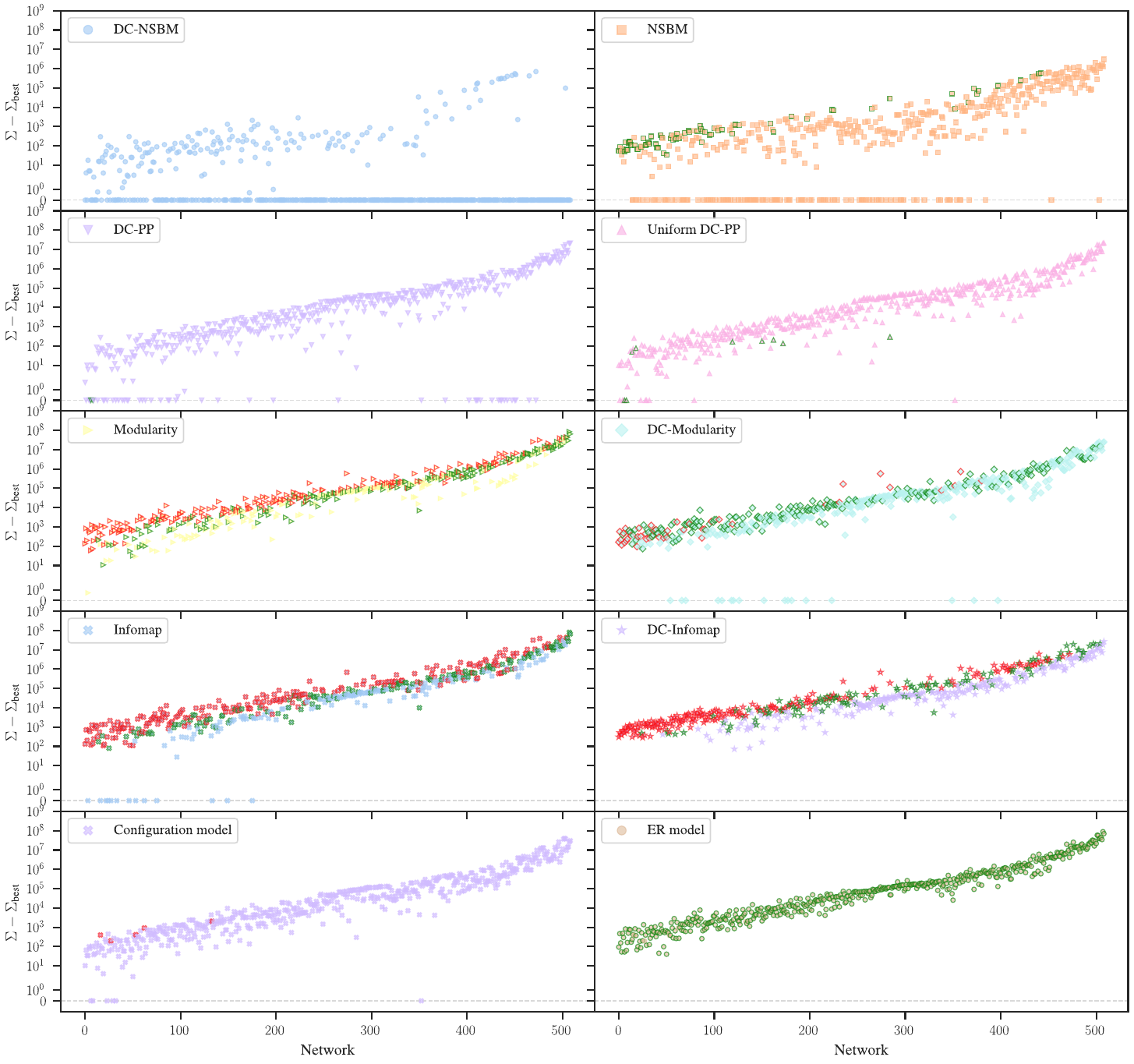}

  \caption{Difference in description length values according to the best
  model, obtained with several models for 509 empirical networks,
  ordered according to number of edges. A value of zero indicates that
  the respective model is the most compressive for the particular
  network. Symbols highlighted in red (green) correspond to description
  length values that are larger than the Erd\H{o}s-Rényi model
  (configuration model). \label{fig:empirical}}
\end{figure*}

In Fig.~\ref{fig:nsbm}(a) and (b) we compare samples from modularity's
implicit model and the NSBM. Contrary to the former, the NSBM is
completely nonparametric and yields more realistic problem instances
that combine structure with disorder at several scales. Although they
have an extremely varied number and composition of groups, and mixing
patterns between them, the sampled instances always deviate from a
maximally random graph
--- they are always compressible. Indeed, the structural regularity of a
lower level of the hierarchy is generated with some amount of randomness
and regularity from the level above, and so on recursively, attributing
the samples with a mixture or randomness and regularity at multiple
scales. This larger diversity of samples from the NSBM comes precisely
from its more agnostic character when it is used for inference, since in
this case we make fewer commitments about the structure of the data ---
with respect to the number of groups, how uniformly distributed they are
and the preference of connections between them --- before the data is
actually seen. Importantly, as we will shortly demonstrate, once these
patterns are actually identified, the resulting description length tends
to be very close to the optimal one~\cite{peixoto_nonparametric_2017}.

Due to its more general character, the NSBM generates the kind of
regular community structure expected by modularity only with a
relatively low probability, and hence provides a strictly sub-optimal
encoding for networks that are sampled from this model. However, as
Fig.~\ref{fig:nsbm}(c) shows, the KL divergence
$D_{\text{KL}}(P_Q||P_{\text{NSBM}})$ grows only logarithmically with
$N$, meaning that it can nevertheless efficiently describe networks
sampled from this model. The opposite situation, however, is quite
different: As Fig.~\ref{fig:nsbm}(d) shows, the reversed KL divergence
$D_{\text{KL}}(P_{\text{NSBM}}||P_Q)$ grows log-linearly with $N$,
meaning that modularity's model is very inefficient at encoding samples
from the NSBM.

It is important to remember that, instead of compression directly, typically the
primary objective in community detection is simply to uncover latent
community assignments. Although these objectives are intimately related
--- as we already discussed, the optimal accuracy is always obtained
with the true generative model, which is also the only one that can
achieve maximal compression --- a method might still be maximally
successful at uncovering the correct community labels while providing
strictly inferior compression. We show this in Fig.~\ref{fig:nsbm}(e),
with the maximum overlap $\omega(\hat\bb,\bb)$ between the inferred and
true partitions, $\hat\bb$ and $\bb$ respectively, defined as
\begin{equation}
  \omega(\hat\bb,\bb) = \max_{\mu}\;\frac{1}{N} \sum_i\delta_{\hat b_i,\mu(b_i)},
\end{equation}
where $\mu(r)$ is a bijection between the labels of $\hat\bb$ and $\bb$,
for problem instances sampled from modularity's model, and inferred both
with modularity maximization and the NSBM. In all cases (which consist
only of $\beta>\beta^*$, otherwise the overlap is always zero) the
overlap is maximal with $\omega(\hat\bb,\bb) =1$, showing that both
methods uncover the exact same partition for these easy
instances. Again, the opposite situation is quite different: with
problem instances sampled from the NSBM, the accuracy of modularity
maximization tends to zero, while the NSBM performs significantly
better; although not perfectly --- there is no guarantee of perfect
recovery in these harder instances, only optimality.

It is not surprising that modularity maximization can neither compress
nor correctly uncover the true assignments of samples from the NSBM,
since those will not necessarily correspond to assortative
communities. Our central point here is there is a lack of trade-off: the
NSBM performs just as well for obvious assortative instances, while
still being able to accommodate more general structures that are harder
to detect.

Note that in the discussion above we did not have to make any reference
to particular domains of application. The lack of trade-off is a general
principle that must hold for mixtures in general, and can be articulated
simply using fundamental concepts of mixing patterns between
groups. Although one could expect networks belonging to different
domains having different kinds of mixing patterns, the above arguments
tell us that the superiority of hierarchical mixtures should transcend
various domains. We evaluate this hypothesis in the following.

\section{Empirical networks}\label{sec:empirical}

\begin{figure}
  \includegraphics[width=\columnwidth]{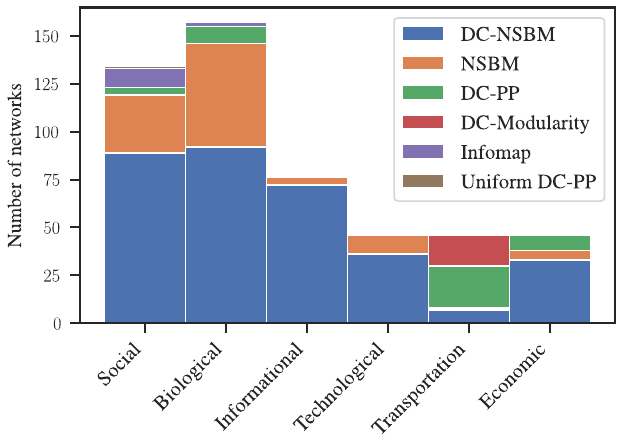}
    \caption{Number of networks for which each model provides the
    smallest description length, as indicated in the legend, across
    the different domains.\label{fig:domain}}
\end{figure}

\begin{figure*}
  \begin{tabular}{cc}
    (a) Fraction of relative compressions &
    (b) Average compression ratio\\
    \includegraphics[width=\columnwidth]{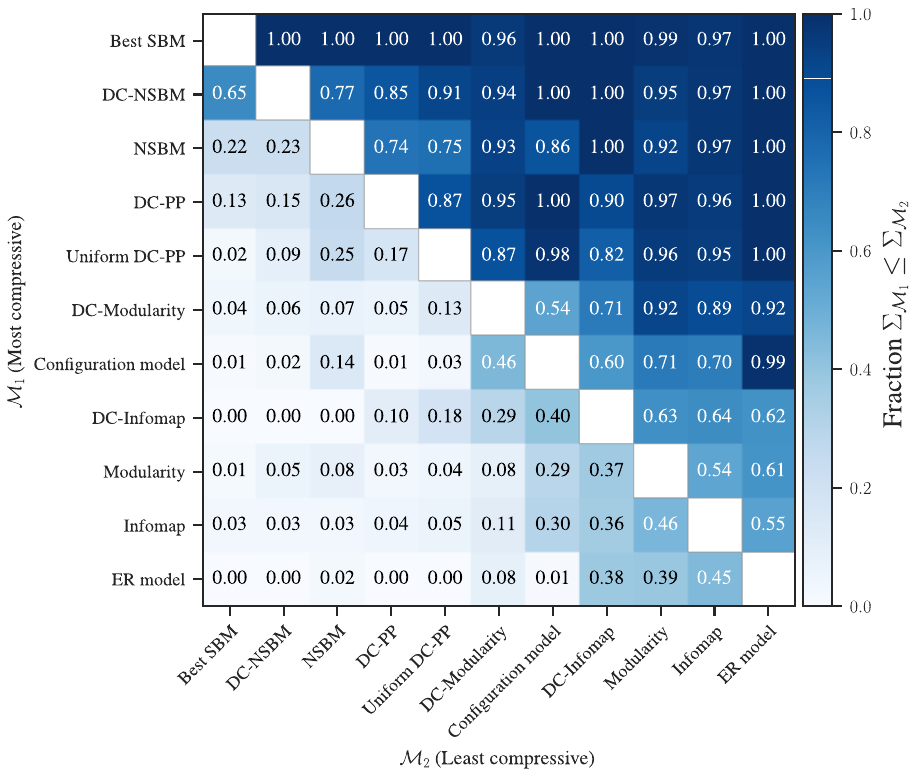}&
    \includegraphics[width=\columnwidth]{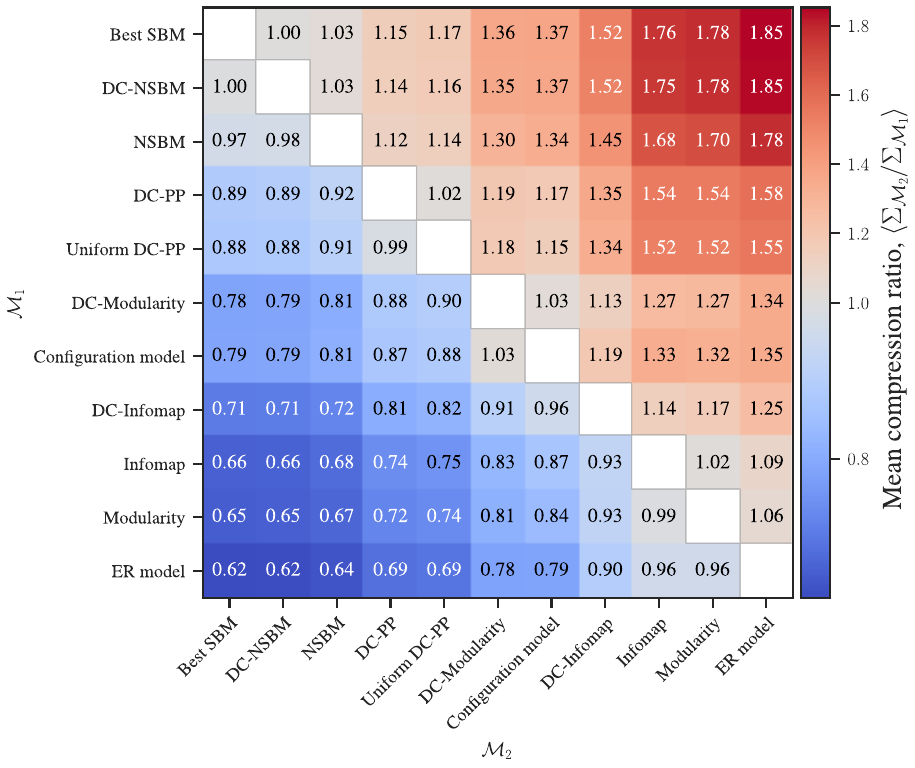}
  \end{tabular}
  \caption{(a) Fraction of networks in our corpus where a given model $\mathcal{M}_1$ (vertical axis) achieves equal or better
    compression than the alternative model $\mathcal{M}_2$ (horizontal
    axis). (b) Average compression ratio between models
    $\avg{\Sigma_{\mathcal{M}_2}/\Sigma_{\mathcal{M}_1}}$ across all
    networks. In both (a) and (b) the order of models corresponds to the
    SpringRank~\cite{de_bacco_physical_2018}, computed using the
    respective pairwise comparisons.\label{fig:mcomp}}
\end{figure*}

\begin{figure}
  \includegraphics[width=\columnwidth]{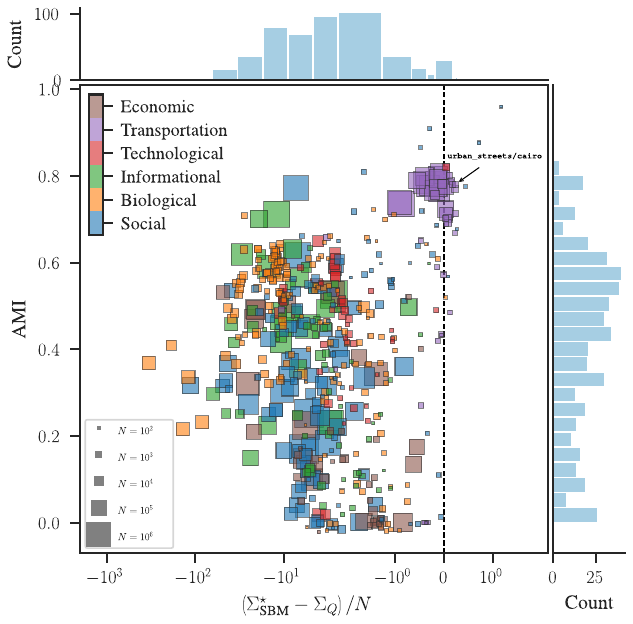}\\
  \includegraphics[width=\columnwidth]{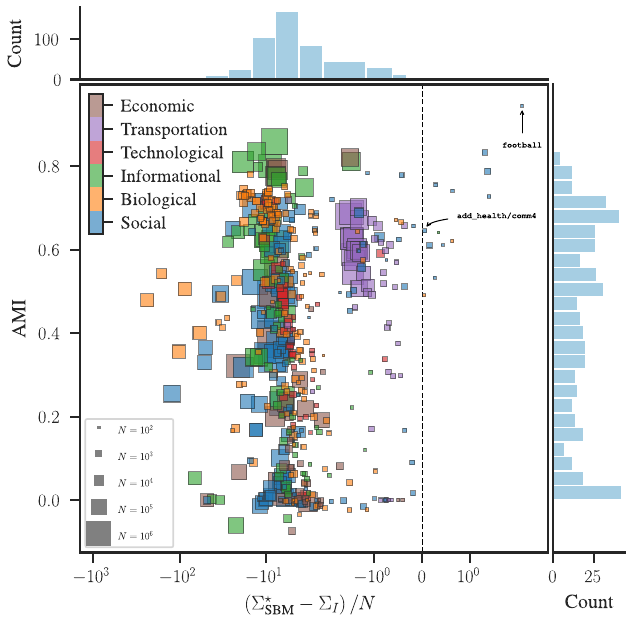}

  \caption{Adjusted mutual information (AMI) between the partitions
  inferred with the best fitting SBM and either modularity (top) or
  Infomap (bottom) as a function of the description length difference
  between models, $\Sigma_{\text{SBM}}^{\star}$, and $\Sigma_{Q}$ or
  $\Sigma_{I}$, divided by the number of nodes in the network. The
  symbol colors indicate the domain, and the size the number of nodes in
  the network, as shown in the legends. The text annotations refer to
  the networks shown in Fig.~\ref{fig:exceptions}\label{fig:ami}}
\end{figure}
\begin{figure}
  \begin{tabular}{cc}
    \multicolumn{2}{c}{(a) Streets of Cairo ($\text{AMI}=0.78$)}\\
    \multicolumn{2}{c}{\smaller \texttt{(urban\_streets/cairo)}}\\
    \includegraphics[width=.5\columnwidth]{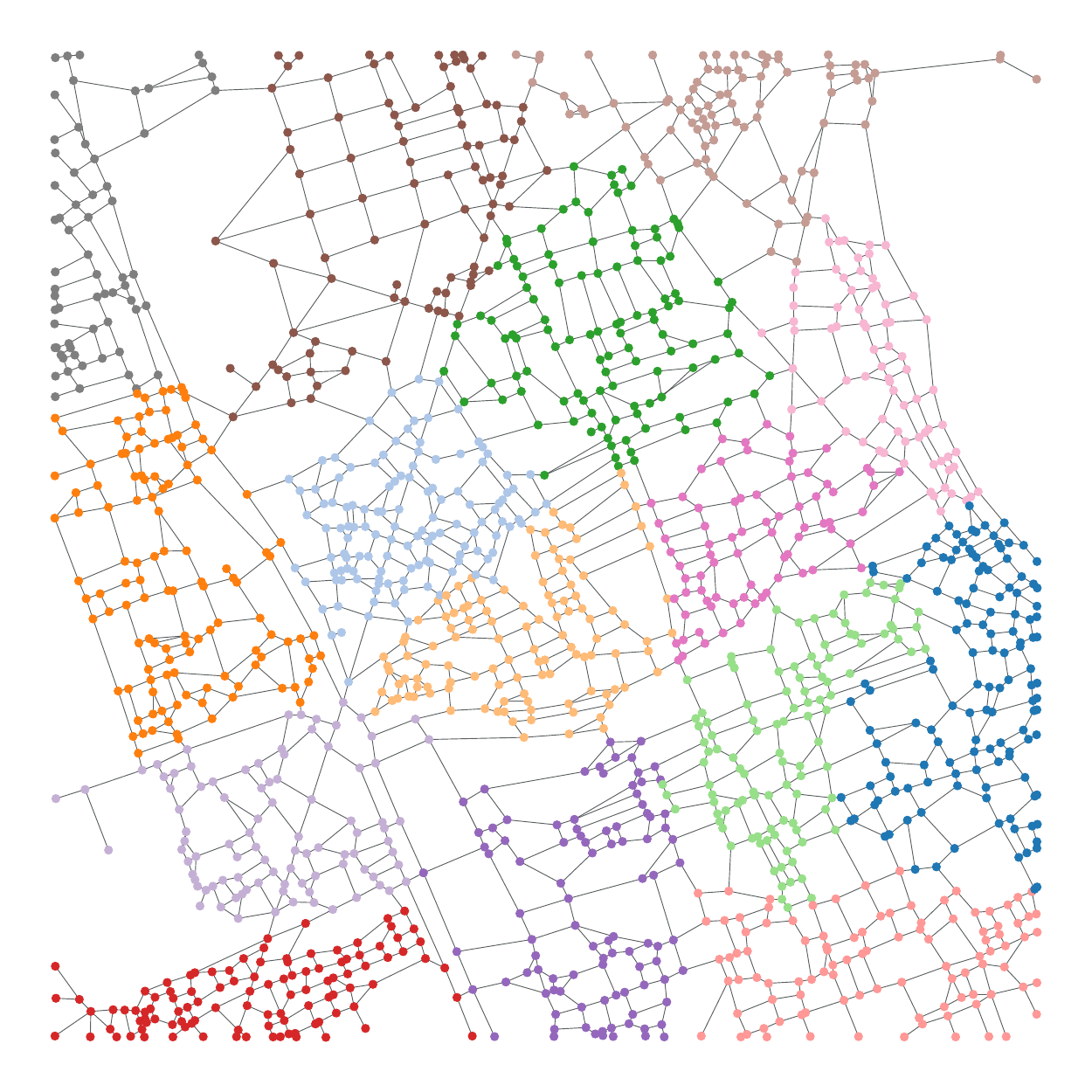}&
    \includegraphics[width=.5\columnwidth]{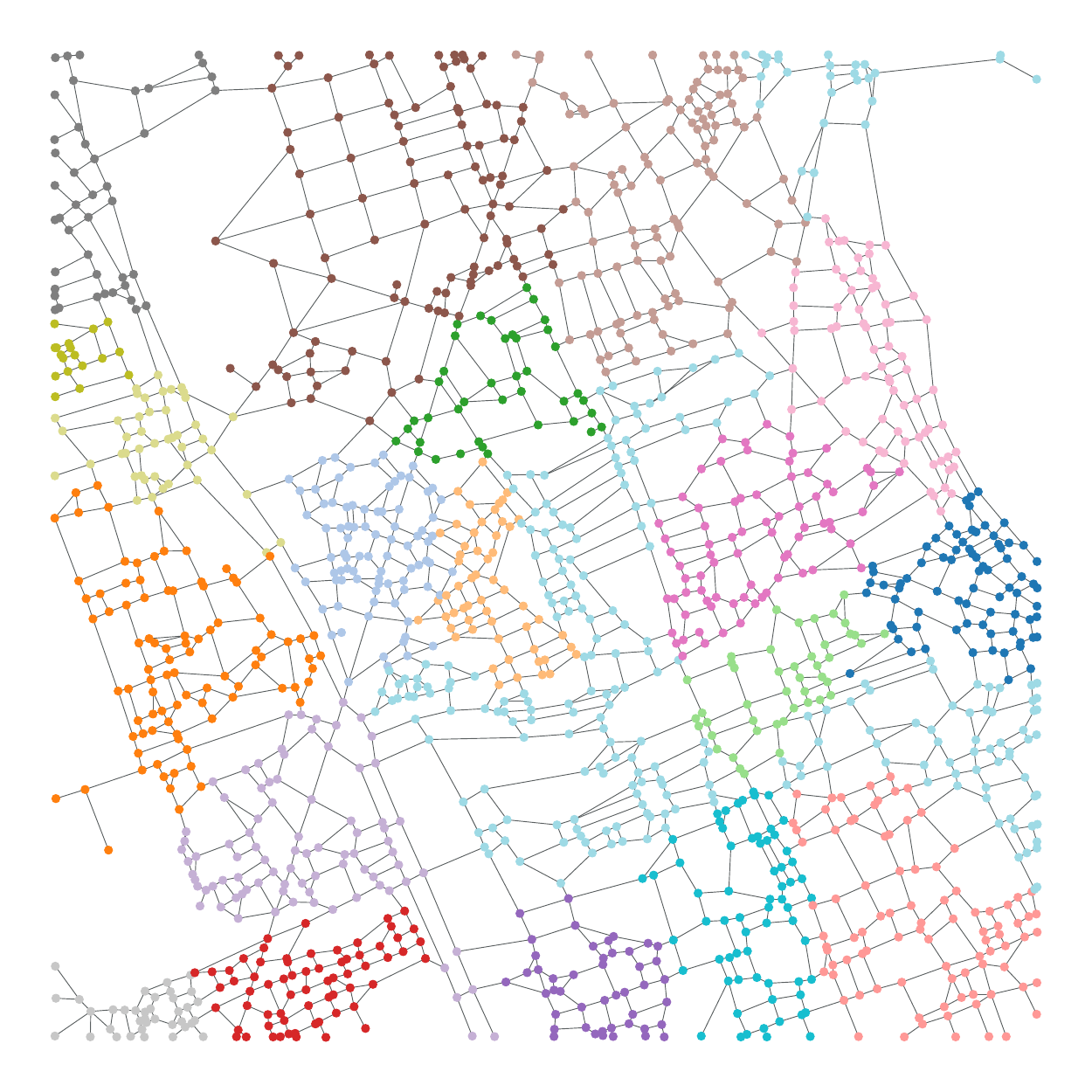}\\
    $\Sigma_{\text{DC-PP}} = 14178$ nats &
    $\Sigma_{\text{I}} = 13825$ nats \\[.5em]
    \multicolumn{2}{c}{(b) High-school friendships ($\text{AMI}=0.65$)}\\
    \multicolumn{2}{c}{\smaller \texttt{(add\_health/comm4)}}\\
    \includegraphics[width=.5\columnwidth]{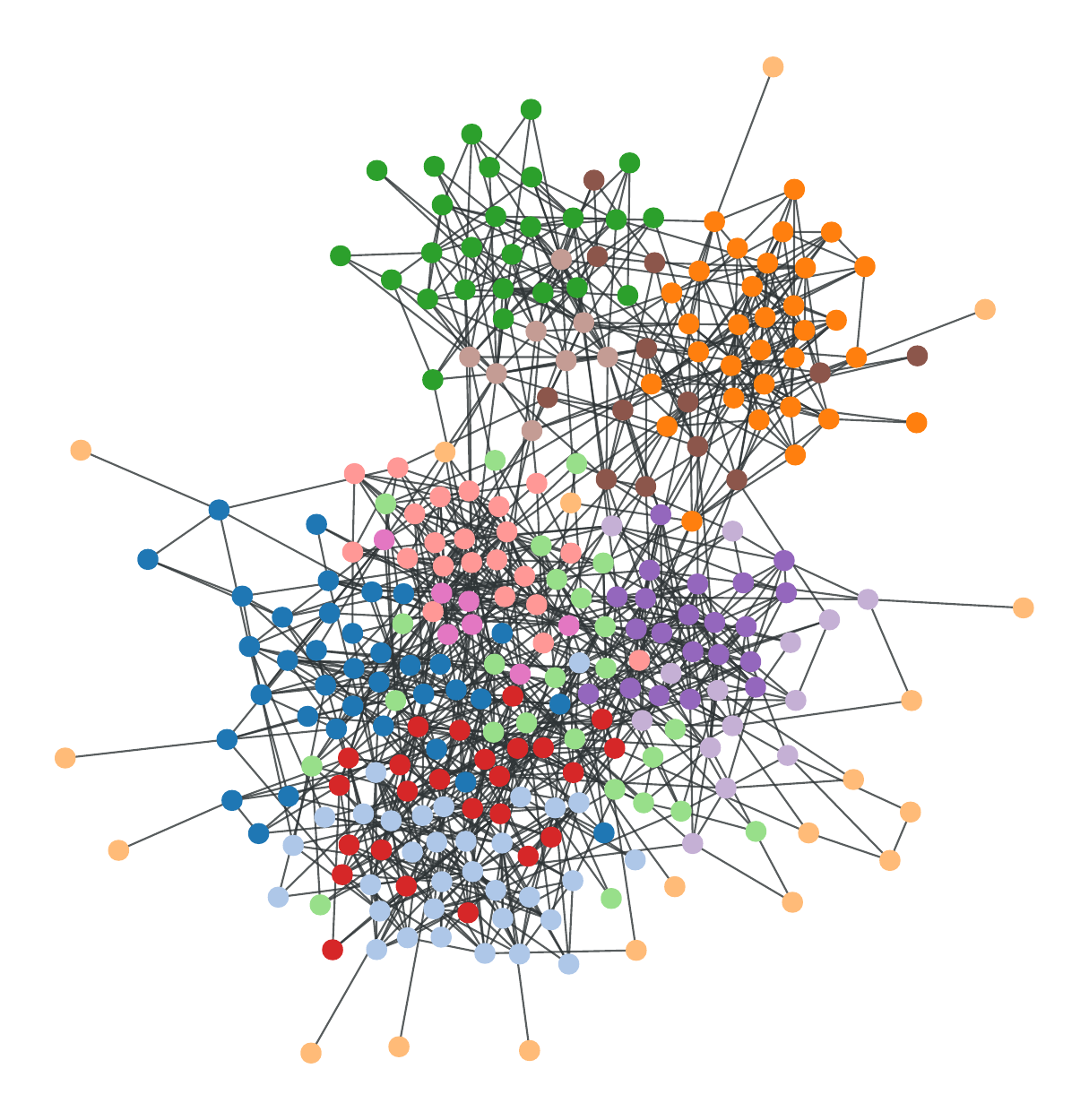}&
    \includegraphics[width=.5\columnwidth]{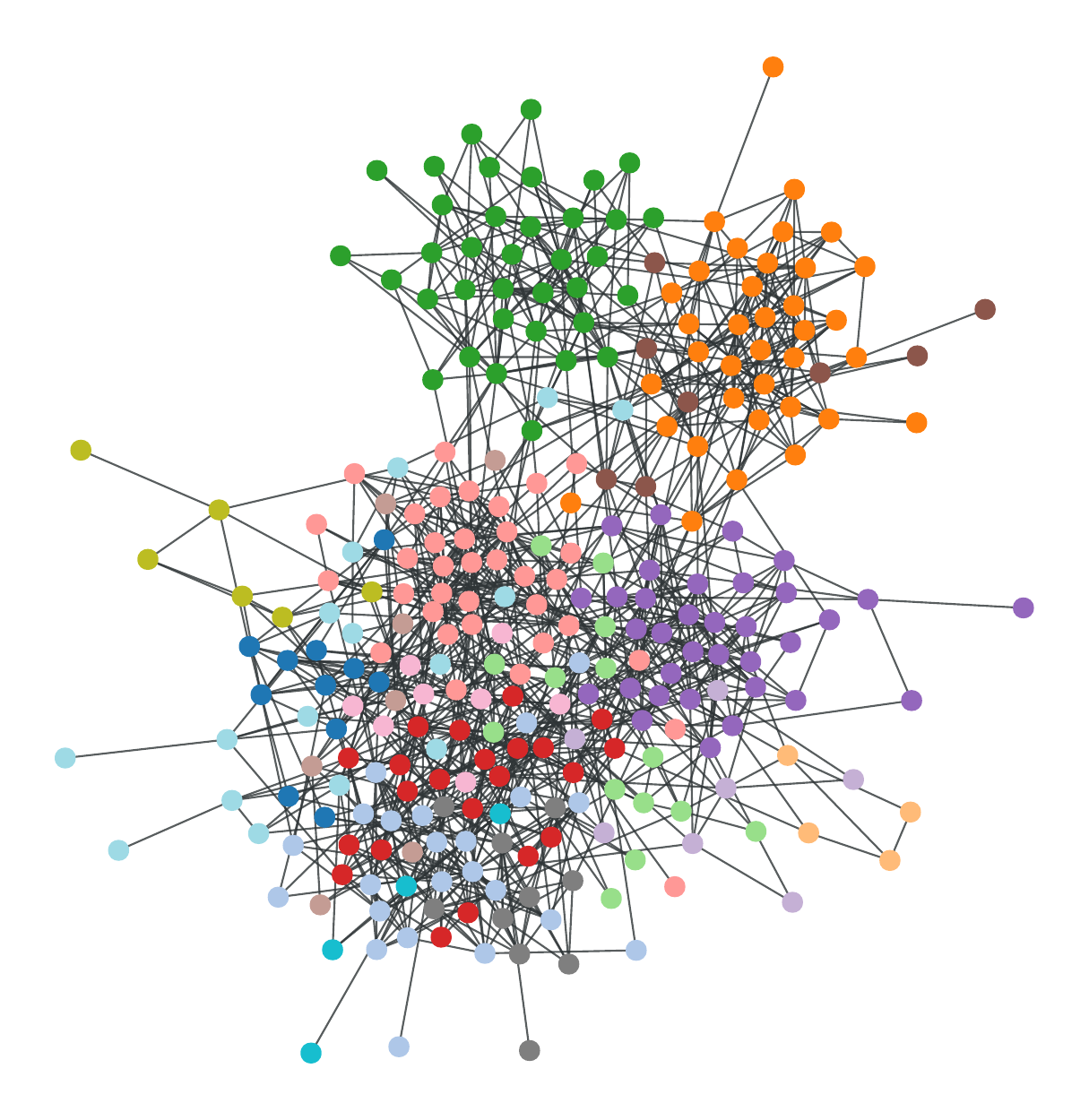}\\
    $\Sigma_{\text{NSBM}} = 4637$ nats &
    $\Sigma_{\text{Q}} = 4623$ nats\\[.5em]
    \multicolumn{2}{c}{(b) American college football ($\text{AMI}=0.94$)}\\
    \multicolumn{2}{c}{\smaller \texttt{(football)}}\\
    \includegraphics[width=.5\columnwidth]{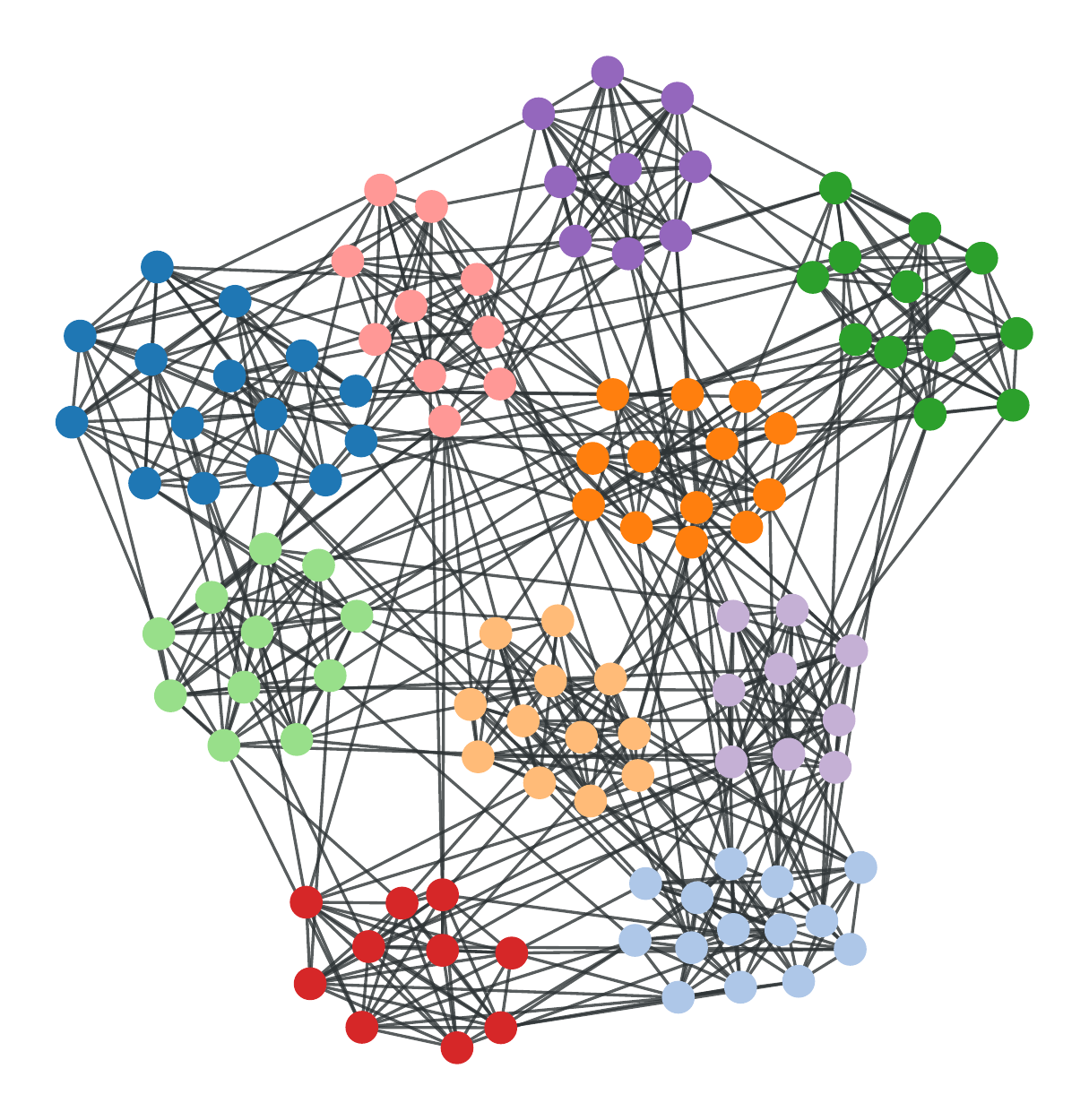}&
    \includegraphics[width=.5\columnwidth]{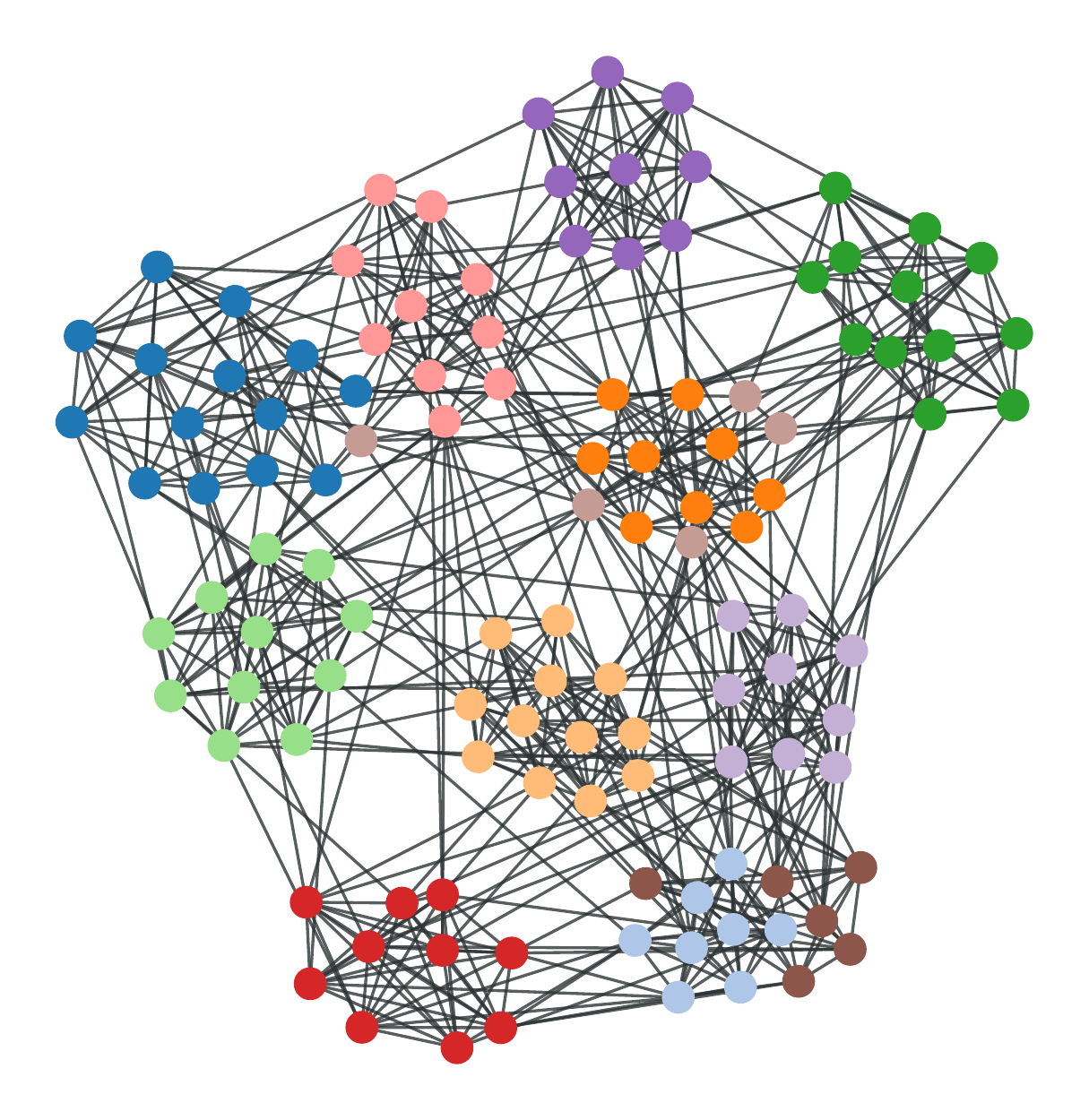}\\
    $\Sigma_{\text{NSBM}} = 1733$ nats &
    $\Sigma_{\text{I}} = 1479$ nats
  \end{tabular}

  \caption{Examples of exceptional networks where either modularity
  maximization or Infomap yield description length values, $\Sigma_Q$
  and $\Sigma_I$ respectively, that are smaller than what is obtained
  with any of the SBM variants. In all cases the partitions obtained are
  shown as node colors, and the adjusted mutual information (AMI)
  between them is given in the panel title, which also shows the
  corresponding Netzschleuder~\cite{peixoto_netzschleuder_2020} codename
  used in Fig.~\ref{fig:ami}.
  \label{fig:exceptions}}
\end{figure}

\begin{figure*}
  \includegraphics[width=.45\textwidth]{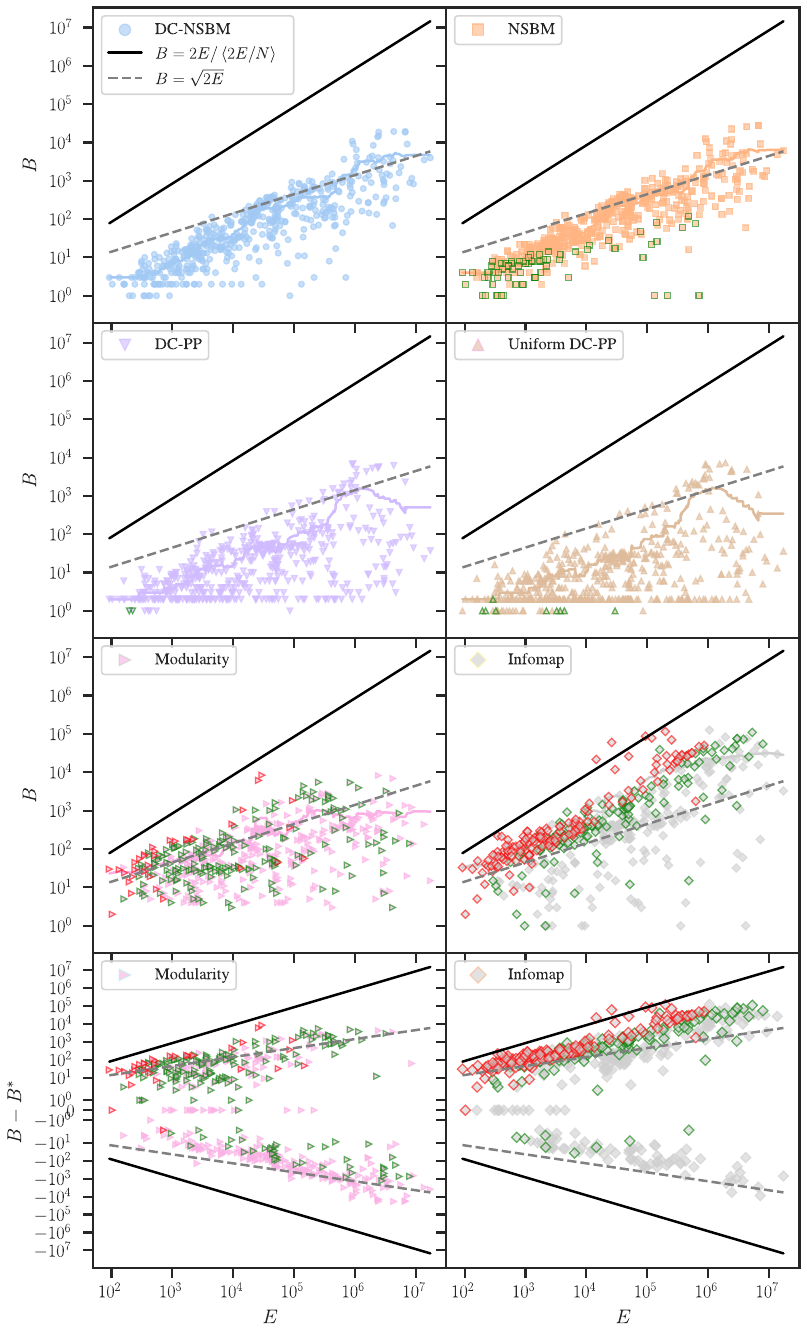}
  \includegraphics[width=.45\textwidth]{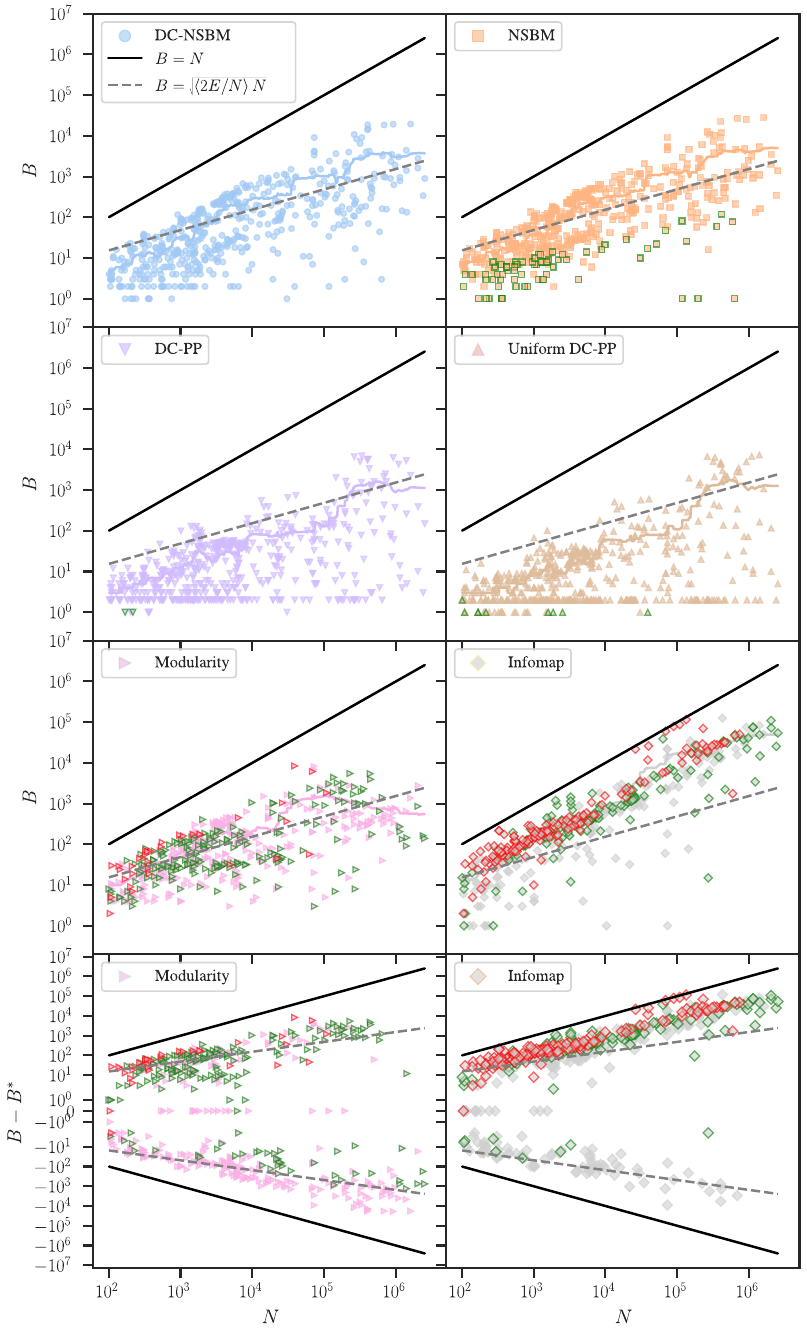}

  \caption{Number of groups $B$ as function of number of edges $E$ (left
  panel) and number of nodes $N$ (right panel) according to each method
  as indicated by the legend, for every network in our corpus. The
  symbols in red indicate partitions for which the description length
  value is larger than the Erd\H{o}s-Rényi model, and likewise for those
  in green for the configuration model. The solid lines correspond to
  moving averages, and the black solid and dashed lines are slopes as
  indicated in the legend. \label{fig:ngroups}}
\end{figure*}

The arguments above mean that we should expect that methods that are
optimal for general mixtures of models should perform just as well as
those that are specialized for any of the mixture components. However,
when encountering networks in the real world, we can confidently assume
that they are not in fact sampled from any model we can articulate
exactly --- even though it is often easy to determine that they are
structured (e.g. either via statistical tests designed to reject the
uniform null model, or simply by compressing it with any model). In
these structured ``out-of-distribution'' cases, we are, strictly
speaking, simultaneously out of scope of the NFL theorem and of the
situation considered previously, where the sample comes from one of the
models being considered.

Despite this, we should expect to be much closer to the scenario
considered in the previous section than that of the NFL theorem, as soon
as our models under considerations can serve as reasonable
approximations of the data~\cite{olhede_network_2014}. Here we test this
hypothesis on a corpus of 509 structurally diverse empirical networks,
from many domains of science, and across several orders of magnitude in
size, as summarized in Fig.~\ref{fig:empirical_desc}.

For each of these networks, we find the partition according to maximum
modularity, Infomap (see Appendix~\ref{app:infomap}), and well as
various versions of the SBM: the NSBM, its degree-corrected version
(DC-SBM), the non-uniform degree-corrected planted partition model
(DC-PP), and its uniform version~\cite{zhang_statistical_2020}. For
modularity and Infomap we then compute their implicit description
lengths, using also the degree-corrected alternatives. We also compute
the description length for the configuration and Erd\H{o}s-Rényi models
as baselines.

In Fig.~\ref{fig:empirical} we show for each model and network the
difference in description length according to the best model for each
network --- a value of zero thus means that the specific model is the
best one for that network.  We can see clear performance gains for the
SBM variants, with the DC-NSBM and the NSBM having the best compression
in the majority of cases, and the DC-PP also performing well, primarily
on small networks. As shown in Fig.~\ref{fig:domain}, this is also true
when each domain is considered separately --- with the exception of
transportation networks, where the DC-PP provides an improved
compression than the DC-NSBM for a larger fraction of cases.  Most cases
where other algorithms achieve the best compression are smaller
networks, which may be due to the fact that we have used lower bounds to
estimate the partition function for these alternative models, giving
them a slight advantage. Alternatively, the communication cost for more
complex models in these cases may outweigh the corresponding
improvements in fit to the data if these happen to be better described
by the more specific constraints of the implicit generative models of
either modularity or Infomap.

We compare the relative compression of the models in a different manner
in Fig.~\ref{fig:mcomp}, plotting the fraction of cases where a
given model $\mathcal{M}_1$ achieves equal or better compression than
the alternative model $\mathcal{M}_2$, as well as their average
compression ratio for all networks. Based on these pairwise comparisons
we ranked each model according to the
SpringRank~\cite{de_bacco_physical_2018} algorithm, which is reflected
in the ordering of Fig.~\ref{fig:mcomp}. We can also see here that the
SBM variants are much more compressive than the other algorithms, even
for the models that do not achieve the lowest overall compression. The
row of the heatmap labelled ``Best SBM'' takes the best compression
among all SBM variants for each network, which is almost completely
unmatched in its compression when compared to all other algorithms,
performing the worst relative to the degree-corrected modularity, where
superior compression is achieved for 96\% of the empirical networks. We
can also see that the configuration and Erd\H{o}s-Rényi models provide
superior compression to modularity and Infomap in a large fraction of
instances.  This inflated description in comparison to a maximally
random baseline indicates a massive amount of overfitting in the results
produced by these algorithms --- i.e. the structures found are better
justified by being the outcome of purely random fluctuations.

Although the SBM variations, and in particular the NSBM, provide a
description superior to the alternatives for the large majority of
networks considered, there are in fact a few exceptions where either
modularity or Infomap do provide a better description. As we discussed
in the previous section, this is expected when the networks are closer
to the typical ones generated by the implicit generative models of these
methods, which have a specific relationship between the number of groups
and the strength of the community structure. Note that we should not be
tempted to attribute the existence of these minority cases as a
necessary outcome of a supposed trade-off that comes as an unavoidable
consequence of the NFL theorem, as suggested in
Refs.~\cite{ghasemian_evaluating_2019,ghasemian_stacking_2020}. As
discussed previously, the NFL theorem is only valid when problem
instances are sampled uniformly at random, resulting almost exclusively
in incompressible networks --- a hypothesis that we can confidently
reject for all networks considered in our corpus.
Furthermore, even when considering the maximally uniform case, the
NFL theorem does not imply any actual trade-off, only that
all algorithms must perform equally poorly in the asymptotic totality of
instances. Besides, the negation that all algorithms perform equally well
when averaged over all cases does not necessarily imply that a single
algorithm must perform strictly better in all of them --- it would be
sufficient that some algorithms perform better than others on average,
precisely as our results and those of 
Refs.~\cite{ghasemian_evaluating_2019,ghasemian_stacking_2020} show.

Indeed, we can see evidence of a systematic hierarchy between
community detection algorithms when we compare the description lengths
with the actual partitions found. In Fig.~\ref{fig:ami} we show for
every network in our corpus the difference between the best description
length per node found with any version of the SBM and the one found with either
modularity or Infomap (the best from the degree-corrected and
non-degree-corrected versions) together with the adjusted mutual
information (AMI)~\cite{vinh_information_2009} between their respective
partitions. In both cases, we see that for networks where either
modularity or Infomap provide a better description (which are often
relatively small or very sparse networks), they yield partitions that
are very similar to the SBM inference. Examples of such instances can be
seen in Fig.~\ref{fig:exceptions}, where we can see that both methods
tend to agree substantially on the network divisions.

The fact that modularity and Infomap tend to agree with the SBM whenever
they yield compressive answers is also a statement about the partial
similarities between these algorithms. Indeed, as we argued previously,
both modularity and Infomap are approximately equivalent to the
inference of versions of the SBM with very particular constraints
imposed on its parameters. Therefore, neither algorithm can exploit
features in the network that deviate from the same underlying SBM
assumption. When we compare them a posteriori, we can only tell which
SBM parametrization is relatively better justified according to the
evidence in the data.

Clearly, the fact that modularity and Infomap amount to particular SBM
parametrizations should not be used as a justification for their use as
reliable inference methods. The implicit priors are so strongly
committed to particular patterns that they will be dredged out of pure
randomness, resulting in description lengths that are not only most of
the time significantly larger than the properly agnostic SBMs, but
very often even larger than maximally random networks.

We can investigate further the tendency of modularity and Infomap to
overfit by comparing how many groups are obtained with each method, as
shown in Fig.~\ref{fig:ngroups}. For modularity maximization, we can
observe its tendency of both overfit and underfit depending on the
circumstance, since most networks have a number of groups smaller than
the resolution limit, i.e. $B < \sqrt{2E}$ --- except those with more
than one component, where this limit does not apply. Despite this
limitation, a large fraction of the results are less compressive than
the maximally random baselines, indicating substantial overfitting. For
Infomap the overfitting is more extreme, with the number of groups found
scaling linearly with the number of nodes. This corresponds exactly to
the implicit prior for the number of groups in Infomap which strongly
prefers a characteristic group size that is independent of the number of
nodes, as shown in Figs.~\ref{fig:Lprior} and~\ref{fig:infomap_feasible}.

It is important to emphasize that even when the description lengths of 
modularity and Infomap are smaller than one of the maximally random
baselines, this does not necessarily mean that method is not
overfitting, since the partition found can still amount to a substantial
amount of randomness. We can assess this by comparing
the number of groups obtained with the model version that yields the
smallest description length, as shown in the bottom row of
Fig.~\ref{fig:ngroups}. Indeed we can see that modularity tends to both
under- and overfit for a comparable fraction of the networks, although
the larger tendency is to underfit, while with Infomap the overwhelming
tendency is to overfit, and return a much larger number of groups than
the most compressive partition.

We observe also that all SBM flavors manage to find a number of groups
in a range that does not necessarily conform to a $\sqrt{E}$ or
$\sqrt{N}$ scaling --- a lack of constraint that is theoretically
prescribed~\cite{peixoto_hierarchical_2014,zhang_statistical_2020}. This
dispels the notion that this scaling is a fundamental limitation of
community detection methods in general, as suggested in Ref.~\cite{ghasemian_evaluating_2019}.
Importantly, this lack of resolution limit of the NSBM and PP models
comes together with a regularization against overfitting, unlike what we
observe for Infomap.

\section{Discussion}\label{sec:conclusion}

In this paper we have presented a framework for identifying the implicit
generative model associated with an arbitrary community detection
algorithm, allowing us to compare descriptive and inferential methods on
the same scale by computing their associated description lengths for a
network and corresponding partition. This method also allows us to
compute the implicit priors on the objective value and number of groups
associated with a community detection objective, giving insights into
the intrinsic biases in existing algorithms. We demonstrate the use of
our method for the widely used modularity and Infomap objectives,
showing that they are biased towards overfitting due to strong priors
favoring high objective values and a large number of groups. We also
find that the implicit models for a wide range of methods, including
modularity and Infomap, correspond asymptotically to restricted
instances of the stochastic block model (SBM). By exploiting the latent
compression associated with community detection algorithms, we were able
to compare these methods on real and synthetic data, demonstrating that
in these structured problem instances certain algorithms (more
expressive variants of the SBM) are systematically favored over others
(variants of modularity and Infomap).

Since it provides a universal scale on which we can assess the capacity of
a model to capture structural regularities in network data, the
description length provides a principled measure to compare the
performance of community detection algorithms without the need for
``ground truth'' labels --- unknowable information for
empirical networks~\cite{peel_ground_2017}. The empirical experiments
here show that by evaluating algorithms using this measure we can reveal
a clear breakdown of the implications of the NFL theorem for real,
structured problem instances. This weakens the practical and conceptual
pertinence of the NFL theorem, which equates all possible community
detection algorithms in terms of performance, but applies only to
unstructured problem instances.

Part of the results in this work confirm what has been found by
Ghasemian et al~\cite{ghasemian_evaluating_2019} with respect to
modularity maximization and Infomap overfitting in a link prediction
task for a diverse set of smaller networks, and the regularized SBM
performing better on average (although
Ref.~\cite{ghasemian_evaluating_2019} omitted the NSBM, which can be
shown to perform strictly better, and substantially so for larger
networks~\cite{peixoto_nonparametric_2017}). This is not unexpected, since
it is known that algorithmic learning procedures where the objective is
to obtain a succinct representation of data (called broadly ``Occam learning''
in the machine learning theory literature) are in general equivalent
to learning procedures where the objective is to choose a predictive model
with low generalization error [known as ``probably approximately
correct'' (PAC) learning]~\cite{kearns_introduction_1994}. Because of
this, we can in principle expect a MDL approach to yield compatible
results with link prediction in suitable limits. However, there is an
important caveat that prevents this equivalence from being exact; namely, the
nominal task would correspond to predicting an entire new network from
past observations of a complete network. Instead, in a more realistic
link prediction scenario one attempts to predict a subset of the
possible edges by observing the remaining network, which is commonly
sparse. In this situation we cannot guarantee that a sufficient data
limit exists, regardless of how large the network is --- the removal
of a fraction of the edges always destroys important information which
could be used to improve the detection of the community labels. Because
of this, discrepancies between both approaches can exist, with link
prediction having a tendency to
overfit when used as a model selection criterion~\cite{valles-catala_consistencies_2018}. Therefore, the results
we present in this work have a more definitive character than those of
Ref.~\cite{ghasemian_evaluating_2019}, since ours make use of the whole
data.

We have applied our method for analytically computing the description length of
modularity and Infomap by exploiting the fact that their objective
functions can be written in terms of the microcanonical SBM
parameters. Our analytical calculations are possible for a much wider set of
objective functions that can also be described in the same manner. We
speculate that a significant fraction of community detection algorithms
proposed in the literature are, like Infomap and modularity, also
equivalent to the inference of constrained versions of the SBM, as has
been suggested by others~\cite{young_universality_2018}. Objective functions
that cannot be written in terms of the microcanonical SBM require different analytical approaches than those we considered in this work, or at the very least can be treated numerically. However, it is conceivable that even these kinds of objectives amount to generative models that are well approximated by particular SBM parametrizations. This would have wider implications for the general nature of SBM-based approaches, and the systematic superiority of their nonparametric formulations. In case particular objectives yield implicit models that deviate significantly from the SBM class, this could be used to formulate a broader unified family of community detection methods. We view the task of a broad unification of community detection methods within an inferential framework as a promising avenue of future research.


\appendix

\section{Robustness of description length to quality function transformations}\label{app:g_x}

As described in the main text, for any quality function $W(\A,\bb)$, we
can attribute a generative model given by
\begin{equation}
  P(\A,\bb|g,f) = \frac{\ee^{g(W(\A,\bb)) + f(\A)}}{Z(g,f)},
\end{equation}
with $Z(g,f)=\sum_{\A,\bb}\ee^{g(W(\A,\bb))+f(\A)}$, and where $g(x)$ is
any strictly increasing function, and $f(\A)$ is an arbitrary weight
attributed to a given network, independent of how its nodes are
partitioned. For any choice of $g(x)$ and $f(\A)$, the
maximum-a-posteriori (MAP) estimate of a partition $\bb$ is equivalent
to the maximization of the quality function $W(\A,\bb)$. In the main
text we used the maximum entropy ansatz to justify the choices $g(x) =
\beta x$, and $f(\A)=0$. This ansatz is well justified, since it
corresponds to a maximum ignorance about modelling aspects that are not
directly specified by the quality function. This is specially true for
the choice of $f(\A)$, which amounts to an arbitrary suppression of
networks independently of how they are partitioned, as discussed in the
main text. However, we may wonder if other choices of $g(x)$ could in
principle have a strong effect in the obtained description lengths,
resulting in compatible generative models that are more favorable to
compression --- and hence more plausible --- than the one based on the
maximum entropy assumption. Here we show that our results are in fact
invariant to any other choice of $g(x)$, and that meaningful compression
can only be achieved by actually changing the quality function
$W(\A,\bb)$ in nontrivial ways.

In the case $f(\A)=0$, without loss of generality $g(x)\to\beta g(x)$ we
can write
\begin{equation}
  P(\A,\bb|\beta, g) = \frac{\ee^{\beta g(W(\A,\bb))}}{Z(\beta)},
\end{equation}
with $Z(\beta)=\sum_{\A',\bb'}\ee^{\beta g(W(\A',\bb'))}$. In this case,
the description length is given by
\begin{equation}
  \Sigma(\A, \bb) = \min_{\beta}\; -\beta g(W(\A,\bb)) + \ln Z(\beta).\label{eq:dl_g}
\end{equation}
We can decompose
\begin{equation}
   Z(\beta) = \int \ee^{\beta g(W)}\Xi(g(W))\,\dd g(W),
\end{equation}
with $\Xi(g(W))$ being the density of states,
\begin{equation}
  \Xi(g(W)) = \sum_{\A, \bb}\delta(g(W(\A,\bb)) - g(W)),
\end{equation}
which counts how many configurations have a particular value of
$g(W(\A,\bb))$.  Since the function $g(x)$ is strictly increasing and
hence invertible, it cannot affect the density of states other than via
a scaling, i.e.
\begin{align}
  \Xi(g(W)) = \frac{\bar\Xi(W)}{g'(W)},
\end{align}
where we must have $g'(W) > 0$, and $\bar\Xi(W)$ is the density of
states for $W(\A,\bb)$, with
\begin{equation}
  \bar\Xi(W) = \sum_{\A, \bb}\delta(W(\A,\bb) - W).
\end{equation}
(The term $g'(W)$ comes from the scaling of Dirac's delta,
i.e. $\delta(h(x)) = \delta(x-x_0)/|h'(x_0)|$, where $x_0$ is the root
of $h(x)$.) Based on this, we can write
\begin{align}
  Z(\beta) &= \int \ee^{\beta g(W)}\frac{\bar\Xi(W)}{g'(W)}\,\dd g(W),\\
           &= \int \ee^{\beta g(W)}\bar\Xi(W)\,\dd W.
\end{align}
In general, we have that the entropic density $\ln\bar\Xi(W)$ is
extensive, i.e.  $\ln\bar\Xi(W) = O(N\ln N)$, which means we can use the
Laplace approximation for $N \gg 1$,
\begin{equation}
  Z(\beta) \approx \sqrt{\frac{2\pi}{|\Delta^*|}} \ee ^{\beta g(W^*) + \ln\bar\Xi(W^*)},
\end{equation}
with $\Delta^* = \frac{\partial^2}{\partial W^2}[\beta g(W) + \ln\bar\Xi(W)]|_{W=W^*}$, and
\begin{equation}
  W^* = \underset{W}{\operatorname{arg\, max}}\; \beta g(W) + \ln\bar\Xi(W).
\end{equation}
From this we have
\begin{equation}
  \ln Z(\beta) = \max_{W}\; \beta g(W) + \ln\bar\Xi(W) + O(\ln N),
\end{equation}
which we use to obtain the following asymptotic value for the
description length,
\begin{equation}
  \Sigma(\A, \bb) \approx \min_{\beta}\max_{W}\; -\beta g(W(\A,\bb)) + \beta g(W) + \ln\bar\Xi(W).
\end{equation}
Setting derivatives with respect to $\beta$ and $W$ to zero, we
find that the saddle point is obtained for
\begin{align}
  g(W) &= g(W(\A,\bb)) \\
  \beta &= \frac{\bar\Xi'(W)}{g'(W)\bar\Xi(W)}.
\end{align}
Using the fact that $g(x)$ is invertible, the first equation above
corresponds to $W=W(\A,\bb)$. Substituting this in
the above we have
\begin{equation}
  \Sigma(\A, \bb) \approx \ln\bar\Xi(W(\A,\bb)). \label{eq:dl_micro}
\end{equation}
The important conclusion from Eq.~\ref{eq:dl_micro} is that the actual
value of the description length is \emph{completely independent} of the
function $g(x)$ as long as it is strictly increasing. For any valid
choice of $g(x)$, the description length will approach the entropy
density of the quality function at the value given by $\A$ and
$\bb$. Therefore, compression cannot be achieved by arbitrary
transformations of the quality function which preserve the same
optimization problem.

Furthermore, Eq.~\ref{eq:dl_micro} corresponds to an asymptotic
equivalence to the microcanonical ensemble where only instances with a
particular value of the quality function are allowed, i.e.
\begin{equation}
  P(\A,\bb | W) = \frac{\delta_{W(\A,\bb),W}}{Z(W)},
\end{equation}
where $Z(W) = \sum_{\A,\bb}\delta_{W(\A,\bb),W} \to \Xi(W)$ for
$N\to\infty$. Clearly, the microcanonical ensemble is completely
invariant to transformations $W(\A,\bb)\to g(W(\A,\bb))$ with $g(x)$
strictly increasing, since we must always have
$\delta_{W(\A,\bb),W}=\delta_{g(W(\A,\bb)),g(W)}$ in this case.

\section{Maximum entropy favors the planted partition model}\label{app:max-ent}

Here we demonstrate that for a certain class of objective functions that
include both modularity and Infomap, the maximum entropy ensemble for a
fixed value of the quality function corresponds to the planted partition
model with uniform group sizes.

We begin by writing the SBM entropy~\cite{peixoto_entropy_2012} for
$n_r\gg 1$ and $e_{rs} \ll n_rn_s$,
\begin{multline}
  \ln \Omega(\bm e,\bm n, B) \approx
  -\frac{1}{2}\sum_{rs}e_{rs}\ln \frac{e_{rs}}{n_rn_s} + \frac{1}{2}\sum_{rs}e_{rs}\\
  + N\ln N - \sum_r n_r \ln n_r.
\end{multline}
We wish to maximize the entropy while enforcing the constraints
\begin{align}
  \sum_rn_r &= N\\
  \sum_{rs}e_{rs} &= 2E\\
  W(\bm e,\bm n) &= W^*,
\end{align}
with Lagrange multipliers $\lambda$, $\mu$, and $\gamma$, respectively, i.e.
\begin{equation}
  \Lambda = \ln \Omega(\bm e,\bm n, B) - \lambda\sum_rn_r -\mu\sum_{rs}e_{rs}-\gamma W(\bm e, \bm n).
\end{equation}
Taking $\partial \Lambda/\partial n_r = \partial \Lambda/\partial e_{rs}
= 0$, we have
\begin{align}
    n_r &= \exp\left[\frac{e_r}{n_r} - \gamma \frac{\partial W(\bm e, \bm n)}{\partial n_r} -1 - \lambda \right] \label{eq:max_ent_n}\\
    e_{rs} &= n_rn_s\exp\left[-\gamma(1+\delta_{rs})\frac{\partial W(\bm e, \bm n)}{\partial e_{rs}}-2\mu\right].\label{eq:max_ent_e}
\end{align}
Now let us consider the special case where the quality function can be written as
\begin{equation}
  W(\bm e, \bm n) = f(\textstyle\sum_re_{rr}) + \displaystyle\sum_rg(e_r, e_{rr}),\label{eq:w_class}
\end{equation}
for some $f(x)$ and $g(x,y)$ --- this is precisely the case for both
modularity and Infomap (see Sec.~\ref{app:infomap}). In this situation we have $\partial W(\bm e, \bm
n)/\partial n_r = 0$ and
\begin{multline}
  \frac{\partial W(\bm e, \bm n)}{\partial e_{rs}} =\\
  \begin{cases}
    f'(\sum_re_{rr}) + g'_x(e_r,e_{rr}) + g'_y(e_r,e_{rr}) &\text{ if } r = s,\\
    g'_x(e_r,e_{rr}) + g'_x(e_s,e_{ss}) &\text{ if } r \neq s,
  \end{cases}
\end{multline}
where we used the shorthand notation $g'_x(x,y)=\partial
g(x',y')/\partial x'|_{x'=x,y'=y}$, and which substituting in
Eqs.~\ref{eq:max_ent_n} and~\ref{eq:max_ent_e} allows us to find the
planted partition solution
\begin{align}
  n_r &= \frac{N}{B},\label{eq:pp_n}\\
  e_{rs} &= \frac{2E_{\text{in}}}{B}\delta_{rs} + \frac{2(E - E_{\text{in}})}{B(B-1)}(1-\delta_{rs}),\label{eq:pp_e}
\end{align}
where $E_{\text{in}}$ is the solution of
\begin{equation}
  E_{\text{in}} = \frac{E-E_{\text{in}}}{B-1}\exp\left\{-2\gamma\left[f'(2E_{\text{in}}) + g'_y(2E/B,2E_{\text{in}}/B)\right]\right\},
\end{equation}
with $\gamma$ chosen so that $W(\bm e, \bm n) = W^*$. The above
calculations can be repeated using the degree-corrected ensemble, for
which the same result is obtained. Therefore, for any quality function
that can be written as Eq.~\ref{eq:w_class}, the corresponding maximum
entropy ensemble amounts to a particular SBM given by Eqs.~\ref{eq:pp_n}
and~\ref{eq:pp_e}.

\begin{figure*}
  \begin{tabular}{ccc}
  $\avg{k}=2$ & $\avg{k}=5$ & $\avg{k}=10$ \\
  \includegraphics[width=.3\textwidth]{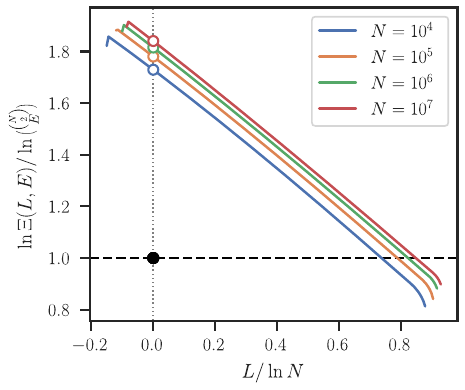} &
  \includegraphics[width=.3\textwidth]{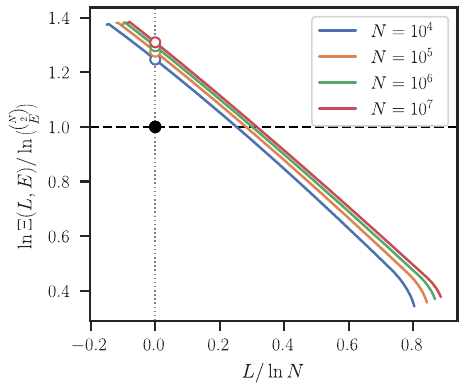} &
  \includegraphics[width=.3\textwidth]{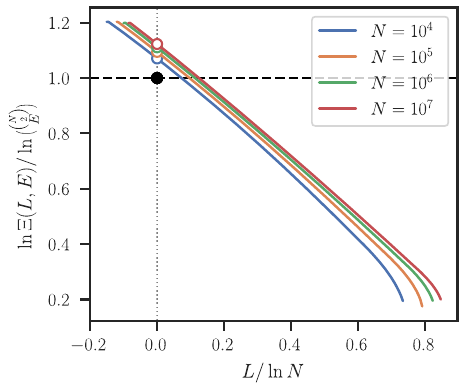}\\
  \includegraphics[width=.3\textwidth]{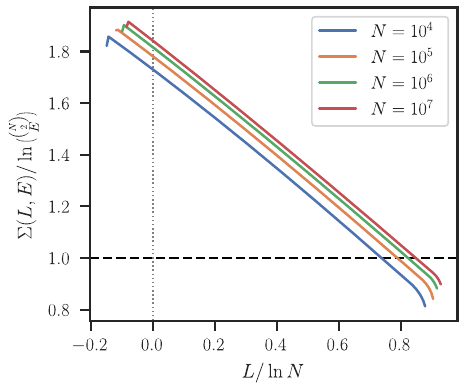} &
  \includegraphics[width=.3\textwidth]{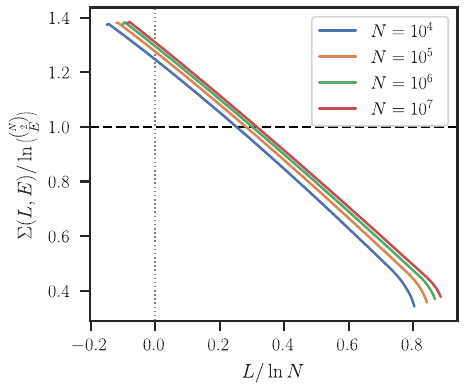}&
  \includegraphics[width=.3\textwidth]{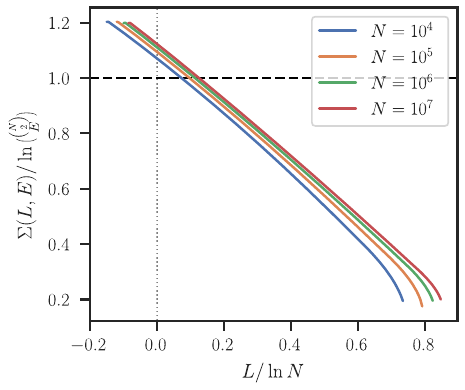}
  \end{tabular}
  \caption{Density of states $\Xi(L, E)$ (top row)
  and description length $\Sigma(L, E)$ (bottom row) as a
  function of value of Infomap score $L$, for different number of nodes $N$
  and average degree values, $\avg{k}=2, 5, 10$, from left to right. The
  values are shown relative to the ER baseline. \label{fig:Ldl}}
\end{figure*}

\begin{figure}
  \begin{tabular}{cc}
    \begin{overpic}[width=.5\columnwidth]{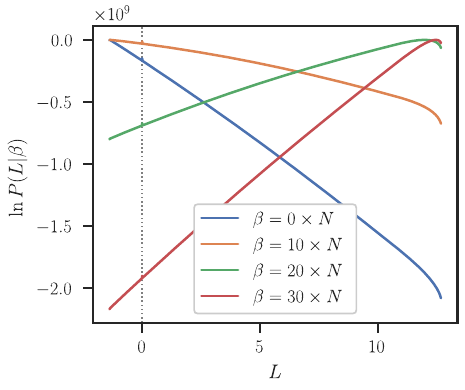}
      \put(0, 86){\smaller (a)}
    \end{overpic} &
    \begin{overpic}[width=.5\columnwidth]{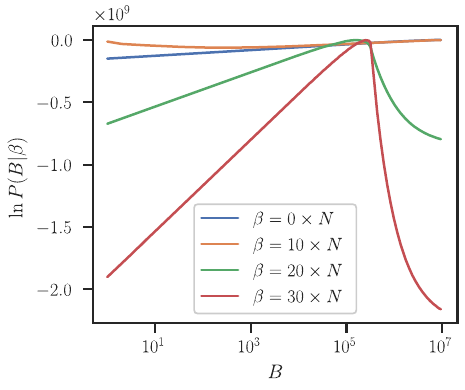}
      \put(0, 86){\smaller (b)}
    \end{overpic} \\
    \begin{overpic}[width=.5\columnwidth]{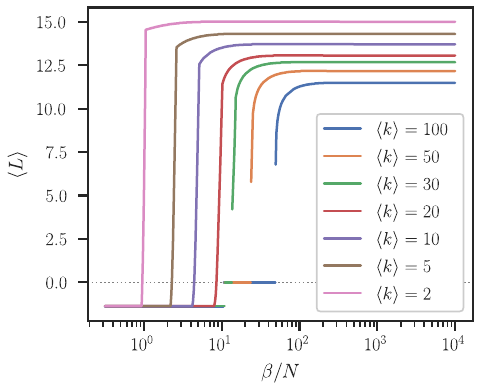}
      \put(0, 86){\smaller (c)}
    \end{overpic} &
    \begin{overpic}[width=.5\columnwidth]{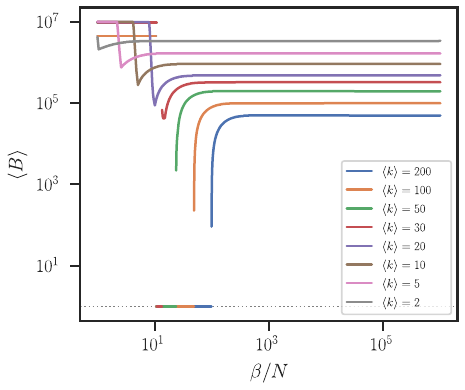}
      \put(0, 86){\smaller (d)}
    \end{overpic} \\
    \begin{overpic}[width=.5\columnwidth]{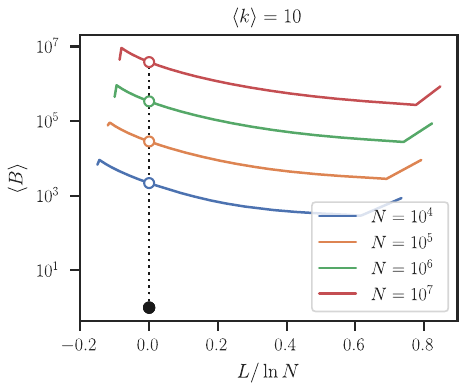}
      \put(0, 86){\smaller (e)}
    \end{overpic} &
    \begin{overpic}[width=.5\columnwidth]{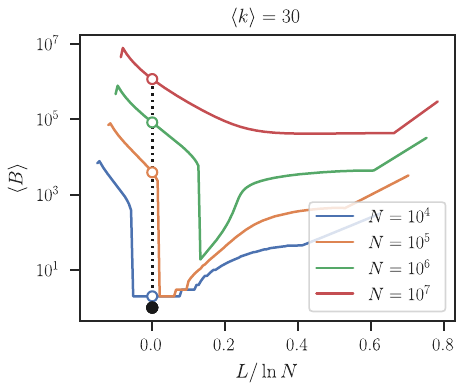}
      \put(0, 86){\smaller (f)}
    \end{overpic} \\
    \begin{overpic}[width=.5\columnwidth]{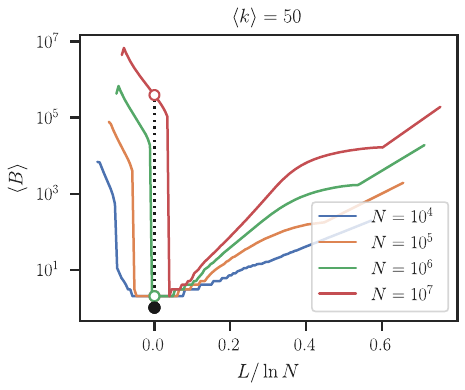}
      \put(0, 86){\smaller (g)}
    \end{overpic} &
    \begin{overpic}[width=.5\columnwidth]{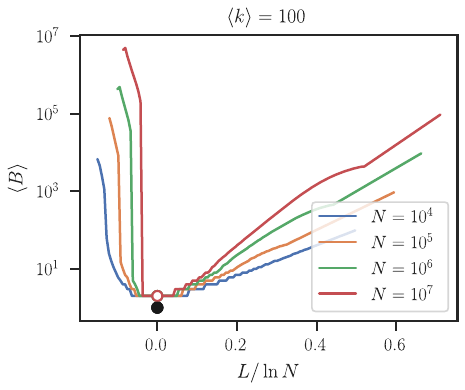}
      \put(0, 86){\smaller (h)}
    \end{overpic}
  \end{tabular}
    \caption{Top row: Implicit prior distribution for the
  value of Infomap quality function $L$ (a), and number of groups $B$
  (b), for different values of $\beta$, $N=10^7$ and
  $\avg{k}=30$. Second row: Average values of $L$ (c) and $B$ (d) as a
  function of $\beta$, and different average degrees $\avg{k}$. Bottom
  rows (e) to (h): Average value of $B$ as a function of $L$ for
  different values of $N$ and $\avg{k}$.\label{fig:Lprior}}
\end{figure}

\begin{figure}
  \begin{tabular}{cc}
    \includegraphics[width=.5\columnwidth]{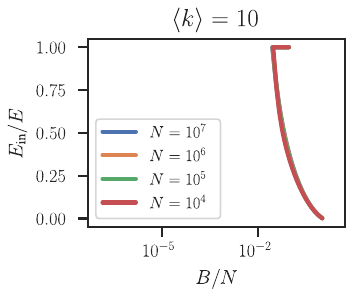}&
    \includegraphics[width=.5\columnwidth]{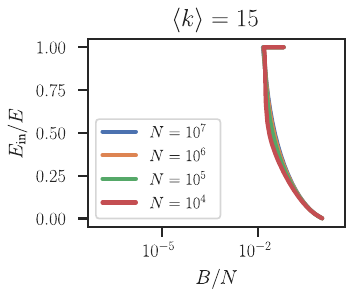}\\
    \includegraphics[width=.5\columnwidth]{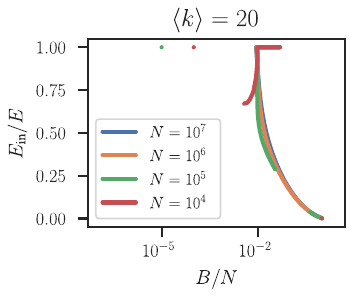}&
    \includegraphics[width=.5\columnwidth]{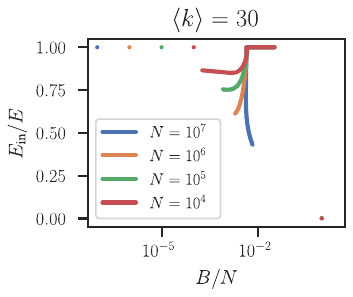}
  \end{tabular} \caption{Feasible realizations of the Infomap
  model. Each curve corresponds to the $(E_{\text{in}}, B)$ values
  achieved with $\beta$ in the range $[0,\infty]$ for a specific value
  of $N$, as indicated in the legend, and various values of the average
  degree $\avg{k}=2E/N$. The isolated points correspond to discontinuous
  transitions both for high and low values of $\beta$.
  \label{fig:infomap_feasible}}
\end{figure}

\section{The Infomap objective}\label{app:infomap}
The Infomap quality function~\cite{rosvall_maps_2008} is given by
\begin{multline}\label{eq:imap-objective}
  L(\A, \bb) = -\left(1-\sum_r\frac{e_{rr}}{2E}\right)\ln \left(1-\sum_r\frac{e_{rr}}{2E}\right) \\
  + 2\sum_r\frac{e_r-e_{rr}}{2E}\ln\left(\frac{e_r-e_{rr}}{2E}\right) - H(\bm{k})\\
  -\sum_r\left(\frac{2e_r-e_{rr}}{2E}\right)\ln \left(\frac{2e_r-e_{rr}}{2E}\right),
\end{multline}
where $H(\bm{k})$ is the entropy of the normalized degree
distribution. (We have flipped the sign so that the optimal partition is obtained through maximizing the objective, consistent with Eq.~\ref{eq:Wopt}.) Since it only depends on $\bm{A}$ but not $\bm{b}$, we can
ignore this degree entropy term, since it will disappear when doing the
normalization, to obtain an objective that only depends on the SBM
parameters.

The computation of the density of states is analogous to modularity,
since in this case the density of states will also favor uniform group
sizes and density (see Sec.~\ref{app:max-ent}),
with the only difference that for the planted partition we have:
\begin{multline}
\label{eq:imapppm}
L(E, E_{\text{in}}, B) = -\frac{E-E_{\text{in}}}{E}\ln \frac{E-E_{\text{in}}}{E}\\
+ 2\frac{E-E_{\text{in}}}{E}\ln \frac{E-E_{\text{in}}}{EB}\\
- \frac{2E-E_{\text{in}}}{E}\ln\frac{2E-E_{\text{in}}}{EB}.
\end{multline}
Following the same procedure as before, we can now invert
Eq.~\ref{eq:imapppm} for $E_{\text{in}}$ to give $E_{\text{in}}(E,L,B)$,
which can be inserted into either Eq.~\ref{eq:omega} or
Eq.~\ref{eq:omega_dc} to obtain the description length according to
Eq.~\ref{eq:W_dl}. Unlike the modularity function, the inversion of
Eq.~\ref{eq:imapppm} cannot be done in closed form, so it needs to be
performed numerically.

Fig.~\ref{fig:Ldl} shows the density of states and description length
values as a function of $L$. Unlike modularity, the relationship between
these two quantities is almost linear. In Fig.~\ref{fig:Lprior} we see
the implicit priors for $L$ and the number of groups $B$ --- we also
observe a transition from low to high values with $\beta$, which
although abrupt is continuous, unlike what is obtained for
modularity. In the case of Infomap, what is noteworthy is a qualitative
dependence on the network density --- only if the average degree is
sufficiently large does the prior for $B$ allows for finite values, as
seen in Fig.~\ref{fig:Lprior}d, otherwise the mean is always at
$\left<B\right>=O(N)$ (the precise value of $\avg{k}$ at which this
transition happens is size-dependent). This transition is reflected in
the expected value of $B$ as a function of $L$, which displays a minimum
at $L=0$ only for sufficiently dense networks, besides a discontinuity
at $L=0$, since for this value only a partition in $B=1$ groups is
allowed. This overall picture is entirely consistent with the observed
tendency of the method to find spurious groups in fully random networks
whenever they are sufficiently
sparse~\cite{lancichinetti_community_2009,kawamoto_comparative_2018}.

In terms of the optimal problem instances, for Infomap the situation is
comparable to modularity (see Sec.~\ref{sec:optimal-instances}). As we
show in Fig.~\ref{fig:infomap_feasible}, the lack of an additional
parameter analogous to the resolution $\gamma$ of modularity means that
the value of $\beta$ can only select values on a line in the
$(E_{\text{in}},B)$ plane. We can observe two regimes:
1. For sufficiently sparse networks, although a wide range of $E_{\text{in}}$ can be reached, we
cannot meaningfully talk about an undetectable regime because $B$ is
proportional to $N$ --- all instances are easy; 2. For denser networks,
a discontinuous transition is observed between a
$(E_{\text{in}},B)=(E,1)$ value and another range of values far away
from the detectability transition. The transition between these regimes
is size dependent, such that as the number of nodes increases, then even
denser networks are required for the transition between the above two
regimes to be seen.

\section{Objectives dependent on SBM parameters}\label{app:objectives}

In this Appendix we discuss the application of our calculations to other
community detection objectives and replicate the analysis of
Sec.s~\ref{sec:generative} and~\ref{sec:optimal-instances} for the
Infomap objective.

The inferential framework presented in this paper is applicable to any
community detection method, whether or not it depends on an explicit
objective function $W(\bm e, \bm n)$ that can be written as a function
of the parameters $\{\bm e, \bm n\}$ of the microcanonical SBM (or ${\bm
e, \bm n, \bm k}$ for the degree-corrected case), but the calculations
involved for these methods may be more demanding (see
Sec~\ref{sec:generative} and Appendix~\ref{app:others} for a
discussion). However, the analytical method employed in this paper to
estimate the description length of modularity is directly applicable to
any community detection method with an objective function of the form
$W(\bm e, \bm n)$, of which there are many: Besides generalized
modularity~\cite{reichardt_statistical_2006} and
Infomap~\cite{rosvall_maps_2008}, this covers also
surprise~\cite{aldecoa_deciphering_2011}, coverage
significance~\cite{gaertler2007significance}, performance
significance~\cite{ohkubo2006nonadditive}, $q$-state Potts
model~\cite{reichardt2004detecting},
significance~\cite{traag2013significant},
conductance~\cite{leskovec_empirical_2010}, and
OSLOM~\cite{lancichinetti_finding_2011}. Degree corrected variants of
any such method can also be directly cast into our framework, using the
density of states for the microcanonical degree-corrected SBM
(Eq.~\ref{eq:DCSBM-dos}). Additionally, objectives without a model
selection mechanism for the number of clusters (which constitute a large
portion of existing objectives) can be cast into our framework by simply
restricting the set of allowed partitions $\bb$ in the partition
function $Z(\beta)$ to be of a particular size $B$, or by using the
approximation in Eq.~\ref{eq:dl-approx-logXi}.

We also emphasize that the majority of community detection algorithms
used in practice aim to maximize modularity or its generalized
form~\cite{fortunato_community_2010}, which is easily accommodated
within our framework as discussed in the main manuscript. These include
methods based on greedy algorithms, hierarchical clustering, simulated
annealing, extremal optimization, spectral optimization, genetic
algorithms, and quadratic
programming~\cite{fortunato_community_2010}. These algorithms may differ
in their final obtained value for $W$, the first term in the description
length of Eq.~\ref{eq:dl-general}, but the partition function of the
second term in Eq.~\ref{eq:dl-general} will remain the same as it only
depends on the modularity objective $W$. One can then see that optimal
compression among modularity maximizing algorithms is obtained by
whichever method returns the highest value of the modularity for a given
network.

\section{Other types of community detection methods}\label{app:others}

Some community detection methods available in the literature are neither
deterministic nor rely on the optimization of any quality function. A good
example of this is the label propagation algorithm~\cite{raghavan_near_2007},
defined as a the result of a dynamical process: Given an initial labelling of
the nodes into groups (usually each node in its own group), one proceeds by
updating the labels of each node in random sequence by the value corresponding
to the majority of its neighbors (with ties resolved uniformly at random). Once
a fixed point is reached, the algorithm stops.

This type of algorithm can also be cast into our inferential framework without
any problems, since it direcly defines a posterior distribution of partitions,
\begin{equation}
  P(\bb | \A) = w_{\bb,\A},
\end{equation}
where $w_{\bb,\A}$ is the frequency with which partition $\bb$ is the
output of the algorithm for network $\A$. This posterior distribution is
equivalent to the one obtained with a generative model given by
\begin{equation}
  P(\A, \bb) = \frac{w_{\bb,\A}}{\sum_{\A',\bb'}w_{\bb',\A'}},
\end{equation}
which has a description length given by
\begin{equation}
  \Sigma(\A, \bb) = -\ln w_{\A,\bb} + \ln \sum_{\A',\bb'}w_{\bb',\A'}.
\end{equation}
Therefore, it is possible to extend our analysis to this class of
problems as well, as long as $w_{\A,\bb}$ can be reliably estimated.

The analytical tools and numerical methods required to estimate these
description lengths are different from the ones considered in this work,
and are arguably more technically demanding. Nevertheless, a possible
avenue of future work is to find efficient ways to estimate these
quantities in practical settings.

\bibliography{bib,bib2}

\end{document}